\documentclass[preprint,12pt]{elsarticle}
\usepackage{amssymb}
\usepackage{bm}

\usepackage{lineno}
\modulolinenumbers[5]

\usepackage{hyperref}

\hypersetup{
  colorlinks=true,
  linkcolor=blue,
  citecolor=blue,
  urlcolor=blue,
}

\usepackage{multicol} 

\usepackage{framed}
\usepackage{multirow}

\usepackage{geometry}

\usepackage{amsmath}

\usepackage[normalem]{ulem}

\usepackage{color} 
\usepackage[usenames,dvipsnames,table]{xcolor}

\DeclareRobustCommand{\Eraser}{\bgroup\markoverwith{\textcolor{Red}{\rule[.5ex]{2pt}{0.4pt}}}\ULon}
\DeclareRobustCommand{\Eraseg}{\bgroup\markoverwith{\textcolor{PineGreen}{\rule[.5ex]{2pt}{0.4pt}}}\ULon}
\DeclareRobustCommand{\Eraseb}{\bgroup\markoverwith{\textcolor{Cyan}{\rule[.5ex]{2pt}{0.4pt}}}\ULon}
\DeclareRobustCommand{\Eraseo}{\bgroup\markoverwith{\textcolor{Orange}{\rule[.5ex]{2pt}{0.4pt}}}\ULon}

\makeatletter
\@addtoreset{equation}{section}
\makeatother
\graphicspath{{../figure}}

\allowdisplaybreaks

\begin{document}

\begin{frontmatter}

%------------------------------------------------------------------------------
\title{
Numerical method for the magnetic vector potential in incompressible magnetohydrodynamic flows 
and the conservation properties of magnetic helicity
}
%------------------------------------------------------------------------------

%------------------------------------------------------------------------------
%% Authors per affiliation:
\author[mymainaddress]{Hideki Yanaoka\corref{mycorrespondingauthor}}
\cortext[mycorrespondingauthor]{Corresponding author}
\ead{yanaoka@iwate-u.ac.jp}

%% Address:
\address[mymainaddress]{Department of Systems Innovation Engineering, 
Faculty of Science and Engineering, Iwate University, 
4-3-5 Ueda, Morioka, Iwate 020-8551, Japan}

%------------------------------------------------------------------------------
%% Abstract
\begin{abstract}
Analyzing magnetohydrodynamic (MHD) flows requires accurate predictions of 
the Lorentz force and energy conversion. 
Total energy, cross-helicity, and magnetic helicity can be used 
to investigate energy conservation properties in inviscid MHD flows. 
However, the conservation property of magnetic helicity has not been fully clarified 
using the magnetic vector potential equation. 
This study presents a numerical method to simultaneously relax 
magnetic vector and electric potentials for incompressible MHD flows 
using a conservative finite difference scheme 
that discretely conserves total energy. 
First, it was proven that the transport equations of total energy, cross-helicity, 
and magnetic helicity can be discretely derived 
from the equations of momentum, magnetic flux density, and magnetic vector potential, 
thereby elucidating the conservation properties of these quantities. 
Subsequently, five models for steady and unsteady problems were analyzed to verify 
the accuracy and convergence of the proposed numerical method. 
Additionally, the computational approach involving the magnetic vector and electric potentials was validated. 
A comparison of the calculated results with exact solutions 
in the analysis of one- and two-dimensional flow models and Hartmann flow 
further validated the numerical method. 
Unsteady analyses of two- and three-dimensional decaying vortices were performed. 
The ideal periodic inviscid MHD flow exhibited good conservation properties for total energy 
and cross-helicity. 
Magnetic helicity was discretely preserved even in three-dimensional flow. 
Furthermore, in viscous flow, the attenuation trends of total energy, cross-helicity, 
and magnetic helicity aligned with the exact solution. 
The numerical method accurately captured the decay trends of energy. 
Thus, the proposed method can facilitate 
the investigation of energy conservation and conversion in compressible MHD flows.
\end{abstract}
%------------------------------------------------------------------------------

%------------------------------------------------------------------------------
%%Research highlights
\begin{highlights} %(85 characters or fewer, including spaces.）
\item A conservative finite difference method for incompressible MHD flows is proposed.
\item Conservation equations for transport quantities are discretely derived.
\item The magnetic vector and electric potentials are relaxed simultaneously.
\item Total energy and cross- and magnetic helicities are discretely conserved.
%\item Accuracy and convergence were verified for five types of flow models.
\end{highlights}
%------------------------------------------------------------------------------

%------------------------------------------------------------------------------
%------------------------------------------------------------------------------

\end{frontmatter}

%\linenumbers

%##############################################################################
\section{Introduction}
%##############################################################################

In magnetohydrodynamic (MHD) flows, 
the generated Lorentz force changes kinetic and magnetic energies, 
resulting in a complex energy conversion. 
When applying flow control using magnetic fields to engineering devices, 
developing computational methods that can accurately predict flow and magnetic fields is essential. 
Hence, MHD flows have been numerically analyzed 
for incompressible and compressible fluids \citep{Toth_2000,Munz_et_al_2000,Liu&Wang_2001,Dedner_et_al_2002,Gawlik_et_al_2011,Ni&Li_2012,Kraus_et_al_2016,
Hu_et_al_2017,Hiptmair_et_al_2018}. 
In the MHD flows, energy is generated via the Lorentz force, 
and is converted to other forms such as Joule heat. 
Owing to this energy conversion, 
energy conservation cannot be verified easily while analyzing MHD flows. 
In a periodic flow with zero kinematic viscosity and magnetic diffusivity, 
the total amounts of total energy, cross-helicity, and magnetic helicity are preserved \citep{Woltjer_1958}. 
Using these transport quantities, one can verify the preservation properties. 
Some studies \citep{Gawlik_et_al_2011,Kraus_et_al_2016,Hu_et_al_2017,Hiptmair_et_al_2018} 
have proposed a structure-preserving numerical method for MHD flows. 
Using this method, excellent energy and helicity conservation properties have been demonstrated 
\citep{Gawlik_et_al_2011, Kraus_et_al_2016}; 
however, only a few studies have mentioned the conservation properties 
of magnetic helicity \citep{Kraus_et_al_2016}.

Another significant problem in analyzing MHD flows is satisfying 
the constraint imposed by Gauss's law for magnetism, 
which requires the magnetic flux density to be divergence-free. 
The Faraday equation should be integrated in the time direction 
while satisfying this constraint requirement. 
To date, some methods for satisfying the constraint condition have been proposed 
\citep{Brackbill&Barnes_1980,Evans&Hawley_1988,Munz_et_al_2000,Dedner_et_al_2002} 
and validated \citep{Toth_2000}. 
Hu et al. \citep{Hu_et_al_2017} and Hiptmair et al. \citep{Hiptmair_et_al_2018} 
elucidated the characteristics of the divergence-free condition 
for velocity and magnetic fields using a structure-preserving numerical method. 
Dumbser et al. \citep{Dumbser_et_al_2019} and Fambri \citep{Fambri_2021} 
proposed a semi-implicit finite-volume solver for compressible MHD flows, 
in which the divergence-free condition of magnetic flux density is discretely satisfied. 
However, the effects of Lorentz force discretization 
on energy conservation have not been clarified.

Several methods for introducing magnetic vector potential have been proposed 
to satisfy the constraint condition of magnetic flux density 
\citep{Lodrillo&Del_Zanna_2001, Helzel_et_al_2011}. 
In the numerical analysis method proposed by Lodrillo and Del Zanna \citep{Lodrillo&Del_Zanna_2001}, 
the definition points of magnetic flux density $\bm{B}$ 
and magnetic vector potential $\bm{A}$ in a computational cell are different; 
therefore, magnetic helicity $\bm{B} \cdot \bm{A}$ must be identified via interpolation. 
In an ideal inviscid MHD flow, magnetic helicity is conserved; 
however, the effect of this interpolation on the conservation properties of magnetic helicity is unknown. 
Helzel et al. \citep{Helzel_et_al_2011} defined all dependent variables at a cell center; 
hence, the magnetic helicity can be calculated directly without interpolation. 
Therefore, it is conjectured that total energy and magnetic helicity at the cell center are discretely preserved. 
Nonetheless, discretized equations for transport quantities such as total energy have not been presented, 
and the preservation properties of these transport quantities have not been investigated in detail.

The Lorentz force occurs in MHD flows and significantly affects flow and magnetic fields. 
As the Lorentz force is included as a body force in the momentum conservation equation, 
the equation takes a nonconservative form. 
Considering that the Lorentz force is nonconservative, 
Toth \citep{Toth_2000}, Munz et al. \citep{Munz_et_al_2000}, and Dedner et al. \citep{Dedner_et_al_2002} 
transformed it into a conservative form using Ampere's law 
and solved the resulting conservative fundamental equation. 
Ni and Li \citep{Ni&Li_2012} proposed a method for converting the Lorentz force 
into a divergent form using a distance-vector. 
However, even when the momentum equation is discretized, 
the feasibility of converting the Lorentz force 
between nonconserved and conserved forms remains unclear. 
The work done by the Lorentz force changes the associated kinetic and magnetic energies; 
hence, the force significantly affects energy conservation and conversion. 
The Lorentz force increases with the strength of the applied magnetic field, 
resulting in an increase in the Hartmann number. 
For high Hartmann numbers, the Lorentz force exerts more flow. 
Therefore, the Lorentz force and energy conversion should be predicted accurately. 
The author has previously investigated the effects of the discretization of the Lorentz force 
on numerical accuracy \citep{Yanaoka_2023}, 
discovering that the calculation method of Lorentz force significantly affects 
energy conservation properties. 
Total energy is preserved using the conservative finite difference method, 
even in nonuniform grids. 
However, cross-helicity, which is preserved well in uniform grids, 
deteriorates in nonuniform grids. 
Efforts are currently underway to apply the constructed conservative finite difference method 
to MHD flows at low Mach numbers. 
%In ideal inviscid compressible MHD flows, cross-helicity is not preserved. 
%The conserved transport quantities for compressible flows are magnetic helicity and total energy. 
%Therefore, a finite difference method that discretely preserves magnetic helicity must be constructed.
Even in ideal inviscid compressible MHD flows, cross-helicity is not a conserved quantity. 
However, transport quantities of magnetic helicity and total energy are conserved in such flows. 
Therefore, it is essential to develop a finite difference method 
that discretely preserves magnetic helicity and total energy. 
This ensures that numerical simulations accurately reflect the conservation laws 
governing transport quantities, 
maintaining fidelity to the physical behavior of ideal MHD systems.

To this end, the author presents a conservative finite difference method 
that preserves total energy and magnetic helicity. 
The remainder of this paper is organized as follows: 
Section \ref{fundamental_equation} presents the transport equations used in this study. 
In Section \ref{discretization}, 
the discretization method of the Lorentz force 
and discretized form of the fundamental equation for an incompressible MHD flow are drived. 
The conservation equations for total energy, 
cross-helicity, and magnetic helicity are also discretely derived. 
Section \ref{numerical_method} describes a simultaneous relaxation method 
for solving the fundamental equation. 
A method for obtaining the magnetic vector potential is also proposed. 
In Section \ref{verification}, 
several models are analyzed using the present numerical method 
to verify the validity of the computational method for magnetic vector 
and electric potentials. 
Additionally, the conservative properties of total energy, 
cross-helicity, and magnetic helicity and the calculation accuracy are clarified. 
Finally, Section \ref{summary} presents a summary of the results.

%##############################################################################
\section{Equations of transport quantities}
\label{fundamental_equation}
%##############################################################################

%++++++++++++++++++++++++++++++++++++++++++++++++++++++++++++++++++++++++++++++
\subsection{Fundamental equation}
%++++++++++++++++++++++++++++++++++++++++++++++++++++++++++++++++++++++++++++++

The fundamental equations governing an incompressible MHD flow are 
the transport equations for mass, momentum, and magnetic flux density. 
The magnetic flux density must satisfy the solenoidal constraint 
imposed by Gauss's law for magnetism. 
These dimensionless fundamental equations are expressed as
%------------------------------------------------------------------------------
\begin{equation}
  \frac{\partial u_j}{\partial x_j} = 0,
  \label{continuity}
\end{equation}
\begin{equation}
  \frac{Wo^2}{Re} \frac{\partial u_i}{\partial t} 
  + \frac{\partial u_j u_i}{\partial x_j} 
  = - \frac{\partial p}{\partial x_i} 
  + \frac{1}{Re} \frac{\partial^2 u_i}{\partial x_j^2} 
  + \frac{1}{Al^2} \epsilon_{ijk} j_j B_k,
  \label{navier-stokes}
\end{equation}
\begin{equation}
  \frac{\partial B_i}{\partial x_i} = 0,
  \label{divergence_magnetic}
\end{equation}
\begin{equation}
  \frac{Wo^2}{Re} \frac{\partial B_i}{\partial t} 
  + \epsilon_{ijk} \frac{\partial E_k}{\partial x_j} = 0,
  \label{Faraday}
\end{equation}
%------------------------------------------------------------------------------
where $t$ represents time, $u_i$ represents the velocity vector at the coordinate $x_i$, 
$p$ represents the pressure, 
$j_i$ represents the current density, and $B_i$ represents the magnetic flux density, 
and $E_i$ represents the electrical field. 
The term $\epsilon_{ijk} j_j B_k/Al^2$ in Eq. (\ref{navier-stokes}) 
expresses the Lorentz force. 
Regarding the reference values used for nondimensionalization, 
the length is $l_\mathrm{ref}$, velocity is $u_\mathrm{ref}$, 
time is $t_\mathrm{ref}$, and magnetic flux density is $B_\mathrm{ref}$. 
Using these reference values, 
the variables in the fundamental equations are nondimensionalized as follows:
%------------------------------------------------------------------------------
\begin{subequations}
\begin{equation}
  x_i^* = \frac{x_i}{l_\mathrm{ref}}, \quad 
  u_i^* = \frac{u_i}{u_\mathrm{ref}}, \quad 
  p^* = \frac{p}{\rho (u_\mathrm{ref})^{2}}, \quad 
  t^* = \frac{t}{t_\mathrm{ref}}, \quad 
\end{equation}
\begin{equation}
  E_i^* = \frac{E_i}{u_\mathrm{ref} B_\mathrm{ref}}, \quad 
  \psi^* = \frac{\psi}{u_\mathrm{ref} l_\mathrm{ref} B_\mathrm{ref}}, \quad
  j_i^* = \frac{j_i}{B_\mathrm{ref}/(\mu_{m} l_\mathrm{ref})}, \quad 
  B_i^* = \frac{B_i}{B_\mathrm{ref}},
\end{equation}
\end{subequations}
%------------------------------------------------------------------------------
where $*$ represents the nondimensional variable 
and is omitted in the fundamental equations. 
The nondimensional parameters in these fundamental equations are defined as follows: 
$Re$, $Wo$, $Al$, and $Re_m$ represent the Reynolds,  Womersley, 
 Alfv\'{e}n, and magnetic Reynolds numbers, respectively:
%------------------------------------------------------------------------------
\begin{equation}
  Re = \frac{u_\mathrm{ref} l_\mathrm{ref}}{\nu}, \quad 
  Wo = l_\mathrm{ref} \sqrt{\frac{1}{\nu t_\mathrm{ref}}}, \quad 
  Al = \frac{\sqrt{\mu_{m} \rho} u_\mathrm{ref}}{B_\mathrm{ref}}, \quad  
  Re_m = \frac{u_\mathrm{ref} l_\mathrm{ref}}{\nu_{m}},
\end{equation}
%------------------------------------------------------------------------------
where $\rho$ and $\nu$ represent the density and kinematic viscosity of the fluid, respectively, 
and $\mu_m$ represents the magnetic permeability 
related to the magnetic diffusivity $\nu_m$ and electrical conductivity $\sigma$, 
as $\nu_m = 1/(\sigma \mu_m)$.

The current density defined using Ampere's and Ohm's laws, respectively, is as follows:
%------------------------------------------------------------------------------
\begin{equation}
  j_i = \epsilon_{ijk} \frac{\partial B_k}{\partial x_j},
  \label{Ampere}
\end{equation}
\begin{equation}
  j_i = Re_m \left( E_i + \epsilon_{ijk} u_j B_k \right).
  \label{Ohm}
\end{equation}
%------------------------------------------------------------------------------

The magnetic flux density equation is obtained by revising 
Eq. (\ref{Faraday}) using Ohm's law (\ref{Ohm}) as follows:
%------------------------------------------------------------------------------
\begin{equation}
  \frac{Wo^2}{Re} \frac{\partial B_i}{\partial t} 
  + \frac{\partial (u_j B_i - B_j u_i)}{\partial x_j} 
  = - \frac{1}{Re_m} \epsilon_{ijk} \frac{\partial j_k}{\partial x_j}.
  \label{magnetic_field}
\end{equation}
%------------------------------------------------------------------------------
When $Re_m = \infty$, Eq. (\ref{magnetic_field}) is in a conservative form, 
which is the form used by \citep{Toth_2000,Dedner_et_al_2002}. 

The induced electric field is expressed using the electric potential $\psi$ as
%------------------------------------------------------------------------------
\begin{equation}
  E_i = - \frac{\partial \psi}{\partial x_i} 
  - \frac{Wo^2}{Re} \frac{\partial A_i}{\partial t},
  \label{electric_potential}
\end{equation}
%------------------------------------------------------------------------------
where $A_i$ is the magnetic vector potential satisfying $B_i = \epsilon_{ijk} \partial_j A_k$. 
The conservation law of electric charge is given as
%------------------------------------------------------------------------------
\begin{equation}
  \frac{\partial j_i}{\partial x_i} = 0.
  \label{electric_charge_law}
\end{equation}
%------------------------------------------------------------------------------
The Poisson equation for the electric potential can be obtained 
using Eqs. (\ref{Ohm}) and (\ref{electric_charge_law}) as follows:
%------------------------------------------------------------------------------
\begin{equation}
  \frac{\partial}{\partial x_i} 
  Re_m \left( - \frac{\partial \psi}{\partial x_i} 
  - \frac{Wo^2}{Re} \frac{\partial A_i}{\partial t} 
  + \frac{\partial \epsilon_{ijk} u_j B_k}{\partial x_i} \right) = 0.
  \label{electric_potential_poisson}
\end{equation}
%------------------------------------------------------------------------------
Applying the Coulomb gauge $\partial_i A_i = 0$ yields the following Poisson's equation:
%------------------------------------------------------------------------------
\begin{equation}
  \frac{\partial^2 \psi}{\partial x_i^2} 
  = \frac{\partial \epsilon_{ijk} u_j B_k}{\partial x_i}.
\end{equation}
%------------------------------------------------------------------------------
Applying Ampere's (\ref{Ampere}) and Ohm's laws (\ref{Ohm}) 
to Eq. (\ref{electric_potential}), 
the equation for the magnetic vector potential is obtained as follows:
%------------------------------------------------------------------------------
\begin{equation}
  \frac{Wo^2}{Re} \frac{\partial A_i}{\partial t} 
  + \epsilon_{ijk} B_j u_k = - \frac{\partial \psi}{\partial x_i} 
  - \frac{1}{Re_m} j_i,
  \label{vector_potential}
\end{equation}
%------------------------------------------------------------------------------

When the magnetic permeability is constant, 
the Lorentz force in Eq. (\ref{navier-stokes}) 
can be revised from a nonconservative to a conservative form as follows:
%------------------------------------------------------------------------------
\begin{align}
  F_i &= \frac{1}{Al^2} \epsilon_{ijk} j_j B_k \nonumber \\ 
  &= \frac{1}{Al^2} \left[ 
  \frac{\partial B_j B_i}{\partial x_j} 
  - \frac{1}{2} \frac{\partial B_j^2}{\partial x_i} 
  - \frac{\partial B_j}{\partial x_j} B_i \right],
  \label{Lorentz}
\end{align}
%------------------------------------------------------------------------------
where $B_i^2/(2Al^2)$ is the magnetic pressure. 
The last term is zero according to the solenoidal constraint. 
The momentum equation is transformed using Eq. (\ref{Lorentz}), 
and the terms of the equation, other than the final term, are expressed 
in the conservative form as follows:
%------------------------------------------------------------------------------
\begin{equation}
  \frac{Wo^2}{Re} \frac{\partial u_i}{\partial t} 
  + \frac{\partial}{\partial x_j} 
  \left( u_j u_i - \frac{1}{Al^2} B_j B_i \right) 
  = - \frac{\partial P}{\partial x_i} 
  + \frac{1}{Re} \frac{\partial^2 u_i}{\partial x_j^2} 
  - \frac{1}{Al^2} B_i \frac{\partial B_j}{\partial x_j},
  \label{navier-stokes2}
\end{equation}
%------------------------------------------------------------------------------
where $P$ is a component obtained by adding a magnetic pressure component to pressure 
as follows:
%------------------------------------------------------------------------------
\begin{equation}
  P = p + \frac{1}{2} \frac{1}{Al^2} B_i^2.
\end{equation}
%------------------------------------------------------------------------------
The final term of Eq. (\ref{navier-stokes2}) becomes zero 
when the solenoidal constraint $\partial_j B_j = 0$ is satisfied. 
Equation (\ref{Lorentz}) indicates that the direction of Lorentz force is 
perpendicular to the magnetic field 
if the solenoidal constraint is satisfied. 
If $\partial_j B_j$ is not zero, a nonphysical Lorentz force proportional to 
$\partial_j B_j$ also occurs in the direction parallel to the magnetic field. 
Therefore, the time should be advanced while satisfying the solenoidal constraint.

Equation (\ref{Lorentz}) represents the transformation of the Lorentz force. 
The Lorentz forces in (\ref{navier-stokes}) and (\ref{navier-stokes2}) are 
in the nonconservative and conservative forms, respectively. 
If the conversion of the Lorentz force in Eq. (\ref{Lorentz}) holds discretely, 
the nonconservative Lorentz force can be converted to a conservative form. 
Therefore, Eq. (\ref{navier-stokes}) can be transformed 
into Eq. (\ref{navier-stokes2}) using discretized Eq. (\ref{Lorentz}). 
Thus, if the constraint condition of the magnetic flux density is satisfied, 
Eq. (\ref{navier-stokes}) becomes conservative, 
and momentum is conserved for $Re = Re_m = \infty$. 
Moreover, the work done by the Lorentz force changes 
the kinetic and magnetic energies; this change affects energy conservation properties. 
If the Lorentz force cannot be transformed discretely as in Eq. (\ref{Lorentz}), 
the discrete forms of the Lorentz forces in Eqs. (\ref{navier-stokes}) 
and (\ref{navier-stokes2}) are different. 
The form of the Lorentz force can affect the conservation of energy and momentum 
and the conversion of energy. 
A previous study \citep{Yanaoka_2023} clarified 
that even when the Lorentz force in Eq. (\ref{navier-stokes}) is discretized, 
the nonconservative Lorentz force is converted to the conservative form, 
and the transformation between the nonconservative and conservative forms is established. 
In Section \ref{discretization}, 
the discretization method of the Lorentz force is described 
such that this transformation holds. 
Analyzing various models confimed 
that the calculation stability in both nonconservative 
and conservative forms is the same in uniform grids. 
However, as the transformation is not established on nonuniform grids, 
the calculation using a conservative form that does not discretely satisfy 
the preservation of total energy becomes unstable. 
Therefore, this study adopts a method that uses a nonconservative form of the Lorentz force.

%++++++++++++++++++++++++++++++++++++++++++++++++++++++++++++++++++++++++++++++
\subsection{Energy equations}
%++++++++++++++++++++++++++++++++++++++++++++++++++++++++++++++++++++++++++++++

Here, the equations for kinetic, magnetic, and total energies are derived. 
Each energy is nondimensionalized using $u_\mathrm {ref}^2$. 
The Lorentz force can be analytically transformed from a nonconservative to a conservative form 
using Eq. (\ref{Lorentz}). 
The following equation uses Eq. (\ref{navier-stokes}), 
which expresses the Lorentz force in a nonconservative form. 
Using the nonconservative Lorentz force, 
The author demonstrates that the work done by the Lorentz force cancels out the work 
appearing in the magnetic energy equation. 
By calculating the inner product of the velocity $u_i$ 
and Eq. (\ref{navier-stokes}), 
the transport equation for the kinetic energy $K=u_i u_i/2$ is obtained as follows:
%------------------------------------------------------------------------------
\begin{equation}
  \frac{Wo^2}{Re} \frac{\partial K}{\partial t} 
  + \frac{\partial u_j K}{\partial x_j} 
  + K \frac{\partial u_j}{\partial x_j} 
  = - \frac{\partial u_i p}{\partial x_i} + p \frac{\partial u_i}{\partial x_i} 
  + \frac{1}{Re} \left[ \frac{\partial}{\partial x_j} 
    \frac{\partial K}{\partial x_j} 
  - \frac{\partial u_i}{\partial x_j} \frac{\partial u_i}{\partial x_j} \right] 
  +\frac{1}{Al^2}  u_i \epsilon_{ijk} j_j B_k.
  \label{energy_K}
\end{equation}
%------------------------------------------------------------------------------
If the magnetic field is not applied, the last term, $u_i \epsilon_{ijk} j_j B_k$, 
of the above equation will be zero. 
If the continuity equation (\ref{continuity}) is satisfied, 
the third and second terms on the left- and right-hand sides, respectively, will be zero. 
The fourth term on the right side is the viscous dissipation term, 
not a conservative form. 
The fifth term caused by the Lorentz force is also a nonconservative form.

Subsequently, by calculating the inner product of the Faraday equation (\ref{Faraday}) 
and magnetic flux density $B_i$, 
the following transport equation for the magnetic energy $M = B_i B_i/(2Al^2)$ is obtained:
%------------------------------------------------------------------------------
\begin{equation}
  \frac{Wo^2}{Re} \frac{\partial M}{\partial t} 
  = - \frac{1}{Al^2} \left( \frac{\partial \epsilon_{ijk} E_j B_k}{\partial x_i} 
  + u_i \epsilon_{ijk} j_j B_k + \frac{1}{Re_m} j_i^2 \right).
  \label{energy_M}
\end{equation}
%------------------------------------------------------------------------------
The second term on the right side resulting from the Lorentz force 
and the third term corresponding to Joule heat are nonconservative.

The sum of Eqs. (\ref{energy_K}) and (\ref{energy_M}) 
yields the equation for the total energy $E_t = K + M$ as
%------------------------------------------------------------------------------
\begin{align}
  \frac{Wo^2}{Re} \frac{\partial E_t}{\partial t} 
  + \frac{\partial u_j K}{\partial x_j} 
  + K \frac{\partial u_j}{\partial x_j} 
  &= - \frac{\partial u_i p}{\partial x_i} + p \frac{\partial u_i}{\partial x_i} 
  + \frac{1}{Re} \left[ \frac{\partial}{\partial x_j} 
    \frac{\partial K}{\partial x_j} 
  + \frac{\partial u_i}{\partial x_j} \frac{\partial u_i}{\partial x_j} \right] 
  \nonumber \\
  &
  - \frac{1}{Al^2} \left( \frac{\partial \epsilon_{ijk} E_j B_k}{\partial x_i} 
  + \frac{1}{Re_m} j_i^2 \right).
  \label{energy_Et}
\end{align}
%------------------------------------------------------------------------------
In Eq. (\ref{energy_Et}), 
the work terms $u_i \epsilon_{ijk} j_j B_k$ in Eqs. (\ref{energy_K}) and (\ref{energy_M}) 
cancel each other. 
However, energy is exchanged between the velocity and magnetic fields 
through this term. 
The Lorentz forces appearing in Eqs. (\ref{energy_K}) and (\ref{energy_M}) 
should be obtained by the same discretization and interpolation. 
If the calculation method of the Lorentz force is inconsistent, 
the energy conversion cannot be captured correctly.

Assuming zero kinematic and magnetic viscosities, 
Eq. (\ref{energy_Et}) is expressed as follows:
%------------------------------------------------------------------------------
\begin{equation}
  \frac{\partial E_t}{\partial t} 
  + \frac{\partial u_j K}{\partial x_j} 
  + K \frac{\partial u_j}{\partial x_j} 
  = - \frac{\partial u_i p}{\partial x_i} + p \frac{\partial u_i}{\partial x_i} 
  - \frac{1}{Al^2} \frac{\partial \epsilon_{ijk} E_j B_k}{\partial x_i},
  \label{energy_inviscid_Et}
\end{equation}
%------------------------------------------------------------------------------
where $Wo = \sqrt{Re}$ is set to remove $Re$. 
From the above equation, if the continuity equation (\ref{continuity}) is satisfied, 
the transport equation for the total energy $E_t$, 
which is the sum of the kinetic and magnetic energies, is conservative; 
that is, Eq. (\ref{energy_inviscid_Et}) states that the total energy is conserved.

%++++++++++++++++++++++++++++++++++++++++++++++++++++++++++++++++++++++++++++++
\subsection{Equations of cross-helicity and magnetic helicity}
%++++++++++++++++++++++++++++++++++++++++++++++++++++++++++++++++++++++++++++++

In an ideal inviscid, incompressible MHD flow, 
the total energy $E_t$ is a conserved quantity. 
Moreover, the cross-helicity $H_c = u_i B_i/Al$ is preserved. 
The cross-helicity is nondimensionalized by $u_\mathrm{ref}^2$. 
Using the momentum equation (\ref{navier-stokes}) 
and magnetic flux density equation (\ref{magnetic_field}), 
if $Al \partial_t H_c = B_i \partial_t u_i + u_i\partial_t B_i$  is calculated, 
the following cross-helicity equation is obtained:
%------------------------------------------------------------------------------
\begin{align}
  \frac{\partial H_c}{\partial t} 
  &= \frac{1}{Al} \left[ 
  - \frac{\partial u_j u_i B_i}{\partial x_j} 
  - \frac{\partial p B_i}{\partial x_i}  + p \frac{\partial B_i}{\partial x_i} 
  + \frac{\partial}{\partial x_j} \left( \frac{1}{2} u_i u_i B_j \right) 
  \nonumber \right. \\
  & \quad \left. 
  - \left( \frac{1}{2} u_i u_i \right) \frac{\partial B_j}{\partial x_j} 
  + u_i \left(- B_i \frac{\partial u_j}{\partial x_j} 
              + u_i \frac{\partial B_j}{\partial x_j} \right) \right],
  \label{cross_helicity_eq}
\end{align}
%------------------------------------------------------------------------------
where $\partial_j u_j = 0$ and $\partial_j B_j = 0$ in incompressible flows, 
Therefore, the third and fifth, sixth, and seventh terms on the right side are zeros. 
The cross-helicity is expressed as a conservative equation, 
and the total amount of cross-helicity is conserved 
under the assumption of periodic flow:
%------------------------------------------------------------------------------
\begin{align}
  \frac{\partial H_c}{\partial t} 
  &= - \frac{1}{Al} \left[ 
    \frac{\partial u_j u_i B_i}{\partial x_j} 
  + \frac{\partial p B_i}{\partial x_i}  
  - \frac{\partial}{\partial x_j} \left( \frac{1}{2} u_i u_i B_j \right) \right].
  \label{cross_helicity_inviscid_eq}
\end{align}
%------------------------------------------------------------------------------

The magnetic helicity $H_m = B_i A_i/Al^2$ is also preserved. 
It is nondimensionalized by $l_\mathrm{ref} u_\mathrm{ref}^2$. 
Using Eqs. (\ref{Faraday}) and (\ref{vector_potential}) for $Re_m = \infty$, 
if $Al^2 \partial_t H_m = A_i \partial_t B_i + B_i\partial_t A_i$ is calculated, 
the following magnetic helicity equation is obtained:
%------------------------------------------------------------------------------
\begin{equation}
  \frac{\partial H_m}{\partial t} 
  = \frac{1}{Al^2} \left[ 
  - \frac{\partial}{\partial x_j} (\epsilon_{jki} E_k A_i) 
  + 2 B_i (\epsilon_{ijk} u_j B_k) 
  - \frac{\partial B_i \psi}{\partial x_i} 
  + \psi \frac{\partial B_i}{\partial x_i} \right],
  \label{magnetic_helicity_eq}
\end{equation}
%------------------------------------------------------------------------------
where the final term is zero because $\partial_i B_i = 0$. 
Additionally, as $\epsilon_{ijk} u_j B_k$ and $B_i$ are orthogonal, 
$\epsilon_{ijk} (u_j B_k) B_i = 0$. 
The magnetic helicity is expressed as a conservative equation, 
and the total amount of magnetic helicity is conserved 
under the assumption of periodic flow:
%------------------------------------------------------------------------------
\begin{equation}
  \frac{\partial H_m}{\partial t} 
  = - \frac{1}{Al^2} \left[ 
    \frac{\partial}{\partial x_j} (\epsilon_{jki} E_k A_i) 
  + \frac{\partial B_i \psi}{\partial x_i} \right].
  \label{magnetic_helicity_ideal_eq}
\end{equation}
%------------------------------------------------------------------------------

%##############################################################################
\section{Discretization of transport equation}
\label{discretization}
%##############################################################################

For periodic inviscid incompressible flows without applied magnetic fields, 
the transport quantity, such as the kinetic energy, 
must be discretely conserved \citep{Morinishi_1996a,Morinishi_1998}. 
The generation of nonphysical kinetic energy leads to computational instability. 
Additionally, the transformation between the conservative and nonconservative forms 
of convection terms must be discretely satisfied 
\citep{Morinishi_1996a,Morinishi_1998}. 
In a flow field without an applied magnetic field, 
a fully conservative finite difference method, 
in which the transport quantity is discretely conserved in the spatiotemporal direction, 
has been proposed. 
The transformation between conservative and nonconservative forms 
of the advection term has been established \citep{Ham_et_al_2002,Morinishi_2009,Morinishi_2010}. 
In this study, ihe fully conservative finite difference method is applied to analyze MHD flows, 
as in \citep{Ham_et_al_2002,Morinishi_2009,Morinishi_2010}. 
The implicit midpoint rule for the time derivative 
and the second-order central difference for the spatial derivative are applied. 
Previous research \citep{Yanaoka_2023} presented a method for discretizing equations 
for transport quantities such as total energy. 
However, no detailed discretization equations have been formulated yet. 
In the following subsection, the discretization of each transport quantity equation is described comprehensively.

%++++++++++++++++++++++++++++++++++++++++++++++++++++++++++++++++++++++++++++++
\subsection{Definitions of finite difference and interpolation operations}
%++++++++++++++++++++++++++++++++++++++++++++++++++++++++++++++++++++++++++++++

The Cartesian coordinates $x_m$ in the physical space are transformed into 
the computational space $\xi_m$ for discretization in a nonuniform grid. 
The relationship $x_m = x_m(\xi_m)$ is assumed between both spaces. 
By letting dependent variables such as velocity, pressure, 
and magnetic flux density be $\Phi$, 
the first derivative can be converted as follows:
%------------------------------------------------------------------------------
\begin{equation}
  \frac{\partial \Phi}{\partial x_1} 
    = \frac{1}{J} \frac{\partial (J \xi_{1,1} \Phi)}{\partial \xi_1}, \quad 
  \frac{\partial \Phi}{\partial x_2} 
    = \frac{1}{J} \frac{\partial (J \xi_{2,2} \Phi)}{\partial \xi_2}, \quad 
  \frac{\partial \Phi}{\partial x_3} 
    = \frac{1}{J} \frac{\partial (J \xi_{3,3} \Phi)}{\partial \xi_3},
\end{equation}
%------------------------------------------------------------------------------
where $J$ is the Jacobian defined as $J = x_{1,1} x_{2,2} x_{3,3}$. 
$\xi_{i,i}$ is given as
%------------------------------------------------------------------------------
\begin{equation}
  \xi_{1,1} = \frac{1}{J} x_{2,2} x_{3,3}, \quad 
  \xi_{2,2} = \frac{1}{J} x_{3,3} x_{1,1}, \quad 
  \xi_{3,3} = \frac{1}{J} x_{1,1} x_{2,2}.
\end{equation}
%------------------------------------------------------------------------------

The variables at a cell center $(i, j, k)$ are defined as $\Phi_{i,j,k}$ and $\Psi_{i,j,k}$. 
For the $x$ ($\xi_1$)-direction, 
the second-order central difference equation and interpolation for the variable $\Phi$ 
and the permanent product for two variables are given, respectively, as follows: 
\citep{Morinishi_1996a,Morinishi_1998}:
%------------------------------------------------------------------------------
\begin{equation}
  \left. \frac{\partial \Phi}{\partial x_1} \right|_{i,j,k} 
  = \delta_{\xi_1} \Phi 
  = \frac{1}{J} \frac{(J \xi_{1,1} \bar{\Phi}^{\xi_1})_{i+1/2,j,k} 
                    - (J \xi_{1.1} \bar{\Phi}^{\xi_1})_{i-1/2,j,k}}{\Delta \xi_1},
\end{equation}
\begin{equation}
  \left. \bar{\Phi}^{\xi_1} \right|_{i+1/2,j,k} 
  = \frac{\Phi_{i,j,k} + \Phi_{i+1,j,k}}{2},
\end{equation}
\begin{equation}
  \left. \widetilde{\Phi \Psi}^{\xi_1} \right|_{i+1/2,j,k} 
  = \frac{\Phi_{i,j,k} \Psi_{i+1,j,k} 
        + \Phi_{i+1,j,k} \Psi_{i,j,k}}{2},
\end{equation}
%------------------------------------------------------------------------------
where $\Delta \xi_1$ is the grid spacing in the computational space. 
The definitions of the $x_2$ ($\xi_2$)- and $x_3$ ($\xi_3$)-directions are identical. 
The Jacobian is defined at a cell center. 
The index $j$ representing the direction of the finite difference $\delta_{\xi_j}$ 
is considered a tensor component and follows the summation convention. 
The indices $j$ of the interpolation $\bar{\Phi}^{\xi_j}$ 
and permanent product $\widetilde{\Phi \Psi}^{\xi_j}$ do not follow the convention. 
The indices $j$ change simultaneously with the indices of the tensor components 
in the same term.
Derivative terms that are not directly related to conservation properties, 
such as momentum and total energy, are discretized 
without coordinate transformation, as follows:
%------------------------------------------------------------------------------
\begin{equation}
  \left. \frac{\partial \Phi}{\partial x_1} \right|_{i,j,k} 
  = \delta_{x_1} \Phi 
  = \frac{\bar{\Phi}^{x_1}_{i+1/2,j,k} - \bar{\Phi}^{x_1}_{i-1/2,j,k}}{\Delta x_{1 i}},
\end{equation}
%------------------------------------------------------------------------------
where $\Delta x_{1 i} = x_{i+1/2}-x_{i-1/2}$ is the grid spacing. 
If a variable at time level $n$ is defined as $\Phi^n$, 
the derivative and interpolation of the variable for time are similarly expressed as follows:
%------------------------------------------------------------------------------
\begin{equation}
  \left. \frac{\partial \Phi}{\partial t} \right|^{n+1/2} 
  = \delta_t \Phi 
  = \frac{\Phi^{n+1} - \Phi^{n}}{\Delta t},
\end{equation}
\begin{equation}
  \Phi^{n+1/2} = \bar{\Phi}^t = \frac{\Phi^{n+1} + \Phi^{n}}{2},
\end{equation}
%------------------------------------------------------------------------------
where $\Delta t$ is a time increment. 
For derivations in the subsequent subsections, 
the following discrete relational formula is used \citep{Morinishi_1996a,Morinishi_1998}:
%------------------------------------------------------------------------------
%\begin{equation}
%  \delta_{\xi_i} \Psi \bar{\Phi}^{\xi_i} 
%  = \overline{\Psi \delta_{\xi_i} \bar{\Phi}^{\xi_i}}^{\xi_i} 
%  + \Phi \delta_{\xi_i} \Psi,
%\end{equation}
\begin{equation}
 \bar{\Psi}^t \delta_t \Phi + \bar{\Phi}^t \delta_t \Psi = \delta_t \Psi \Phi,
\end{equation}
\begin{equation}
  \frac{1}{2} \delta_t \Phi^2 
  = \bar{\Phi}^t \delta_t \Phi.
\end{equation}
%------------------------------------------------------------------------------

%++++++++++++++++++++++++++++++++++++++++++++++++++++++++++++++++++++++++++++++
\subsection{Discretization of the Lorentz force}
%++++++++++++++++++++++++++++++++++++++++++++++++++++++++++++++++++++++++++++++

This study uses a staggered grid. 
The velocities, $u_1$, $u_2$, and $u_3$, are defined 
at the cell interfaces, $(i+1/2, j, k)$, $(i, j+1/2, k)$, and $(i, j, k+1/2)$, respectively. 
As with the velocity field, 
the magnetic flux densities, $B_1$, $B_2$, and $B_3$, are defined 
at the cell interfaces, $(i+1/2, j, k)$, $(i, j+1/2, k)$, and $(i, j, k+1/2)$, respectively. 
The definition point of the electric field is different from that of the magnetic field. 
The current densities, $j_1$, $j_2$, and $j_3$, are defined at the midpoints of the cell edge, 
$(i, j+1/2, k+1/2)$, $(i+1/2, j, k+1/2)$, and $(i+1/2, j+1/2, k)$, respectively. 
The electric field $E_i$ is similar. 
The method of spatially shifting the definition points of the electric 
and magnetic fields is similar to that described in \citep{Yee_1966}. 
However, when the electric potential is obtained from the charge conservation law 
(\ref{electric_charge_law}) using Ohm's law, 
the current densities, $j_1$, $j_2$, and $j_3$, are defined 
at the cell interfaces in the same manner as the velocity. 
Scalar quantities such as pressure and energy are defined 
at the cell center $(i, j, k)$.

In this study, the nonconservative Lorentz force is obtained 
through the weighted interpolation of magnetic flux and current densities 
using the Jacobian \citep{Yanaoka_2023}. 
The nonconservative Lorentz force is expressed discretely as follows:
%------------------------------------------------------------------------------
\begin{equation}
  F_i = \frac{1}{Al^2} \frac{1}{\bar{J}^{\xi_i}} 
  \overline{\epsilon_{ijk} \overline{\bar{J}^{\xi_i}}^{\xi_k} 
  j_j \bar{B_k}^{\xi_i}}^{\xi_k}.
  \label{compact_interpolation_F^nc}
\end{equation}
%------------------------------------------------------------------------------
The conservative Lorentz force is expressed discretely as follows:
%------------------------------------------------------------------------------
\begin{equation}
  F_i = \frac{1}{Al^2} \frac{1}{\bar{J}^{\xi_i}} \left[ 
  \delta_{\xi_j} \left( \overline{J \xi_{j,j} B_j}^{\xi_i} \bar{B_i}^{\xi_j} \right) 
  - \frac{1}{2} \overline{\delta_{\xi_i} (J \xi_{i,i} B_j^2)}^{\xi_j} 
  - B_i \overline{\delta_{\xi_j} J \xi_{j,j} B_j}^{\xi_i} \right].
  \label{compact_interpolation_F^c}
\end{equation}
%------------------------------------------------------------------------------
If the divergence-free condition $\partial_i B_i = 0$ for the magnetic flux density 
is satisfied discretely, the Lorentz force given by Eq. (\ref{compact_interpolation_F^c}) 
becomes conservative. 
In the case of uniform grids, the Lorentz force can be transformed 
from a nonconservative to a conservative form, 
and the transformation of the Lorentz force holds discretely \citep{Yanaoka_2023}. 
Current density is used in Eq. (\ref{compact_interpolation_F^nc}); 
therefore, the current density that satisfies the charge conservation law should be used. 
If the current density $j_i$ obtained from Ampere's law is defined at the cell interface, 
the charge conservation law is satisfied at the cell center. 
If the current density is defined at the cell interface, 
the surrounding 12 magnetic flux densities $B_1$ are required 
to obtain the current density $j_2$ at the cell interface $(i, j+1/2, k)$. 
The surrounding four current densities $j_2$ are required 
to calculate the Lorentz force $F_1$. 
Twenty-seven surrounding $B_1$ are required for calculating $F_1$. 
Therefore, the Lorentz force is obtained via interpolation 
using numerous magnetic flux densities, 
which results in grid dependence and a decrease in accuracy. 
In this study, the Lorentz force is calculated using the current density 
defined at the midpoint of the cell edge. 
The two surrounding $B_1$ are required to calculate $j_2$ at the point $(i+1/2, j, k+1/2)$, 
and two surrounding $j_2$ to calculate $F_1$. 
Three surrounding $B_1$ are required for calculating $F_1$. 
Thus, the Lorentz force can be obtained via compact interpolation. 
The current density (\ref{Ampere}) is discretized as follows:
%------------------------------------------------------------------------------
\begin{equation}
  j_i = \epsilon_{ijk} \delta_{x_j} B_k.
  \label{compact_interpolation_J}
\end{equation}
%------------------------------------------------------------------------------
The charge conservation law $\partial_i j_i = 0$ is satisfied 
at the grid point $(i+1/2, j+1/2, k+1/2)$ as follows:
%------------------------------------------------------------------------------
\begin{equation}
  \frac{\partial j_i}{\partial x_i} 
  = \delta_{x_1} (\delta_{x_2} B_3 - \delta_{x_3} B_2) 
  + \delta_{x_2} (\delta_{x_3} B_1 - \delta_{x_1} B_3) 
  + \delta_{x_3} (\delta_{x_1} B_2 - \delta_{x_2} B_1) = 0.
\end{equation}
%------------------------------------------------------------------------------
In this study, compact interpolation refers to calculating the Lorentz force 
via the interpolation defined by Eq. (\ref{compact_interpolation_F^nc}) 
using Eq. (\ref{compact_interpolation_J}) \citep{Yanaoka_2023}.

Using Eq. (\ref{compact_interpolation_F^nc}), 
the Lorentz force is converted from a nonconservative to a conservative form 
if the divergence-free condition of the magnetic flux density is satisfied. 
Therefore, even in the discretized formula, 
the Lorentz force transformation formula (\ref{Lorentz}) approximately holds. 
Additionally, the Lorentz force can be calculated using the current density 
that satisfies the charge conservation law. 
Conversely, when the Lorentz force is converted to the conservative form, 
the Lorentz force $\epsilon_{ijk} j_j B_k/Al^2$ is approximately calculated in a nonuniform grid. 
When the nonconservative form of the Lorentz force is obtained 
from Eq. (\ref{compact_interpolation_F^nc}) in the nonuniform grid, 
total energy is conserved. 
The Lorentz force in Eq. (\ref{compact_interpolation_F^c}) is discretized 
to satisfy the constraint $\partial_j B_j = 0$ of the magnetic flux density in Eq. (\ref{Lorentz}). 
Therefore, the transformation of the Lorentz force does not hold for nonuniform grids. 

An alternative approach for calculating the current density is presented in \citep{Yanaoka_2023} 
to compare the compact interpolation method for the Lorentz force. 
Ampere's law (\ref{Ampere}) must discretely satisfy 
the charge conservation law $\partial_i j_i = 0$. 
The current densities, $j_1$, $j_2$, and $j_3$, are defined 
at the cell interfaces, $(i+1/2, j, k)$, $(i, j+1/2, k)$, 
and $(i, j, k+1/2)$, respectively. 
The equation (\ref{Ampere}) for the current density is discretized as
%------------------------------------------------------------------------------
\begin{equation}
  j_i = \epsilon_{ijk} \delta_{x_j} 
  \overline{\overline{\bar{B_k}^{x_k}}^{x_i}}^{x_j}.
  \label{wide-range_interpolation_J}
\end{equation}
%------------------------------------------------------------------------------
The magnetic flux densities $B_i$ in the current density 
are obtained via interpolation in the $x_1$-, $x_2$-, and $x_3$-directions. 
Calculating the divergence of the current density indicates 
that the discretized charge conservation law $\partial_i j_i = 0$ 
is satisfied at the cell center $(i, j, k)$ as follows:
%------------------------------------------------------------------------------
\begin{align}
  \delta_{x_i} j_i 
  &= \delta_{x_i} \epsilon_{ijk} \delta_{x_j} 
  \overline{\overline{\bar{B_k}^{x_k}}^{x_i}}^{x_j} 
  \nonumber \\
  &= 
    \delta_{x_1} \left( 
    \delta_{x_2} \overline{\overline{\bar{B_3}^{x_3}}^{x_1}}^{x_2} 
  - \delta_{x_3} \overline{\overline{\bar{B_2}^{x_2}}^{x_1}}^{x_3} \right) 
  + \delta_{x_2} \left( 
    \delta_{x_3} \overline{\overline{\bar{B_1}^{x_1}}^{x_2}}^{x_3} 
  - \delta_{x_2} \overline{\overline{\bar{B_3}^{x_3}}^{x_2}}^{x_1} \right) 
  \nonumber \\
  &
  + \delta_{x_3} \left( 
    \delta_{x_1} \overline{\overline{\bar{B_2}^{x_2}}^{x_3}}^{x_1} 
  - \delta_{x_2} \overline{\overline{\bar{B_1}^{x_1}}^{x_3}}^{x_2} \right) 
  = 0.
\end{align}
%------------------------------------------------------------------------------
Using the current density of Eq. (\ref{wide-range_interpolation_J}), 
the Lorentz force can be obtained as follows:
%------------------------------------------------------------------------------
\begin{equation}
  F_i = \frac{1}{Al^2} \epsilon_{ijk} 
  \overline{\bar{j_j}^{x_j}}^{x_i} \overline{\bar{B_k}^{x_k}}^{x_i}.
  \label{wide-range_interpolation_F}
\end{equation}
%------------------------------------------------------------------------------
However, this method uses 12 magnetic flux densities $B_1$ 
to determine the current density $j_2$ in $F_1$. 
Therefore, the accuracy may decrease owing to interpolation. 
In this study, wide-range interpolation refers to the method for calculating the Lorentz force 
via the interpolation method defined by Eq. (\ref{wide-range_interpolation_F}) 
using Eq. (\ref{wide-range_interpolation_J}). 
When using wide-range interpolation, 
the nonconservative form of the Lorentz force cannot be discretely converted 
to the conservative form, in contrast to when using Eq. (\ref{compact_interpolation_F^nc}). 
Therefore, in this study, the Lorentz force is calculated 
using interpolation by Eq. (\ref{compact_interpolation_F^nc}). 
An earlier study \citep{Yanaoka_2023} has clarified 
the difference in energy conservation properties 
based on the calculation method of the Lorentz force.

%++++++++++++++++++++++++++++++++++++++++++++++++++++++++++++++++++++++++++++++
\subsection{Discretization of the continuity and momentum equations}
%++++++++++++++++++++++++++++++++++++++++++++++++++++++++++++++++++++++++++++++

This study uses the fully conservative finite difference method 
proposed in \citep{Morinishi_2009,Morinishi_2010} 
for discretization of the mass and momentum conservation equations. 
Using the same discretization method, 
Eqs. (\ref{continuity}) and (\ref{navier-stokes}) are discretized as
%------------------------------------------------------------------------------
\begin{equation}
  \frac{1}{J} \delta_{\xi_j} U_j = 0,
  \label{continuity_fdm}
\end{equation}
\begin{align}
  \frac{Wo^2}{Re} \delta_t u_i 
  + \frac{1}{\bar{J}^{\xi_i}} 
    \delta_{\xi_j} \overline{\bar{U_j}^t}^{\xi_i} \overline{\bar{u_i}^t}^{\xi_j} 
  &= - \frac{1}{\bar{J}^{\xi_i}} J \xi_{i,i} \delta_{\xi_i} \bar{p}^t 
  + \frac{1}{Re} \frac{1}{\bar{J}^{\xi_i}} \delta_{\xi_j} J \xi_{j,j} \delta_{x_i} \bar{u_i}^t 
  \nonumber \\
  &
  + \frac{1}{Al^2} \frac{1}{\bar{J}^{\xi_i}} 
  \overline{\epsilon_{ijk} \overline{\bar{J}^{\xi_i}}^{\xi_k} 
  \bar{j_j}^t \overline{\bar{B_k}^t}^{\xi_i}}^{\xi_k}, 
  \label{navier-stokes_fdm}
\end{align}
%------------------------------------------------------------------------------
respectively, where $U_j$ is the mass flux defined as
%------------------------------------------------------------------------------
\begin{equation}
  U_j = J \xi_{j,j} u_j.
\end{equation}
%------------------------------------------------------------------------------
The compatibility of the convective term \citep{Morinishi_1998} is maintained 
by calculating the interpolated value $\bar{U_i}^{\xi_i}$ 
using the contravariant velocity $\xi_{i,i}u_i$ 
and discretizing the convection term.

%++++++++++++++++++++++++++++++++++++++++++++++++++++++++++++++++++++++++++++++
\subsection{Discretization of Faraday's equation}
\label{Faraday_fdm}
%++++++++++++++++++++++++++++++++++++++++++++++++++++++++++++++++++++++++++++++

Here, the author describes the discretization of Faraday's equation (\ref{Faraday}) 
and verifies that the magnetic flux density equation (\ref{magnetic_field}) 
can be discretely derived from Eq. (\ref{Faraday}) 
using compact interpolation \citep{Yanaoka_2023}. 
Equation (\ref{Faraday}) is discretized as follows:
%------------------------------------------------------------------------------
\begin{align}
  \frac{Wo^2}{Re} \delta_t B_i 
  &= - \epsilon_{ijk} \delta_{x_j} \bar{E_k}^t 
  = - \epsilon_{ijk} \frac{1}{\bar{J}^{\xi_i}} \delta_{\xi_j} J \xi_{j,j} \bar{E_k}^t 
  \nonumber \\
  &= 
  - \epsilon_{ijk} \frac{1}{\bar{J}^{\xi_i}} \delta_{\xi_j} J \xi_{j,j} 
    \left( \frac{1}{Re_m} \bar{j_k}^t - \epsilon_{klm} 
    \overline{\bar{u_l}^t}^{\xi_m} \overline{\bar{B_m}^t}^{\xi_l} \right) 
  \nonumber \\
  &= 
    \frac{1}{\bar{J}^{\xi_i}} \delta_{\xi_j} J \xi_{j,j} 
    \left( \epsilon_{ijk} \epsilon_{klm} 
    \overline{\bar{u_l}^t}^{\xi_m} \overline{\bar{B_m}^t}^{\xi_l} \right) 
  - \frac{1}{Re_m} \frac{1}{\bar{J}^{\xi_i}} 
    \epsilon_{ijk} \delta_{\xi_j} J \xi_{j,j} \bar{j_k}^t 
  \nonumber \\
  &= 
  - \frac{1}{\bar{J}^{\xi_i}} \delta_{\xi_j} \left( 
    J \xi_{j,j} \overline{\bar{u_j}^t}^{\xi_i} \overline{\bar{B_i}^t}^{\xi_j} 
  - J \xi_{j,j} \overline{\bar{B_j}^t}^{\xi_i} \overline{\bar{u_i}^t}^{\xi_j} \right) 
  - \frac{1}{Re_m} \frac{1}{\bar{J}^{\xi_i}} 
    \epsilon_{ijk} \delta_{\xi_j} J \xi_{j,j} \bar{j_k}^t, 
  \label{magnetic_flux_density_fdm}
\end{align}
%------------------------------------------------------------------------------
where $\epsilon_{ijk}\epsilon_{klm} = \delta_{il}\delta_{jm}-\delta_{im}\delta_{jl}$ is used. 
Evidently, Eqs. (\ref{Faraday}) and (\ref{magnetic_field}) can be discretely transformed into each other. 
Furthermore, when $Re_m = \infty$, 
the discretization equation of Eq. (\ref{magnetic_field}) also has a conservative form.

The discretization method for the convection terms, 
$\partial_{\xi_j} J \xi_{j,j} u_j B_i$ and $- \partial_{\xi_j} J \xi_{j,j} B_j u_i$, 
in this equation is different from that for the convection terms in the momentum equation (\ref{navier-stokes}). 
When discretizing the convection term $\partial_{\xi_j} J \xi_{j,j} u_j u_i$, 
the interpolated value 
$\overline{\bar{U_j}^t}^{\xi_i} \overline{\bar{u_i}^t}^{\xi_j}$ is used 
in Eq. (\ref{navier-stokes_fdm}) to satisfy the transformation of the convection terms \citep{Morinishi_1998}. 
In the conservation form, $\partial_{\xi_j} B_j B_i$, of the Lorentz force, 
the interpolated value $\overline{J \xi_{j,j}  B_j}^{\xi_i} \bar{B_i}^{\xi_j}$ 
is used in Eq. (\ref{compact_interpolation_F^c}). 
In Eq. (\ref{magnetic_flux_density_fdm}), the interpolated value 
$(J \xi_{j,j}) \overline{\bar{u_j}^t}^{\xi_i} \overline{\bar{B_i}^t}^{\xi_j}$ is used. 
As with the momentum equation, 
Eq. (\ref{Faraday}) can be discretized 
using each contravariant component of the velocity and magnetic flux density. 
However, the magnetic energy equation (\ref{energy_M}) cannot be derived 
discretely from Faraday's equation (\ref{Faraday}). 

Calculating the divergence of the formula (\ref{magnetic_flux_density_fdm}) 
at the cell center reveals that the time variation of $\partial_i B_i$ 
is discretely zero, as follows:
%------------------------------------------------------------------------------
\begin{align}
  \frac{Wo^2}{Re} \delta_t \delta_{x_i} B_i 
  &= - \delta_{x_i} \epsilon_{ijk} \delta_{x_j} \bar{E_k}^t 
  \nonumber \\
  &= 
  - \delta_{x_1} \left( \delta_{x_2} \bar{E_3}^t - \delta_{x_3} \bar{E_2}^t \right) 
  - \delta_{x_2} \left( \delta_{x_3} \bar{E_1}^t - \delta_{x_1} \bar{E_3}^t \right) 
  - \delta_{x_3} \left( \delta_{x_1} \bar{E_2}^t - \delta_{x_2} \bar{E_1}^t \right) 
  = 0.
  \label{time_variation_constraint}
\end{align}
%------------------------------------------------------------------------------

%++++++++++++++++++++++++++++++++++++++++++++++++++++++++++++++++++++++++++++++
\subsection{Discretization of the magnetic vector potential equation}
%++++++++++++++++++++++++++++++++++++++++++++++++++++++++++++++++++++++++++++++

As the magnetic vector potential $A_i$ is defined 
as $\check{B_i} = \epsilon_{ijk}\partial_j A_k$, 
it must satisfy the constraint (\ref{divergence_magnetic}) of magnetic flux density. 
Similarly to the current density $j_i$, 
the magnetic flux densities, $\check{B_1}$, $\check{B_2}$, and $\check{B_3}$, 
associated with the magnetic vector potential $A_i$ are defined 
at the midpoint of the cell edge, $(i, j+1/2, k+1/2)$, $(i+1/2, j, k+1/2)$, 
and $(i+1/2, j+1/2, k)$, respectively. 

Equation (\ref{vector_potential}) is discretized as follows:
%------------------------------------------------------------------------------
\begin{equation}
  \frac{Wo^2}{Re} \delta_t A_i 
  + \epsilon_{ijk} \overline{\check{\bar{B_j}^t} \overline{\bar{u_k}^t}^{\xi_i}}^{\xi_k} 
  = - \frac{1}{\bar{J}^{\xi_i}} \delta_{\xi_i} J \xi_{i,i} \bar{\psi}^t 
  - \frac{1}{Re_m} \bar{j_i}^t,
  \label{vector_potential_fdm}
\end{equation}
\begin{equation}
  j_i = \epsilon_{ijk} \delta_{x_j} \check{B_k},
\end{equation}
\begin{equation}
  \check{B_i} = \epsilon_{ijk} \delta_{x_j} A_k.
\end{equation}
%------------------------------------------------------------------------------
The divergence-free condition $\partial_i B_i = 0$ of the magnetic flux density 
of Eq. (\ref{divergence_magnetic}) is satisfied 
at the grid point $(i+1/2, j+1/2, k+1/2)$ as follows:
%------------------------------------------------------------------------------
\begin{equation}
  \frac{\partial \check{B_i}}{\partial x_i} 
  = \delta_{x_1} (\delta_{x_2} A_3 - \delta_{x_3} A_2) 
  + \delta_{x_2} (\delta_{x_3} A_1 - \delta_{x_1} A_3) 
  + \delta_{x_3} (\delta_{x_1} A_2 - \delta_{x_2} A_1) = 0.
\end{equation}
%------------------------------------------------------------------------------
As described above, 
when Eq. (\ref{vector_potential}) is discretized as Eq. (\ref{vector_potential_fdm}), 
the magnetic vector potential that satisfies the constraint condition 
of magnetic flux density can be obtained. 

Subsequently, the magnetic flux density equation (\ref{magnetic_field}) can be obtained 
by rotating the magnetic vector potential equation (\ref{vector_potential}). 
Coordinate transformations are not required; 
hence, the following discretized magnetic vector potential equation is used:
%------------------------------------------------------------------------------
\begin{equation}
  \frac{Wo^2}{Re} \delta_t A_k 
  = - \epsilon_{klm} \overline{\check{\bar{B_l}^t} \overline{\bar{u_m}^t}^{x_k}}^{x_m} 
  - \delta_{x_k} \bar{\psi} - \frac{1}{Re_m} \bar{j_k}^t.
\end{equation}
%------------------------------------------------------------------------------
Calculating the rotation of the above equation gives 
the discretized equation for the magnetic flux density as follows:
%------------------------------------------------------------------------------
\begin{align}
  \frac{Wo^2}{Re} \delta_t \epsilon_{ijk} \delta_{x_j} A_k 
  = \frac{Wo^2}{Re} \delta_t \check{\bar{B_i}^t} 
  &= 
  - \epsilon_{ijk} \delta_{x_j} \left( 
  \epsilon_{klm} \overline{\check{\bar{B_l}^t} \overline{\bar{u_m}^t}^{x_k}}^{x_m} 
  - \delta_{x_k} \bar{\psi}^t - \frac{1}{Re_m} \bar{j_k}^t \right) 
  \nonumber \\
  &= 
  - \epsilon_{ijk} \epsilon_{klm} \delta_{x_j} \overline{\check{\bar{B_l}^t} \overline{\bar{u_m}^t}^{x_k}}^{x_m} 
  - \frac{1}{Re_m} \epsilon_{ijk} \delta_{x_j} \bar{j_k}^t,
\end{align}
%------------------------------------------------------------------------------
where $\epsilon_{ijk} \epsilon_{klm} = \delta_{il}\delta_{jm}-\delta_{im}\delta_{jl}$. 
The equations for the magnetic flux densities, 
$\check{B_1}$, $\check{B_2}$, and $\check{B_3}$, can be discretely derived 
at the midpoints of the cell edge, $(i, j+1/2, k+1/ 2) $, $(i+1/2, j, k+1/2)$, 
and $(i+1/2, j+1/2, k)$, respectively.

The Lorentz force can also be obtained using the magnetic vector potential. 
The nonconservative Lorentz force is discretely expressed as follows:
%------------------------------------------------------------------------------
\begin{equation}
  F_i = \frac{1}{Al^2} \frac{1}{\bar{J}^{\xi_i}} 
  \overline{\epsilon_{ijk} 
  \overline{J \bar{j_j}^{\xi_j}}^{\xi_i} \overline{\check{B_k}}^{\xi_j}}^{\xi_k}.
  \label{vector_potential_F}
\end{equation}
%------------------------------------------------------------------------------
As the Lorentz force (\ref{vector_potential_F}) is determined by the second-order differential 
of the magnetic vector potential, the above-discretized formula may decrease accuracy. 
Additionally, the conservation of momentum and total energy deteriorate 
in an ideal periodic inviscid MHD flow 
because Eq. (\ref{vector_potential_F}) cannot be discretely transformed into a conservative form of the Lorentz force.

Furthermore, the magnetic energy equation can be derived 
using the magnetic flux density $\check{B_i}$ calculated from the magnetic vector potential $A_i$. 
However, as many transformations of the dependent variable occur, 
numerous interpolations are required. 
Therefore, the magnetic energy equation is derived 
from the discretized Faraday's equation, 
namely, the discretized magnetic flux density equation (\ref{magnetic_flux_density_fdm}), 
as described in Subsection \ref{subsection_magnetic_energy_eq}.

%++++++++++++++++++++++++++++++++++++++++++++++++++++++++++++++++++++++++++++++
\subsection{Derivation of the magnetic energy equation}
\label{subsection_magnetic_energy_eq}
%++++++++++++++++++++++++++++++++++++++++++++++++++++++++++++++++++++++++++++++

The conservation equation (\ref{energy_inviscid_Et}) for the total energy $E_t$ 
is derived from the equations of the kinetic and magnetic energies, 
Eqs. (\ref{energy_K}) and (\ref{energy_M}), respectively. 
Morinishi \citep{Morinishi_1998} reported that kinetic energy is discretely conserved 
when using an appropriate finite difference method. 
If the magnetic energy equation (\ref{energy_M}) can be derived 
discretely from Faraday's equation (\ref{Faraday}), 
the discrete total energy conservation equation can be derived. 
The magnetic energy $M$ is defined at the cell center $(i, j, k)$ as follows:
%------------------------------------------------------------------------------
\begin{equation}
  M = \frac{1}{2 Al^2} \frac{1}{J} \overline{\bar{J}^{\xi_i} B_i B_i}^{\xi_i} 
\end{equation}
%------------------------------------------------------------------------------
Calculating the inner product of the discretized Faraday's equation (\ref{magnetic_flux_density_fdm}) 
with the magnetic flux density $B_i^{n+1/2}$ facilitates 
the derivation of the discretized equation of the magnetic energy 
as described in \citep{Yanaoka_2023}
%------------------------------------------------------------------------------
\begin{align}
  \frac{Wo^2}{Re} \delta_t M 
  &= - \frac{1}{Al^2} \frac{1}{J} 
  \overline{\bar{J}^{x_i} \bar{B_i}^t \epsilon_{ijk} \delta_{x_j} \bar{E_k}^t}^{x_i} 
  \nonumber \\
  &= - \frac{1}{Al^2} \frac{1}{J} 
  \overline{\bar{B_i}^t \epsilon_{ijk} \delta_{\xi_j} J \xi_{j,j} \bar{E_k}^t}^{\xi_i} 
  \nonumber \\
  &= - \frac{1}{Al^2} \frac{1}{J} \left( 
  \delta_{\xi_j} \overline{\epsilon_{ijk} J \xi_{j,j} \bar{E_k}^t \overline{\bar{B_i}^t}^{\xi_j}}^{\xi_i} 
  - \overline{\overline{\epsilon_{ijk} J \xi_{j,j} \bar{E_k}^t \delta_{\xi_j} \bar{B_i}^t}^{\xi_j}}^{\xi_i} 
  \right) 
  \nonumber \\
  &= - \frac{1}{Al^2} \frac{1}{J} \left( 
  \delta_{\xi_j} \overline{\epsilon_{ijk} J \xi_{j,j} \bar{E_k}^t \overline{\bar{B_i}^t}^{\xi_j}}^{\xi_i} 
  + \overline{\overline{\bar{E_k}^t J \epsilon_{ijk} \delta_{x_i} \bar{B_j}^t}^{\xi_i}}^{\xi_j} 
  \right) 
  \nonumber \\
  &= - \frac{1}{Al^2} \frac{1}{J} \left( 
  \delta_{\xi_j} \overline{\epsilon_{ijk} J \xi_{j,j} \bar{E_k}^t \overline{\bar{B_i}^t}^{\xi_j}}^{\xi_i} 
  + \overline{\overline{\bar{E_k}^t J \bar{j_k}^t}^{\xi_i}}^{\xi_j} 
  \right).
\end{align}
%------------------------------------------------------------------------------
As the electric field is defined at the same point as the current density, 
it is given by $E_k = j_k/Re_m - \epsilon_{kij} \bar{u_i}^{\xi_j} \bar{B_j}^{\xi_i}$. 
The above equation can be transformed using $E_k$ as follows:
%------------------------------------------------------------------------------
\begin{align}
  \frac{Wo^2}{Re} \delta_t M 
  &= - \frac{1}{Al^2} \frac{1}{J} \left[ 
  \delta_{\xi_j} \overline{\epsilon_{ijk} J \xi_{j,j} \bar{E_k}^t \overline{\bar{B_i}^t}^{\xi_j}}^{\xi_i} 
  + \overline{\overline{J \left( 
  \frac{1}{Re_m} \bar{j_k}^t - \epsilon_{kij} \overline{\bar{u_j}^t}^{\xi_j} \overline{\bar{B_j}^t}^{\xi_i} 
  \right) \bar{j_k}^t}^{\xi_i}}^{\xi_j} \right] 
  \nonumber \\
  &= - \frac{1}{Al^2} \frac{1}{J} \left[ 
  \delta_{\xi_j} \overline{\epsilon_{ijk} J \xi_{j,j} \bar{E_k}^t \overline{\bar{B_i}^t}^{\xi_j}}^{\xi_i} 
  + \overline{\overline{J \frac{1}{Re_m} \bar{j_k}^t \bar{j_k}^t}^{\xi_i}}^{\xi_j} 
  - \overline{\overline{(J \epsilon_{kij} \overline{\bar{u_i}^t}^{\xi_j} \overline{\bar{B_j}^t}^{\xi_i}) \bar{j_k}^t}^{\xi_i}}^{\xi_j} 
  \right] 
  \nonumber \\
  &= - \frac{1}{Al^2} \frac{1}{J} \left[ 
  \delta_{\xi_j} \overline{\epsilon_{ijk} J \xi_{j,j} \bar{E_k}^t \overline{\bar{B_i}^t}^{\xi_j}}^{\xi_i} 
  + \overline{\overline{\frac{1}{Re_m} J \bar{j_k}^t \bar{j_k}^t}^{\xi_i}}^{\xi_j} 
  + \overline{\overline{\overline{\bar{u_i}^t}^{\xi_k} (J \epsilon_{ijk} \bar{j_j}^t \overline{\bar{B_k}^t}^{\xi_i})}^{\xi_i}}^{\xi_k} 
  \right]. \quad
  \label{energy_M_fdm}
\end{align}
%------------------------------------------------------------------------------
The third term on the right side is the work done by the Lorentz force. 
Additionally, the third term is interpolated using the Jacobian $J$. 
The interpolated form of the Lorentz force is consistent with 
Eq. (\ref{compact_interpolation_F^nc}).

Further, the time derivative term of Eq. (\ref{energy_M}) is considered. 
Applying the implicit midpoint rule to the time derivative affords 
the time derivative of the magnetic energy as follows:
%------------------------------------------------------------------------------
\begin{align}
  \frac{Wo^2}{Re} \frac{1}{Al^2} \bar{B_i}^t \delta_t B_i 
  &= \frac{Wo^2}{Re} \frac{1}{Al^2} 
    \frac{1}{J} \overline{\bar{J}^{\xi_i} \bar{B_i}^t \delta_t B_i}^{\xi_i} 
  \nonumber \\
  &= \frac{Wo^2}{Re} \frac{1}{Al^2} 
    \frac{1}{J} \overline{\bar{J}^{\xi_i} \delta_t B_i^2/2}^{\xi_i} 
  = \frac{Wo^2}{Re} \delta_t M.
  \label{energy_M_time}
\end{align}
%------------------------------------------------------------------------------
The magnetic energy equation (\ref{energy_M}) can be derived discretely 
in both time and space directions.

%++++++++++++++++++++++++++++++++++++++++++++++++++++++++++++++++++++++++++++++
\subsection{Derivation of the total energy equation}
%++++++++++++++++++++++++++++++++++++++++++++++++++++++++++++++++++++++++++++++

Assuming that the kinematic viscosity and magnetic diffusivity are zeros, 
Eq. (\ref{energy_inviscid_Et}) for total energy $E_t = K + M$ holds 
even discretely, and total energy is conserved. 
The conservation properties of momentum and kinetic energy 
have been elucidated in \citep{Morinishi_1998}. 
Herein, the total energy equation is derived using the discretely derived kinetic energy equation. 

The kinetic energy $K$ is defined at the cell center $(i, j, k)$ as follows:
%------------------------------------------------------------------------------
\begin{equation}
  K = \frac{1}{2} \frac{1}{J} \overline{\bar{J}^{\xi_i} u_i u_i}^{\xi_i}.
  \label{kinetic_energy_fdm}
\end{equation}
%------------------------------------------------------------------------------
For inviscid fluids, the discretized Eq. (\ref{energy_K}) is given 
by the inner product of the discretized Eq. (\ref{navier-stokes_fdm}) 
and velocity $u_i^{n+1/2}$ as in \citep{Morinishi_1998}
%------------------------------------------------------------------------------
\begin{align}
  \delta_t K 
  &= - \frac{1}{J} \delta_{\xi_j} \left( 
  \overline{ \overline{\bar{U_j}^t}^{\xi_i} 
  \frac{1}{2} \widetilde{\bar{u_i}^t \bar{u_i}^t}^{\xi_j}}^{\xi_i}  \right) 
  - \frac{1}{J} \overline{\frac{1}{2} \bar{u_i}^t \bar{u_i}^t \overline{\delta_{\xi_j} \bar{U_j}^t}^{\xi_i}}^{\xi_i} 
  \nonumber \\
  & \quad 
  - \frac{1}{J} \delta_{\xi_i} J \xi_{i,i} \bar{u_i}^t \overline{\bar{p}^t}^{\xi_i} 
  + \frac{1}{J} \bar{p}^t \delta_{\xi_i} \bar{U_i}^t 
  + \frac{1}{Al^2} \frac{1}{J} \overline{
  \bar{u_i}^t \overline{\epsilon_{ijk} \overline{\bar{J}^{\xi_i}}^{\xi_k} 
  \bar{j_j}^t \overline{\bar{B_k}^t}^{\xi_i}}^{\xi_k}}^{\xi_i}, 
  \label{inviscid_energy_K_fdm}
\end{align}
%------------------------------------------------------------------------------
where $Wo = \sqrt{Re}$ is set to remove $Re$. 
If the continuity equation (\ref{continuity_fdm}) is satisfied discretely, 
the second and fourth terms on the right side of Eq. (\ref{inviscid_energy_K_fdm}) can be ignored. 
In Eq. (\ref{inviscid_energy_K_fdm}), 
the work from the Lorentz force appears in the last term on the right side.

For $Re_m = \infty$, 
the discretized magnetic energy equation (\ref{energy_M_fdm}) is expressed as follows:
%------------------------------------------------------------------------------
\begin{equation}
  \delta_t M 
  = - \frac{1}{Al^2} \frac{1}{J} \left[ 
  \delta_{\xi_j} \overline{\epsilon_{ijk} J \xi_{j,j} \bar{E_k}^t \overline{\bar{B_i}^t}^{\xi_j}}^{\xi_i} 
  + \overline{\overline{\overline{\bar{u_i}^t}^{\xi_k} (J \epsilon_{ijk} \bar{j_j}^t \overline{\bar{B_k}^t}^{\xi_i})}^{\xi_i}}^{\xi_k} 
  \right].
  \label{ideal_energy_M_fdm}
\end{equation}
%------------------------------------------------------------------------------

The total energy $E_t = K + M$ is defined at the cell center $(i, j, k)$ as follows:
%------------------------------------------------------------------------------
\begin{equation}
  E_t = \frac{1}{2} \frac{1}{J} \left( 
    \overline{\bar{J}^{\xi_i} u_i u_i}^{\xi_i}
  + \frac{1}{Al^2} \overline{\bar{J}^{\xi_i} B_i B_i}^{\xi_i} \right).
\end{equation}
%------------------------------------------------------------------------------
Taking the sum of Eqs. (\ref{inviscid_energy_K_fdm}) and (\ref{ideal_energy_M_fdm}) 
yields the total energy equation as follows:
%------------------------------------------------------------------------------
\begin{align}
  \delta_t E_t 
  &= 
  - \frac{1}{J} \delta_{\xi_j} \left( 
    \frac{1}{2} \overline{ \overline{\bar{U_j}^t}^{\xi_i} 
    \widetilde{\bar{u_i}^t \bar{u_i}^t}^{\xi_j}}^{\xi_i}  \right) 
  - \frac{1}{J} \delta_{\xi_i} J \xi_{i,i} \bar{u_i}^t \overline{\bar{p}^t}^{\xi_i} 
  \nonumber \\
  & 
  + \frac{1}{Al^2} \frac{1}{J} \left[ 
  \overline{
  \bar{u_i}^t \overline{\epsilon_{ijk} \overline{\bar{J}^{\xi_i}}^{\xi_k} 
  \bar{j_j}^t \overline{\bar{B_k}^t}^{\xi_i}}^{\xi_k}}^{\xi_i} 
  - \delta_{\xi_j} \overline{J \xi_{j,j} \epsilon_{ijk} \bar{E_k}^t \overline{\bar{B_i}^t}^{\xi_j}}^{\xi_i} 
  - \overline{\overline{\overline{\bar{u_i}^t}^{\xi_k} (J \epsilon_{ijk} \bar{j_j}^t \overline{\bar{B_k}^t}^{\xi_i})}^{\xi_i}}^{\xi_k} \right].
  \label{total_energy_eq1_fdm}
\end{align}
%------------------------------------------------------------------------------
By applying the implicit midpoint rule to the time derivative, 
the total energy equation (\ref{energy_inviscid_Et}) can be derived discretely 
in both time and space directions \citep{Yanaoka_2023}. 
The work done by the Lorentz force appears in the third and fifth terms on the right side. 
Two terms $\overline{\bar{u_i}^t \overline{\epsilon_{ijk} \overline{\bar{J}^{\xi_i}}^{\xi_k} \bar{j_j}^t \overline{\bar{B_k}^t}^{\xi_i}}^{\xi_k}}^{\xi_i}$ 
and $\overline{\overline{\overline{\bar{u_i}^t}^{\xi_k} (J \epsilon_{ijk} \bar{j_j}^t \overline{\bar{B_k}^t}^{\xi_i})}^{\xi_i}}^{\xi_k}$ 
have the same form of weighted interpolation by the Jacobian 
but with a different interpolation form. 
If these terms approximately cancel each other, 
the total energy is preserved even discretely.

%++++++++++++++++++++++++++++++++++++++++++++++++++++++++++++++++++++++++++++++
\subsection{Derivation of the cross-helicity equation}
%++++++++++++++++++++++++++++++++++++++++++++++++++++++++++++++++++++++++++++++

The cross-helicity $H_c $ is defined at the cell center $(i, j, k)$ as follows:
%------------------------------------------------------------------------------
\begin{equation}
  H_c = \frac{1}{Al} \frac{1}{J} \overline{\bar{J}^{\xi_i} u_i B_i}^{\xi_i}.
\end{equation}
%------------------------------------------------------------------------------
The time derivative of the cross-helicity is expressed discretely as follows:
%------------------------------------------------------------------------------
\begin{equation}
  \delta_t H_c 
  = \frac{1}{Al} \frac{1}{J} \overline{\delta_t \bar{J}^{\xi_i} (u_i B_i)}^{\xi_i} 
  = \frac{1}{Al} \frac{1}{J} 
  \left( \overline{\bar{J}^{\xi_i} \bar{B_i}^t \delta_t u_i}^{\xi_i} 
       + \overline{\bar{J}^{\xi_i} \bar{u_i}^t \delta_t B_i}^{\xi_i} \right).
  \label{cross_helicity_time_fdm}
\end{equation}
%------------------------------------------------------------------------------
By calculating the inner product of the discretized Eq. (\ref{navier-stokes_fdm}) at $Re=\infty$ 
and magnetic flux density $B_i$, 
the first term of Eq. (\ref{cross_helicity_time_fdm}) is obtained as follows:
%------------------------------------------------------------------------------
\begin{align}
  \overline{\bar{J}^{\xi_i} \bar{B_i}^t \delta_t u_i}^{\xi_i} 
  &= - \delta_{\xi_j} \overline{\overline{\bar{U_j}^t}^{\xi_i} \overline{\bar{u_i}^t}^{\xi_j} \overline{\bar{B_i}^t}^{\xi_j}}^{\xi_i} 
  + \overline{\overline{\overline{\bar{U_j}^t}^{\xi_i} \overline{\bar{u_i}^t}^{\xi_j} \delta_{\xi_j} \bar{B_i}^t}^{\xi_j}}^{\xi_i} 
  - \delta_{\xi_i} J \xi_{j,j} \bar{B_i}^t \overline{\bar{p}^t}^{\xi_i} 
  + \bar{p}^t \delta_{\xi_i} J \xi_{j,j} \bar{B_i}^t 
  \nonumber \\
  &
  + \frac{1}{Al^2} \overline{\bar{J}^{\xi_i} \bar{B_i}^t 
    \epsilon_{ijk} \overline{\bar{j_j}^t \overline{\bar{B_k}^t}^{\xi_i}}^{\xi_k} }^{\xi_i}.
  \label{cross_helicity_time_fdm_1}
\end{align}
%------------------------------------------------------------------------------
In the inner product of the Lorentz force and the magnetic flux density, 
because $B_1 F_1 = B_1(j_2 B_3 - j_3 B_2)$ and $B_3 F_3 = B_3(j_1 B_2 - j_2 B_1)$, 
the first term of $B_1 F_1$ and the second term of $B_3 F_3$ cancel each other. 
However, in the discrete inner product, these terms do not strictly cancel. 
The accuracy of the inner product $B_i F_i = 0$ may change with the interpolation method.

By calculating the inner product of the discretized Eq. (\ref{magnetic_flux_density_fdm}) 
at $Re_m = \infty$ and velocity $u_i$, 
the second term of Eq. (\ref{cross_helicity_time_fdm}) is obtained as
%------------------------------------------------------------------------------
\begin{align}
  \overline{\bar{J}^{\xi_i} \bar{u_i}^t \delta_t B_i}^{\xi_i} 
  &= 
  - \overline{\bar{u_i}^t \overline{J \xi_{j,j} \overline{\bar{u_j}^t}^{\xi_j} \delta_{\xi_j} \bar{B_i}^t}^{\xi_j}}^{\xi_i} 
  - \overline{\bar{u_i}^t \bar{B_i}^t J \xi_{j,j} \overline{\delta_{\xi_j} \overline{\bar{u_j}^t}^{\xi_i}}^{\xi_i}}^{\xi_i} 
  \nonumber \\
  &
  + \overline{\frac{1}{2} \delta_{\xi_j} J \xi_{j,j} \overline{\bar{B_j}^t}^{\xi_i} \widetilde{\bar{u_i}^t \bar{u_i}^t}^{\xi_j}}^{\xi_i} 
  + \overline{\frac{1}{2} \bar{u_i}^t \bar{u_i}^t J_{\xi_j,j} \overline{\delta_{\xi_j} \bar{B_j}^t}^{\xi_i}}^{\xi_i}.
  \label{cross_helicity_time_fdm_2}
\end{align}
%------------------------------------------------------------------------------
The second term $\overline{\overline{\overline{\bar{U_j}^t}^{\xi_i} \overline{\bar{u_i}^t}^{\xi_j} \delta_{\xi_j} \bar{B_i}^t}^{\xi_j}}^{\xi_i}$ 
on the right side of Eq. (\ref{cross_helicity_time_fdm_1}) 
and the first term $-\overline{\bar{u_i}^t \overline{J \xi_{j,j} \overline{\bar{u_j}^t}^{\xi_j} \delta_{\xi_j} \bar{B_i}^t}^{\xi_j}}^{\xi_i}$ 
on the right side of Eq. (\ref{cross_helicity_time_fdm_2}) 
have different interpolation forms; 
therefore, these terms approximately cancel but not exactly.
Additionally, if $\partial_i u_i = 0$ is satisfied, 
the second term $-\overline{\bar{u_i}^t \bar{B_i}^t J \xi_{j,j} \overline{\delta_{\xi_j} \overline{\bar{u_j}^t}^{\xi_i}}^{\xi_i}}^{\xi_i}$ 
of Eq. (\ref{cross_helicity_time_fdm_2})is negligible for uniform grids 
but not negligible for nonuniform grids. 
Similarly, if $\partial_i B_i = 0$, the fourth term $\overline{\frac{1}{2} \bar{u_i}^t \bar{u_i}^t J_{\xi_j,j} \overline{\delta_{\xi_j} \bar{B_j}^t}^{\xi_i}}^{\xi_i}$ 
of Eq. (\ref{cross_helicity_time_fdm_2}) is negligible for uniform grids 
but not negligible for nonuniform grids. 
The time derivative of cross-helicity (\ref{cross_helicity_time_fdm}) is expressed as follows:
%------------------------------------------------------------------------------
\begin{align}
  \delta_t H_c 
  &= \frac{1}{Al} \frac{1}{J} \left[ 
  - \delta_{\xi_j} \overline{\overline{\bar{U_j}^t}^{\xi_i} \overline{\bar{u_i}^t}^{\xi_j} \overline{\bar{B_i}^t}^{\xi_j}}^{\xi_i} 
  + \overline{\overline{\overline{\bar{U_j}^t}^{\xi_i} \overline{\bar{u_i}^t}^{\xi_j} \delta_{\xi_j} \bar{B_i}^t}^{\xi_j}}^{\xi_i} 
  - \delta_{\xi_i} J \xi_{j,j} \bar{B_i}^t \overline{\bar{p}^t}^{\xi_i} 
  + \bar{p}^t \delta_{\xi_i} J \xi_{j,j} \bar{B_i}^t 
  \nonumber \right. \\
  & \quad \left. 
  + \frac{1}{Al^2} \overline{\bar{J}^{\xi_i} \bar{B_i}^t 
    \epsilon_{ijk} \overline{\bar{j_j}^t \overline{\bar{B_k}^t}^{\xi_i}}^{\xi_k} }^{\xi_i} 
  - \overline{\bar{u_i}^t \overline{J \xi_{j,j} \overline{\bar{u_j}^t}^{\xi_j} \delta_{\xi_j} \bar{B_i}^t}^{\xi_j}}^{\xi_i} 
  - \overline{\bar{u_i}^t \bar{B_i}^t J \xi_{j,j} \overline{\delta_{\xi_j} \overline{\bar{u_j}^t}^{\xi_i}}^{\xi_i}}^{\xi_i} 
  \nonumber \right. \\
  & \quad \left. 
  + \overline{\frac{1}{2} \delta_{\xi_j} J \xi_{j,j} \overline{\bar{B_j}^t}^{\xi_i} \widetilde{\bar{u_i}^t \bar{u_i}^t}^{\xi_j}}^{\xi_i} 
  + \overline{\frac{1}{2} \bar{u_i}^t \bar{u_i}^t J_{\xi_j,j} \overline{\delta_{\xi_j} \bar{B_j}^t}^{\xi_i}}^{\xi_i} 
  \right].
  \label{cross_helicity_fdm}
\end{align}
%------------------------------------------------------------------------------
If $\partial_i u_i = 0$ and $\partial_i B_i = 0$ are satisfied, 
the second term on the right side of Eq. (\ref{cross_helicity_time_fdm_1}) 
and the first term on the right side of Eq. (\ref{cross_helicity_time_fdm_2}) 
cancel each other, 
and the inner product of the Lorentz force and magnetic flux density is zero. 
Subsequently, the aforementioned equation can be transformed as follows:
%------------------------------------------------------------------------------
\begin{align}
  \delta_t H_c 
  &= - \frac{1}{Al} \frac{1}{J} \left[ 
    \delta_{\xi_j} \overline{\bar{U_j}^{\xi_i} \bar{u_i}^{\xi_j} \bar{B_i}^{\xi_j}}^{\xi_i} 
  + \delta_{\xi_i} J \xi_{j,j} B_i \bar{p}^{\xi_i} 
  - \overline{\frac{1}{2} \delta_{\xi_j} J \xi_{j,j} \bar{B_j}^{\xi_i} \widetilde{u_i u_i}^{\xi_j}}^{\xi_i} 
  \right].
\end{align}
%------------------------------------------------------------------------------
The above equation is approximately conservative. 
Although the cross-helicity equation (\ref{cross_helicity_fdm}) 
is not discretely conservative, 
the conservation property of cross-helicity is well maintained for uniform grids \citep{Yanaoka_2023}.

%++++++++++++++++++++++++++++++++++++++++++++++++++++++++++++++++++++++++++++++
\subsection{Derivation of the magnetic helicity equation}
%++++++++++++++++++++++++++++++++++++++++++++++++++++++++++++++++++++++++++++++

The magnetic helicity $H_m $ is defined at the cell center $(i, j, k)$ as follows:
%------------------------------------------------------------------------------
\begin{equation}
  H_m = \frac{1}{Al^2} \frac{1}{J} \overline{\bar{J}^{\xi_i} B_i A_i}^{\xi_i}.
\end{equation}
%------------------------------------------------------------------------------
The time derivative of the magnetic helicity is expressed discretely as follows:
%------------------------------------------------------------------------------
\begin{equation}
  \delta_t H_m 
  = \frac{1}{Al^2} \frac{1}{J} \overline{\delta_t \bar{J}^{\xi_i} (B_i A_i)}^{\xi_i} 
  = \frac{1}{Al^2} \frac{1}{J} 
  \left( \overline{\bar{J}^{\xi_i} \bar{A_i}^t \delta_t B_i}^{\xi_i} 
       + \overline{\bar{J}^{\xi_i} \bar{B_i}^t \delta_t A_i}^{\xi_i} \right).
  \label{magnetic_helicity_time_fdm}
\end{equation}
%------------------------------------------------------------------------------
By calculating the inner product of the discretized equation (\ref{magnetic_flux_density_fdm}) 
at $Re_m = \infty$ and magnetic vector potential $A_i$, 
the first term of Eq. (\ref{magnetic_helicity_time_fdm}) is obtained as follows:
%------------------------------------------------------------------------------
\begin{align}
  \overline{\bar{J}^{\xi_i} \bar{A_i}^t \delta_t B_i}^{\xi_i} 
  &= - \overline{\bar{J}^{\xi_i} \bar{A_i}^t \epsilon_{ijk} \frac{1}{\bar{J}^{\xi_j}} \delta_{\xi_j} J \xi_{j,j} \bar{E_k}^t}^{\xi_i} 
  = - \epsilon_{ijk} \left( \delta_{\xi_j} J \xi_{j,j} \overline{\bar{A_i}^t}^{\xi_j} \bar{E_k}^t 
  - \overline{\bar{E_k}^t \delta_{\xi_j} J \xi_{j,j} \bar{A_i}^t}^{\xi_j} \right) 
  \nonumber \\
  &= - \delta_{\xi_j} \epsilon_{jki} J \xi_{j,j} \overline{\bar{A_i}^t}^{\xi_j} \bar{E_k}^t 
  + \overline{\epsilon_{kij} \bar{E_k}^t \delta_{\xi_i} J \xi_{i,i} \bar{A_j}^t}^{\xi_j} 
  \nonumber \\
  &= - \delta_{\xi_j} \epsilon_{jki} J \xi_{j,j} \overline{\bar{A_i}^t}^{\xi_j} \bar{E_k}^t 
  + \overline{(\epsilon_{kij} \overline{\bar{u_i}^t}^{\xi_j} \overline{\bar{B_j}^t}^{\xi_i}) (J \check{\bar{B_k}^t})}^{\xi_j}.
\end{align}
%------------------------------------------------------------------------------
By calculating the inner product of the discretized equation (\ref{vector_potential_fdm}) 
at $Re_m = \infty$ and magnetic flux density $B_i$, 
the second term of Eq. (\ref{magnetic_helicity_time_fdm}) is obtained as follows:
%------------------------------------------------------------------------------
\begin{align}
  \overline{\bar{J}^{\xi_i} \bar{B_i}^t \delta_t A_i}^{\xi_i} 
  &= - \overline{
  \epsilon_{ijk} \bar{J}^{\xi_i} \bar{B_i}^t \overline{\check{\bar{B_j}^t} \overline{\bar{u_k}^t}^{\xi_i}}^{\xi_k} 
  }^{\xi_i} 
  - \overline{\bar{B_i}^t \delta_{\xi_i} J \xi_{i,i} \bar{\psi}^t}^{\xi_i} 
  \nonumber \\
  &= - \overline{ 
  \epsilon_{ijk} \bar{J}^{\xi_i} \bar{B_i}^t \overline{\check{\bar{B_j}^t} \overline{\bar{u_k}^t}^{\xi_i}}^{\xi_k} 
  }^{\xi_i} 
  - \delta_{\xi_i} J \xi_{i,i} \bar{B_i}^t \overline{\bar{\psi}^t}^{\xi_i} 
  + \bar{\psi}^t \delta_{\xi_i} J \xi_{i,i} \bar{B_i}^t. 
  \label{magnetic_helocity_eq_fdm}
\end{align}
%------------------------------------------------------------------------------
Therefore, 
the time derivative (\ref{magnetic_helicity_time_fdm}) of the magnetic helicity 
is expressed as follows:
%------------------------------------------------------------------------------
\begin{align}
  \delta_t H_m 
  &= \frac{1}{Al^2} \frac{1}{J} \left[ 
  - \delta_{\xi_j} \epsilon_{jki} J \xi_{j,j} \overline{\bar{A_i}^t}^{\xi_j} \bar{E_k}^t 
  + \overline{(\epsilon_{kij} \overline{\bar{u_i}^t}^{\xi_j} \overline{\bar{B_j}^t}^{\xi_i}) (J \check{\bar{B_k}^t})}^{\xi_j} 
  \right. \nonumber \\
  & \quad \left. 
  - \overline{ 
  \epsilon_{ijk} \bar{J}^{\xi_i} \bar{B_i}^t \overline{\check{\bar{B_j}^t} \overline{\bar{u_k}^t}^{\xi_i}}^{\xi_k} 
  }^{\xi_i} 
  - \delta_{\xi_i} J \xi_{i,i} \bar{B_i}^t \overline{\bar{\psi}^t}^{\xi_i} 
  + \bar{\psi}^t \delta_{\xi_i} J \xi_{i,i} \bar{B_i}^t \right]. 
\end{align}
%------------------------------------------------------------------------------
If $\partial_i B_i = 0$ is satisfied discretely, 
the last term on the right side of the above equation approaches zero asymptotically. 
Additionally, because the two vectors, $B_i$ and $\epsilon_{ijk} B_j u_k$, are orthogonal, 
their inner product is zero. 
However, the inner product of two vectors is not strictly zero 
in the discretized equation. 
Therefore, if $\overline{(\epsilon_{kij} \overline{\bar{u_i}^t}^{\xi_j} \overline{\bar{B_j}^t}^{\xi_i}) (J \check{\bar{B_k}^t})}^{\xi_j}$ 
and $\overline{\epsilon_{ijk} \bar{J}^{\xi_i} \bar{B_i}^t \overline{\check{\bar{B_j}^t} \overline{\bar{u_k}^t}^{\xi_i}}^{\xi_k}}^{\xi_i}$ 
discretely approach zeros, then magnetic helicity is preserved discretely.

%++++++++++++++++++++++++++++++++++++++++++++++++++++++++++++++++++++++++++++++
\subsection{Discretization of Poisson's equation for the electric potential}
\label{Ohm_disctretization}
%++++++++++++++++++++++++++++++++++++++++++++++++++++++++++++++++++++++++++++++

Poisson's equation for electric potential (\ref{electric_potential_poisson}) 
can be transformed using the Coulomb gauge $\partial_i A_i = 0$ to the following:
%------------------------------------------------------------------------------
\begin{align}
  \frac{\partial j_i}{\partial x_i} 
  &= Re_m \frac{\partial}{\partial x_i} 
  \left( - \frac{\partial \psi}{\partial x_i} + \epsilon_{ijk} u_j B_k \right) 
  \nonumber \\
  &= Re_m \left( - \frac{\partial^2 \psi}{\partial x_i^2} 
  + \epsilon_{ijk} B_i \frac{\partial u_k}{\partial x_j} 
  - \epsilon_{ijk} u_i \frac{\partial B_k}{\partial x_j} \right) = 0.
  \label{electric_potential_poisson2}
\end{align}
%------------------------------------------------------------------------------
When Poisson's equation (\ref{electric_potential_poisson2}) 
is discretized, the following analytical relational expression must hold:
%------------------------------------------------------------------------------
\begin{equation}
  \frac{\partial}{\partial x_i} \left( \epsilon_{ijk} u_j B_k \right) 
  = \epsilon_{ijk} B_i \frac{\partial u_k}{\partial x_j} 
  - \epsilon_{ijk} u_i \frac{\partial B_k}{\partial x_j}.
  \label{relation}
\end{equation}
%------------------------------------------------------------------------------
Equation (\ref{relation}) implies that 
the outflow of the convective electric field $\epsilon_{ijk} u_j B_k$ occurs 
when the vortex deforms the magnetic flux density, 
and current density deforms velocity. 
Therefore, the computational method for the current density by interpolation 
is significant in capturing such a phenomenon.

When Eq. (\ref{electric_potential_poisson2}) is discretized, 
the relational expression (\ref{relation}) must be satisfied at the cell center. 
The current density is obtained as follows:
%------------------------------------------------------------------------------
\begin{equation}
  j_i = Re_m \left( - \delta_{x_i} \psi 
  + \overline{\epsilon_{ijk} \bar{u_j}^{x_j} \bar{B_k}^{x_k}}^{x_i} \right).
  \label{current_density_ohm}
\end{equation}
%------------------------------------------------------------------------------
When the current density is obtained via interpolation using the above formula, 
the relational expression (\ref{relation}) is discretely obtained as follows:
%------------------------------------------------------------------------------
\begin{equation}
  \delta_{x_i} \overline{\epsilon_{ijk} \bar{u_j}^{x_j} \bar{B_k}^{x_k}}^{x_i} 
  = \overline{\overline{\bar{B_i}^{x_i}}^{x_j} \epsilon_{ijk} \delta_{x_j} \bar{u_k}^{x_k}}^{x_j} 
  - \overline{\overline{\bar{u_i}^{x_i}}^{x_j} \epsilon_{ijk} \delta_{x_j} \bar{B_k}^{x_k}}^{x_j}.
  \label{relation_fdm}
\end{equation}
%------------------------------------------------------------------------------

When the magnetic vector potential is not calculated, 
the electric potential is obtained by solving the following discretized equation 
of the charge conservation law using Eq. (\ref{current_density_ohm}):
%------------------------------------------------------------------------------
\begin{equation}
  \frac{1}{J} \delta_{\xi_i} J \xi_{j,j} j_i = 0.
\end{equation}
%------------------------------------------------------------------------------
In contrast, when calculating the magnetic vector potential using Eq. (\ref{vector_potential_fdm}), 
the Coulomb gauge $\partial_i A_i = 0$ is used to obtain the electric potential.

%##############################################################################
\section{Numerical method}
\label{numerical_method}
%##############################################################################

In this study, the fully conservative finite difference method 
is applied to the analysis of MHD flows, 
as in \citep{Ham_et_al_2002,Morinishi_2009,Morinishi_2010}. 
The Newton method is used to solve unsteady solutions. 
The implicit midpoint rule is applied to Eqs. (\ref{navier-stokes}), 
(\ref{Faraday}), and (\ref{vector_potential}) as follows:
%------------------------------------------------------------------------------
\begin{subequations}
\begin{equation}
  \frac{Wo^2}{Re} \frac{u_i^{n+1,m+1} - u_i^n}{\Delta t} = H_{u_i}^{n+\lambda,m+1} 
  - \frac{\partial p^{n+\lambda,m+1}}{\partial x_i},
  \label{newton.u}
\end{equation}
\begin{align}
  H_{u_i}^{n+\lambda,m+1} 
  &= - \frac{\partial u_j^{n+\lambda,m+1} u_i^{n+\lambda,m+1}}{\partial x_j} 
  - \frac{\partial p^{n+\lambda,m+1}}{\partial x_i} 
  + \frac{1}{Re} \frac{\partial^2 u_i^{n+\lambda,m+1}}{\partial x_j^2} 
  \nonumber \\
  &
  + \frac{1}{Al^2} \epsilon_{ijk} j_j^{n+\lambda,m+1} B_k^{n+\lambda,m+1},
  \label{newton.H_u}
\end{align}
\end{subequations}
\begin{subequations}
\begin{equation}
  \frac{Wo^2}{Re} \frac{B_i^{n+1,m+1} - B_i^n}{\Delta t} = H_{B_i}^{n+\lambda,m+1},
  \label{newton.B}
\end{equation}
\begin{align}
%  H_{B_i}^{n+\lambda,m+1} 
%  &= - \frac{\partial (u_j^{n+\lambda,m+1} B_i^{n+\lambda,m+1} 
%                     - B_j^{n+\lambda,m+1} u_i^{n+\lambda,m+1})}{\partial x_j} 
%  - \frac{1}{Re_m} \epsilon_{ijk} \frac{\partial j_k^{n+\lambda,m+1}}{\partial x_j},
  H_{B_i}^{n+\lambda,m+1} 
  = - \epsilon_{ijk} \frac{\partial E_k^{n+\lambda,m+1}}{\partial x_j},
  \label{newton.H_B}
\end{align}
\end{subequations}
\begin{subequations}
\begin{equation}
  \frac{Wo^2}{Re} \frac{A_i^{n+1,m+1} - A_i^n}{\Delta t} = H_{A_i}^{n+\lambda,m+1} 
  - \frac{\partial \psi^{n+\lambda,m+1}}{\partial x_i},
  \label{newton.A}
\end{equation}
\begin{equation}
  H_{A_i}^{n+\lambda,m+1} 
  - \epsilon_{ijk} B_j^{n+\lambda,m+1} u_k^{n+\lambda,m+1} 
  - \frac{1}{Re_m} j_i^{n+\lambda,m+1},
  \label{newton.H_A}
\end{equation}
\end{subequations}
\begin{equation}
  u_i^{n+\lambda,m+1} = \lambda u_i^{n+1,m+1} + (1 - \lambda) u_i^n,
\end{equation}
\begin{equation}
  p^{n+\lambda,m+1} = \lambda p^{n+1,m+1} + (1 - \lambda) p^n,
\end{equation}
\begin{equation}
  B_i^{n+\lambda,m+1} = \lambda B_i^{n+1,m+1} + (1 - \lambda) B_i^n,
\end{equation}
\begin{equation}
  A_i^{n+\lambda,m+1} = \lambda A_i^{n+1,m+1} + (1 - \lambda) A_i^n,
\end{equation}
\begin{equation}
  \psi_i^{n+\lambda,m+1} = \lambda \psi_i^{n+1,m+1} + (1 - \lambda) \psi_i^n,
\end{equation}
%------------------------------------------------------------------------------
where $n$ and $m$ indicate the time and Newton iterative levels, respectively. 
Regarding the discretization of the time derivative, 
if $\lambda = 1$, the Euler implicit method is applied to the time derivative. 
If $\lambda = 1/2$, the implicit midpoint rule is applied. 
In this study, the conservation laws of mass and electric charge 
and the constraint of Gauss's law are discretely satisfied 
at the time $n+1$ level. 
Thus, Eq. (\ref{electric_charge_law}) is given 
by applying the Coulomb gauge $\partial_i A_i = 0$ as follows:
%------------------------------------------------------------------------------
\begin{equation}
  \frac{\partial j_i^{n+1,m+1}}{\partial x_i} = 0, \quad 
  j_i^{n+1,m+1} = Re_m \left( \frac{\partial \psi^{n+1,m+1}}{\partial x_i} 
  + \epsilon_{ijk} u_j^{n+1,m+1} B_k^{n+1,m+1} \right).
  \label{newton.psi}
\end{equation}
%------------------------------------------------------------------------------

The Yee scheme \citep{Yee_1966}, specifically designed for numerical simulations 
in which both the electric and magnetic fields are discretized on a staggered grid, is an explicit method 
to solve Maxwell's equations to satisfy the boundary conditions, 
is an explicit method. 
The temporal level of a magnetic field is shifted from that of an electric field 
by half a time step. 
When the flow field is also solved, the total energy is not discretely conserved 
for ideal inviscid MHD flows 
unless all dependent variables are defined at the same temporal level. 
Therefore, a total-energy conservative difference scheme cannot be constructed 
using the Yee method \citep{Yee_1966}. 
As explained in Section \ref{discretization}, 
by applying the implicit midpoint rule to Eq. (\ref{Faraday}), 
the total energy equation can be derived discretely. 
Moreover, total energy is discretely conserved in ideal periodic inviscid MHD flows. 
Considering the applications of MHD flow, 
the present implicit method is efficient for applicative calculations. 
The method proposed in this study is the same as in \citep{Yee_1966}, 
which is used for spatially shifting the definition points of the electric and magnetic fields. 
By adopting such a staggered grid, as explained in Section \ref{discretization}, 
the conservative and nonconservative forms 
of the Lorentz force can be interconverted, and such a transformation is discretely satisfied. 
Furthermore, the magnetic flux density equation (\ref{magnetic_field}) 
can be derived discretely from Faraday's equation (\ref{Faraday}).

By applying the simplified marker and cell (SMAC) method \citep{Amsden&Harlow_1970}, 
Eq. (\ref{newton.u}) is temporally split as follows:
%------------------------------------------------------------------------------
\begin{subequations}
\begin{equation}
  \frac{Wo^2}{Re} \frac{\hat{u_i}^{n+1,m+1} - u_i^n}{\Delta t} = H_{u_i}^{n+\lambda,m+1} 
  - \frac{\partial}{\partial x_i} 
    \left[ \lambda p^{n+1,m} + (1-\lambda) p^{n} \right],
  \label{predict.u}
\end{equation}
\begin{equation}
  \frac{Wo^2}{Re} \frac{u_i^{n+1,m+1} - \hat{u_i}^{n+1,m+1}}{\Delta t} 
  = - \lambda \frac{\partial \Delta p^m}{\partial x_i},
  \label{correct.u.1}
\end{equation}
\begin{equation}
  p^{n+1,m+1} = p^{n+1,m} + \Delta p^m,
  \label{correct.p.1}
\end{equation}
\end{subequations}
%------------------------------------------------------------------------------
where $\hat{u_i}^{n+1,m+1}$ is the predicted value of velocity, 
and $\Delta p^m$ is the pressure correction value. 
The velocity in $H_{u_i}^{n+\lambda,m+1}$ on the right side of Eq. (\ref{predict.u}) 
is defined as $u_i^{n+\lambda,m+1} = \lambda \hat{u_i}^{n+1,m+1} + (1 - \lambda) u_i^n$. 
When calculating the velocity $\hat{u_i}^{n+1,m+1}$, 
the convective term is linearized as 
$\partial_j (u_j^{n+\lambda,m} \hat{u_i}^{n+\lambda,m+1})$ using the $m$-level value. 
The magnetic flux density in the Lorentz force is also linearized as $B_i^{n+\lambda,m}$. 
Once the Newton iteration is completed, 
such a linearized approximation can be ignored, 
and second-order accuracy in the time integration is preserved. 
Taking the divergence of Eq. (\ref{correct.u.1}) 
and using the continuity equation (\ref{continuity}) at the $n+1$ level, 
Poisson's equation for the pressure correction value $\Delta p$ is derived as follows:
%------------------------------------------------------------------------------
\begin{equation}
  \lambda \frac{\partial}{\partial x_i} \frac{\partial \Delta p^m}{\partial x_i} 
  = \frac{Wo^2}{Re} \frac{1}{\Delta t} 
  \frac{\partial \hat{u_i}^{n+1,m+1}}{\partial x_i}.
  \label{poisson.1}
\end{equation}
%------------------------------------------------------------------------------

In the SMAC method \citep{Amsden&Harlow_1970}, 
the right side of Eq. (\ref{poisson.1}) enables self-regulation 
of the velocity divergence error, 
and a stable convergent solution can be obtained using an iterative solver 
such as the successive over-relaxation method. 
However, the iteration of Poisson's equation is time consuming. 
To satisfy the continuity condition, 
the velocity and pressure are relaxed simultaneously, 
as in \citep{Hirt_et_al_1975,Takemitsu_1985,Oki&Tanahashi_1993,Yanaoka&Inafune_2023,Yanaoka_2023}. 
The method in this study does not change the form of Poisson's equation (\ref{poisson.1}). 
Thus, simultaneous relaxation does not affect the stability 
when solving Poisson's equation. 
The simultaneous relaxation of velocity and pressure is performed as follows:
%------------------------------------------------------------------------------
\begin{subequations}
\begin{equation}
  \frac{Wo^2}{Re} \frac{u_i^{n+1,m+1,l+1} - u_i^{n+1,m+1,l}}{\Delta t} 
  = - \lambda \frac{\partial \Delta p^{m,l}}{\partial x_i},
  \label{correct.u.2}
\end{equation}
\begin{equation}
  p^{n+1,m+1,l+1} = p^{n+1,m,l} + \Delta p^{m,l},
  \label{correct.p.2}
\end{equation}
\begin{equation}
  \lambda \frac{\partial}{\partial x_i} \frac{\partial \Delta p^{m,l}}{\partial x_i} 
  = \frac{Wo^2}{Re} \frac{1}{\Delta t} 
  \frac{\partial u_i^{n+1,m+1,l}}{\partial x_i},
  \label{poisson.2}
\end{equation}
\end{subequations}
%------------------------------------------------------------------------------
where the superscript $l$ represents the number of iterations. 
When $l = 1$, let $u_i^{n+1,m+1,l} = \hat{u_i}^{n+1,m+1}$ and 
$p^{n+1,m+1,l} = p^{n+1,m}$. 
In such a scenario, the velocity and pressure are simultaneously relaxed. 
The calculation is repeated up to a predetermined iteration number. 
After the simultaneous relaxation, 
let $u_i^{n+1,m+1} = u_i^{n+1,m+1,l+1}$ and $p^{n+1,m+1} = p^{n+1,m+1,l+1}$. 
Equation (\ref{correct.u.2}) is used as the boundary condition 
to solve Eq. (\ref{poisson.2}). 
Takemitsu \citep{Takemitsu_1985} proposed a similar method 
that simultaneously iterates the velocity correction equation 
and Poisson's equation for the pressure correction. 
However, Poisson's equation for pressure should be solved after correcting the velocity. 
The present numerical method does not require Poisson's equation for obtaining pressure.

In MHD flow analyses, the magnetic flux density must be calculated 
while satisfying its constraint. 
As in \citep{Evans&Hawley_1988,Dumbser_et_al_2019,Yanaoka_2023}, 
Faraday's equation (\ref{Faraday}) is discretized 
such that its divergence is zero. 
Therefore, the magnetic flux density is not corrected, 
in contrast to existing studies \citep{Dedner_et_al_2002,Brackbill&Barnes_1980}. 
The discretization method is described in Subsection \ref{Faraday_fdm}.

The magnetic vector potential is calculated in the same manner as the velocity. 
The principle of the SMAC method \citep{Amsden&Harlow_1970} is applied 
to calculate Eq. (\ref{newton.A}) as follows:
%------------------------------------------------------------------------------
\begin{subequations}
\begin{equation}
  \frac{Wo^2}{Re} \frac{\hat{A_i}^{n+1,m+1} - A_i^n}{\Delta t} = H_{A_i}^{n+\lambda,m+1} 
  - \frac{\partial \psi^{n+\lambda,m}}{\partial x_i}, 
  \label{predict.A}
\end{equation}
\begin{equation}
  \frac{Wo^2}{Re} \frac{A_i^{n+1,m+1} - \hat{A_i}^{n+1,m+1}}{\Delta t} 
  = - \lambda \frac{\partial \Delta \psi^m}{\partial x_i},
  \label{correct.A.1}
\end{equation}
\begin{equation}
  \psi^{n+1,m+1} = \psi^{n+1,m} + \Delta \psi^m,
  \label{correct.psi.1}
\end{equation}
\end{subequations}
%------------------------------------------------------------------------------
where $\hat{A_i}^{n+1,m+1}$ is the predicted value of the magnetic vector potential, 
and $\Delta \psi$ is the correction for $\psi$. 
The magnetic vector potential in $H_{A_i}^{n+\lambda,m+1}$ on the right side of Eq. (\ref{predict.A}) 
is defined as $A_i^{n+\lambda,m+1} = \lambda \hat{A_i}^{n+1,m+1} + (1 - \lambda) A_i^n$. 
When calculating the magnetic vector potential $\hat{A_i}^{n+1,m+1}$, 
the convective term is linearized as 
$\epsilon_{ijk} (\epsilon_{jlm} \partial_l \hat{A}_m^{n+\lambda,m+1}) u_k^{n+\lambda,m}$ 
using the $m$-level value. 
Once the Newton iteration is completed, 
such a linearized approximation can be ignored, 
preserving second-order accuracy in the time integration. 
By applying the Coulomb gauge, 
taking the divergence of Eq. (\ref{correct.A.1}) 
and using the divergence-free condition of the magnetic vector potential at the $n+1$ level, 
Poisson's equation for the correction value $\Delta \psi$ is derived as
%------------------------------------------------------------------------------
\begin{equation}
  \lambda \frac{\partial}{\partial x_i} \frac{\partial \Delta \psi^m}{\partial x_i} 
  = \frac{Wo^2}{Re} \frac{1}{\Delta t} \frac{\partial \tilde{A_i}^{n+1,m+1}}{\partial x_i}.
  \label{poisson.p_A.1}
\end{equation}
%------------------------------------------------------------------------------
The magnetic vector potential can also be calculated via simultaneous relaxation 
similar to the velocity as follows:
%------------------------------------------------------------------------------
\begin{subequations}
\begin{equation}
  \frac{Wo^2}{Re} \frac{A_i^{n+1,m+1,l+1} - A_i^{n+1,m+1,l}}{\Delta t} 
  = - \lambda \frac{\partial \Delta \psi^{m,l}}{\partial x_i},
  \label{correct.A.2}
\end{equation}
\begin{equation}
   \psi^{n+1,m+1,l+1} = \psi^{n+1,m,l} + \Delta \psi^{m,l},
   \label{correct.psi.2}
\end{equation}
\begin{equation}
  \lambda \frac{\partial}{\partial x_i} \frac{\partial \Delta \psi^{m,l}}{\partial x_i} 
  = \frac{Wo^2}{Re} \frac{1}{\Delta t} \frac{\partial A_i^{n+1,m+1,l}}{\partial x_i},
  \label{poisson.psi.2}
\end{equation}
\end{subequations}
%------------------------------------------------------------------------------
where, when $l = 1$, 
let $A_i^{n+1,m+1,l} = \tilde{A_i}^{n+1,m+1}$ and $\psi^{n+1,m+1,l} = \psi^{n+1,m}$. 
The magnetic vector potential $A_i$ and electirc potential $\psi$ 
are then simultaneously relaxed. 
After the simultaneous relaxation, 
let $A_i^{n+1,m+1} = A_i^{n+1,m+1,l+1}$ and $\psi^{n+1,m+1} = \psi^{n+1,m+1,l+1}$. 
Equation (\ref{correct.A.2}) is used as the boundary condition 
to solve Eq. (\ref{poisson.psi.2}). 
If the magnetic vector potential is not calculated, 
the electric potential is obtained by Eq. (\ref{newton.psi}). 
When solving the magnetic vector potential, 
the electric potential is obtained by Eqs. (\ref{correct.psi.2}) and (\ref{poisson.psi.2}).

To analyze steady and unsteady flows, 
the Euler implicit method and implicit midpoint rule are used 
for the time derivative, respectively. 
The biconjugate gradient stabilized method \citep{Vorst_1992} is applied 
to solve simultaneous linear equations. 
These discretized equations are solved by following the subsequent procedure.
%------------------------------------------------------------------------------
\begin{enumerate}[Step 1 :]
\setlength{\leftskip}{1.5em}
\setlength{\itemsep}{0mm}
\item At $m = 1$, let $u_i^{n+1,m} = u_i^n$, $p^{n+1,m} = p^n$, 
      $B_i^{n+1,m} = B_i^n$, $A_i^{n+1,m} = A_i^n$, and $\psi^{n+1,m} = \psi^n$.
\item Solve Eq. (\ref{predict.u}), and predict the velocity $\hat{u_i}^{n+1,m+1}$.
\item Solve the pressure correction value $\Delta p^{m,l}$ 
      using Poisson's equation (\ref{poisson.2}).
\item Correct the velocity $u_i^{n+1,m+1,l+1}$ and pressure $p^{n+1,m+1,l+1}$ 
      using Eqs. (\ref{correct.u.2}) and (\ref{correct.p.2}), respectively. 
      At the end of simultaneous relaxation, set $u_i^{n+1,m+1} = u_i^{n+1,m+1,l+1}$ and 
      $p^{n+1,m+1} = p^{n+1,m+1,l+1}$.
\item Solve the magnetic flux density $B_i^{n+1,m+1}$ using Eq. (\ref{newton.B}). 
      If the magnetic vector potential is calculated, 
      solve Eq. (\ref{predict.A}) 
      and predict the magnetic vector potential $\hat{A_i}^{n+1,m+1}$. 
\item If the magnetic vector potential is calculated, 
      solve the correction $\Delta \psi^{m,l}$ 
      using Poisson's equation (\ref{poisson.psi.2}). 
      Correct the magnetic vector potential $A_i^{n+1,m+1,l+1}$ 
      and electric potential $\psi^{n+1,m+1,l+1}$ 
      using Eq. (\ref{correct.A.2}) and (\ref{correct.psi.2}), respectively. 
      At the end of simultaneous relaxation, 
      set $A_i^{n+1,m+1} = A_i^{n+1,m+1,l+1}$ and 
      $\psi^{n+1,m+1} = \psi^{n+1,m+1,l+1}$.
      If the magnetic vector potential is not calculated, 
      solve the electric potential $\psi^{n+1,m+1}$ from Eq. (\ref{newton.psi}).
\item Repeat from Step 2 to Step 6. 
      After the Newton iteration is completed,
      set $u_i^{n+1} = u_i^{n+1,m+1}$, $p^{n+1} = p^{n+1,m+1}$, 
      $B_i^{n+1} = B_i^{n+1,m+1}$, $A_i^{n+1} = A_i^{n+1,m+1}$, 
      and $\psi^{n+1} = \psi^{n+1,m+1}$.
\item Advance the time step and return to Step 1.
\end{enumerate}
%------------------------------------------------------------------------------

%##############################################################################
\section{Verification of the proposed numerical method}
\label{verification}
%##############################################################################

The validity of the method of simultaneously relaxing 
the magnetic vector and electric potentials is verified herein. 
Further, the conservation properties of total energy 
and magnetic helicity in this numerical method are investigated. 
The coordinate $x_i$, the velocity $u_i$, 
the magnetic flux density $B_i$, and magnetic vector potential $A_i$ are denoted 
as $\bm{x} = (x, y, z)$, $\bm{u} = (u, v, w)$, $\bm{B} = (B_x, B_y, B_z)$, 
and $\bm{A} = (A_x, A_y, A_z)$.

%++++++++++++++++++++++++++++++++++++++++++++++++++++++++++++++++++++++++++++++
\subsection{One-dimensional flow}
%++++++++++++++++++++++++++++++++++++++++++++++++++++++++++++++++++++++++++++++

A steady viscous MHD flow with a known exact solution is analyzed 
to validate the accuracy of the numerical method for magnetic vector and electric potentials. 
Similarly to the previous study \citep{Salah_et_al_2001}, 
as a model of the flow and magnetic fields, 
the vector potential $\bm{\Psi}$ and magnetic vector potential $\bm{A}$, 
which represent the one-dimensional flow and magnetic fields, are given as
%------------------------------------------------------------------------------
\begin{equation}
  \Psi_x = 0, \quad 
  \Psi_y = 0, \quad 
  \Psi_z = - x,
  \label{stream_vector_potential_1}
\end{equation}
%------------------------------------------------------------------------------
and 
%------------------------------------------------------------------------------
\begin{equation}
  A_x = 0, \quad 
  A_y = 0, \quad 
  A_z = \frac{1 - e^{-Re_m (1 - y)}}{1 - e^{-Re_m}},
  \label{magnetic_vector_potential_1}
\end{equation}
%------------------------------------------------------------------------------
respectively. 
Equations (\ref{stream_vector_potential_1}) and (\ref{magnetic_vector_potential_1}) 
satisfy the divergence-free condition even discretely. 
Using Eqs. (\ref{stream_vector_potential_1}) and (\ref{magnetic_vector_potential_1}), 
the velocity and magnetic flux density are obtained 
from the relations $\bm{u} = \nabla \times \bm{\Psi}$ 
and $\bm{B} = \nabla \times \bm{A}$ as follows:
%------------------------------------------------------------------------------
\begin{equation}
  u = 0, \quad v = 1, \quad w = 0,
  \label{velocity_1}
\end{equation}
\begin{equation}
  B_x = \frac{Re_m e^{Re_m y}}{1 - e^{Re_m}}, \quad B_y = 0, \quad B_z = 0.
  \label{magnetic_flux_1}
\end{equation}
%------------------------------------------------------------------------------
Equations (\ref{velocity_1}) and (\ref{magnetic_flux_1}) 
automatically satisfy the divergence-free conditions, $\nabla \cdot \bm{u} = 0$ 
and $\nabla \cdot \bm{B} = 0$, respectively. 
Equations (\ref{velocity_1}) and (\ref{magnetic_flux_1}) are made dimensionless 
using a uniform velocity $V$ in the $y$-direction 
and an average magnetic flux density $B_{av}$. 
From Eq. (\ref{navier-stokes}) for the steady flow with an applied magnetic field, 
the following exact solution for pressure is obtained:
%------------------------------------------------------------------------------
\begin{equation}
  p = \frac{Re_m^2 e^{2 Re_m y}}{2 Al^2 (e^{Re_m} - 1)^2}.
  \label{pressure_1}
\end{equation}
%------------------------------------------------------------------------------
As Eq. (\ref{vector_potential}) yields $\nabla \psi = 0$, 
the electric potential $\psi$ becomes constant.

The computational region in the $x$- and $y$-directions is $[0, L]$, 
and the computational region in the $z$-direction is the grid width $\Delta x$. 
The initial values that satisfy the divergence-free condition 
for velocity and magnetic flux density should be given; 
hence, exact solutions for velocity, pressure, and magnetic flux density are given 
as the initial values. 
Therefore, in this problem, 
the author examined the method of solving the magnetic vector 
and electric potentials 
and confirmed that the divergence-free condition of velocity and magnetic flux density is maintained. 
At the boundary in the $x$- and $z$-directions, 
periodic boundary conditions are applied to all dependent variables. 
At the boundary in the $y$-direction, the Dirichlet conditions for the velocity, 
magnetic flux density, and magnetic vector potential are imposed, 
and a zero gradient of the electric potential is given. 
In this calculation, uniform and nonuniform grids of $N \times N \times 2$ are used, 
and $N = 11$, $21$, and $41$. 
A nonuniform grid is generated using the following function:
%------------------------------------------------------------------------------
\begin{equation}
  y_j = \frac{\tanh (\alpha_y r)}{\tanh (\alpha_y)}, \quad r = \frac{j-1}{N-1},
\end{equation}
%------------------------------------------------------------------------------
where $\alpha_y = 1$. 
The maximum ratio of grid width $\Delta y_j$ is $\Delta y_{j-1}/\Delta y_{j} = 2.34$. 
The grid in the $x$-direction is evenly spaced. 
The reference values used for nondimensionalization are $l_\mathrm{ref} = L$, 
$u_\mathrm{ref} = V$, $t_\mathrm{ref} = L/V$, $B_\mathrm{ref} = B_{av}$, 
$A_\mathrm{ref} = B_{av} L$, and $\psi_\mathrm{ref} = V L B_{av}$. 
The given parameters are as follows: the Reynolds number $Re = 100$, 
the Alfv\'{e}n number $Al = 1$, and the magnetic Reynolds number $Re_m = 1$, $5$, $10$, and $20$. 
The Courant number is defined as $\mbox{CFL} = \Delta t V/\Delta y_\mathrm{min}$ 
using the reference velocity $V$ and minimum grid width $\Delta y_\mathrm{min}$. 
For uniform grids, 
the time step is set to $\Delta t/(L/V) = 0.004$ under all conditions. 
The Courant numbers are CFL = 0.04, 0.08, and 0.16 for $N = 11$, $21$, and $41$, respectively. 
For nonuniform grids, 
the time step is set to $\Delta t//(L/V) = 0.002$ under all conditions. 
The Courant numbers are CFL = 0.034, 0.070, and 0.14 for $N = 11$, $21$, and $41$, respectively.

\begin{figure}[!t]
\begin{minipage}{0.325\linewidth}
\begin{center}
\includegraphics[trim=0mm 0mm 0mm 0mm, clip, width=50mm]{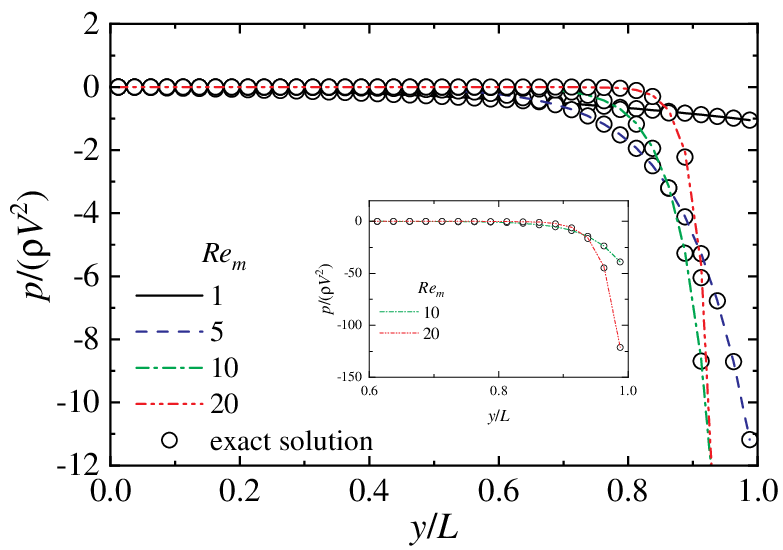} \\
{\small (a) $p$}
\end{center}
\end{minipage}
\begin{minipage}{0.325\linewidth}
\begin{center}
\includegraphics[trim=0mm 0mm 0mm 0mm, clip, width=50mm]{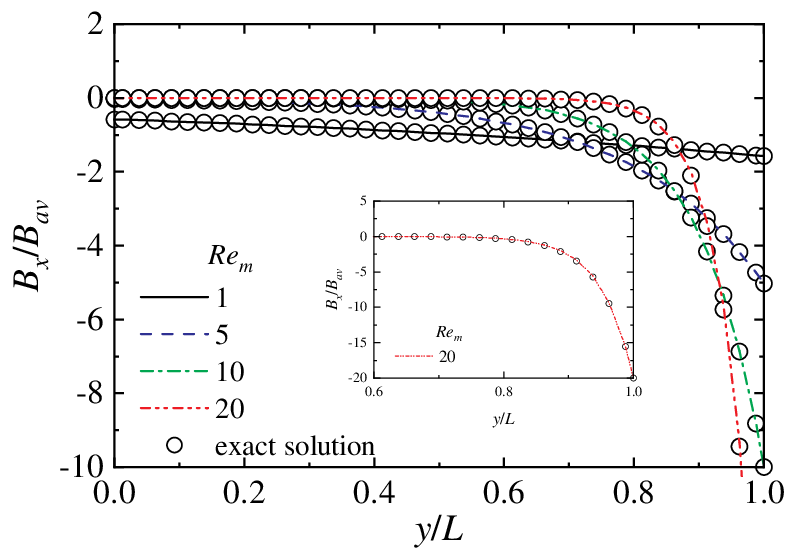} \\
{\small (b) $B_x$}
\end{center}
\end{minipage}
\begin{minipage}{0.325\linewidth}
\begin{center}
\includegraphics[trim=0mm 0mm 0mm 0mm, clip, width=50mm]{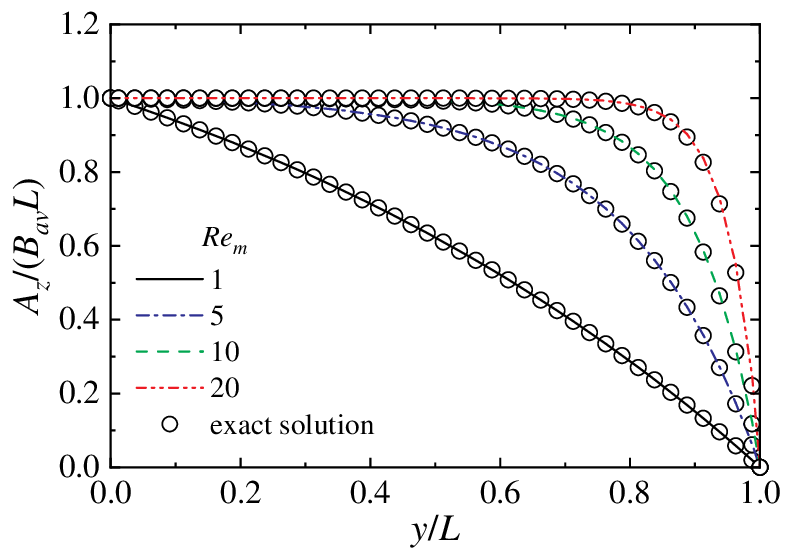} \\
{\small (c) $A_z$}
\end{center}
\end{minipage}
\caption{Distributions of pressure ($p$), magnetic flux density ($B_x$), 
and magnetic vector potential ($A_z$): 
$Re = 100$, $Al = 1$, $Re_m = 1, 5, 10, 20$, $N = 41$ (uniform grid).}
\label{vp2d1_p&Bx&Az}
\end{figure}

\begin{figure}[!t]
\begin{minipage}{0.325\linewidth}
\begin{center}
\includegraphics[trim=0mm 0mm 0mm 0mm, clip, width=50mm]{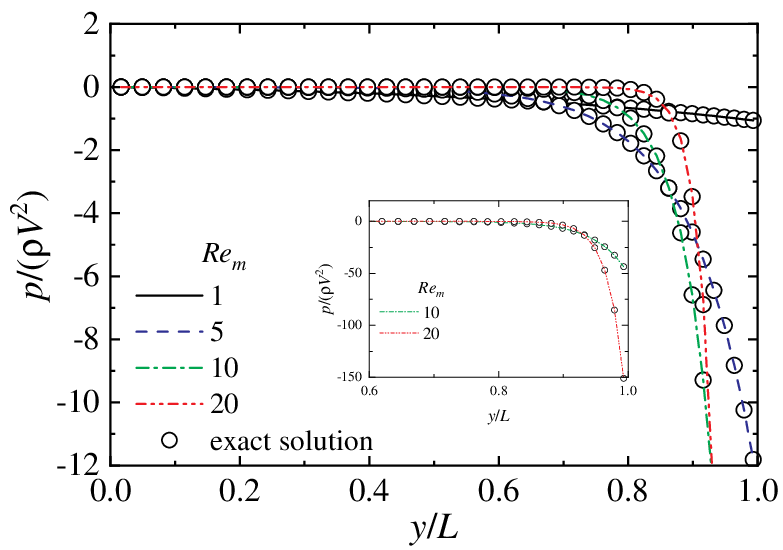} \\
{\small (a) $p$}
\end{center}
\end{minipage}
\begin{minipage}{0.325\linewidth}
\begin{center}
\includegraphics[trim=0mm 0mm 0mm 0mm, clip, width=50mm]{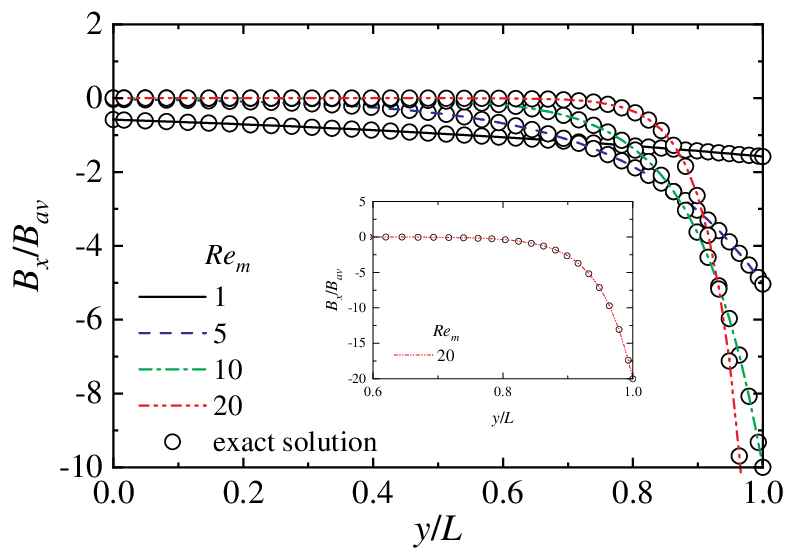} \\
{\small (b) $B_x$}
\end{center}
\end{minipage}
\begin{minipage}{0.325\linewidth}
\begin{center}
\includegraphics[trim=0mm 0mm 0mm 0mm, clip, width=50mm]{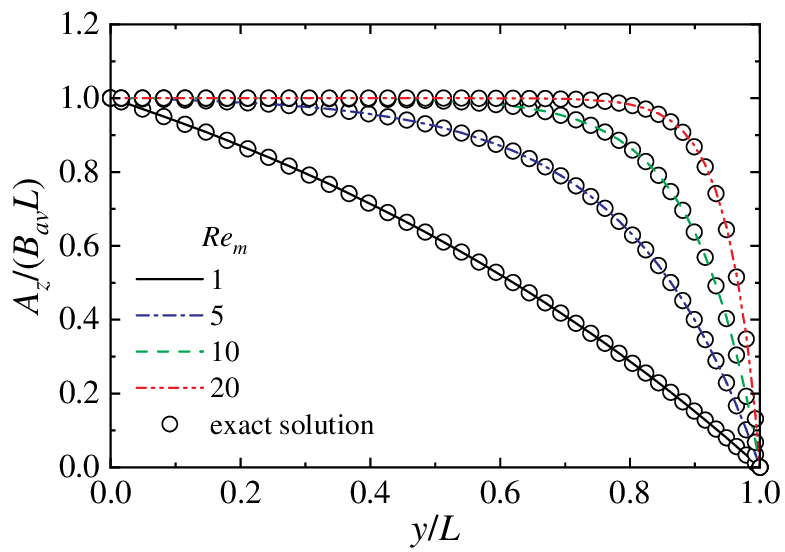} \\
{\small (c) $A_z$}
\end{center}
\end{minipage}
\caption{Distributions of pressure ($p$), magnetic flux density ($B_x$), 
and magnetic vector potential ($A_z$): 
$Re = 100$, $Al = 1$, $Re_m = 1, 5, 10, 20$, $N = 41$ (nonuniform grid).}
\label{vp2d1_p&Bx&Az_n}
\end{figure}

Figure \ref{vp2d1_p&Bx&Az} shows the distributions of pressure $p$, 
magnetic flux density $B_x$, and magnetic vector potential $A_z$ 
obtained using the uniform grid with $N = 41$ grid points. 
This approximate solution supports the exact solution. 
As $Re_m$ increases, the gradient of the distribution near $y/L = 1$ increases, 
and the grid resolution begins to impact the calculation accuracy. 
Figure \ref{vp2d1_p&Bx&Az_n} shows the results obtained using the nonuniform grid with $N=41$. 
As the grid is made finer near $y/L = 1$, a sharp gradient can be captured. 
For all grids, the electric potential is constant, 
and the maximum error of electric potential is zero.

\begin{figure}[!t]
\begin{minipage}{0.5\linewidth}
\begin{center}
\includegraphics[trim=0mm 0mm 0mm 0mm, clip, width=70mm]{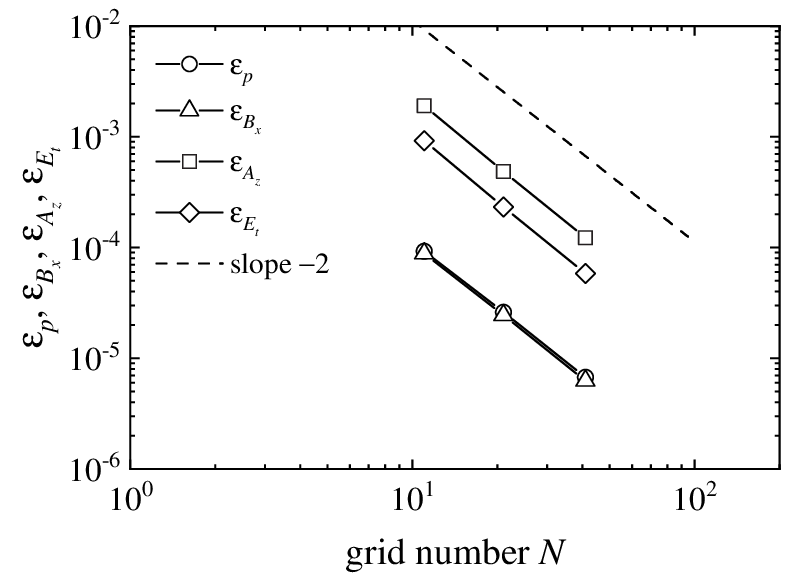} \\
{\small (a) $Re_m=1$}
\end{center}
\end{minipage}
\begin{minipage}{0.5\linewidth}
\begin{center}
\includegraphics[trim=0mm 0mm 0mm 0mm, clip, width=70mm]{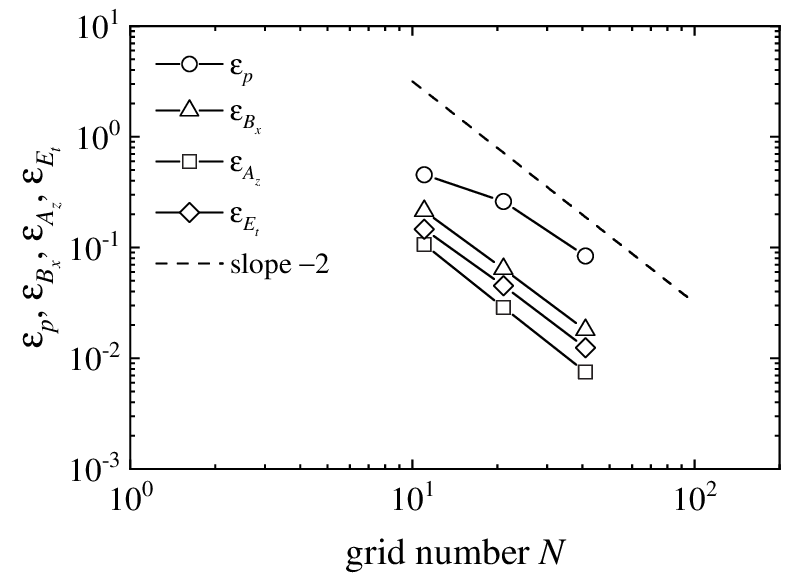} \\
{\small (b) $Re_m=10$}
\end{center}
\end{minipage}
\caption{Maximum errors of pressure, magnetic flux density, 
and magnetic vector potential, and relative error of total energy 
using uniform grid: 
$Re = 100$, $Al = 1$, $Re_m = 1, 10$.}
\label{vp2d1_error}
\end{figure}

\begin{figure}[!t]
\begin{minipage}{0.5\linewidth}
\begin{center}
\includegraphics[trim=0mm 0mm 0mm 0mm, clip, width=70mm]{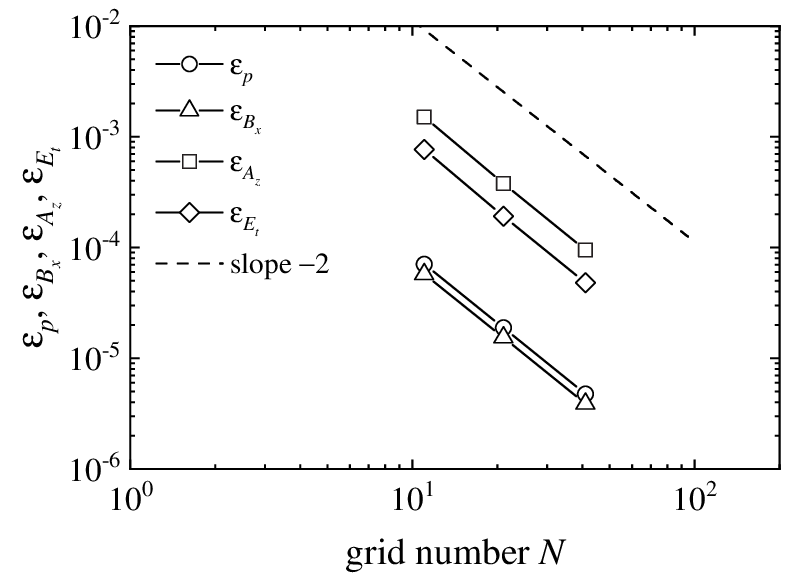} \\
{\small (a) $Re_m=1$}
\end{center}
\end{minipage}
\begin{minipage}{0.5\linewidth}
\begin{center}
\includegraphics[trim=0mm 0mm 0mm 0mm, clip, width=70mm]{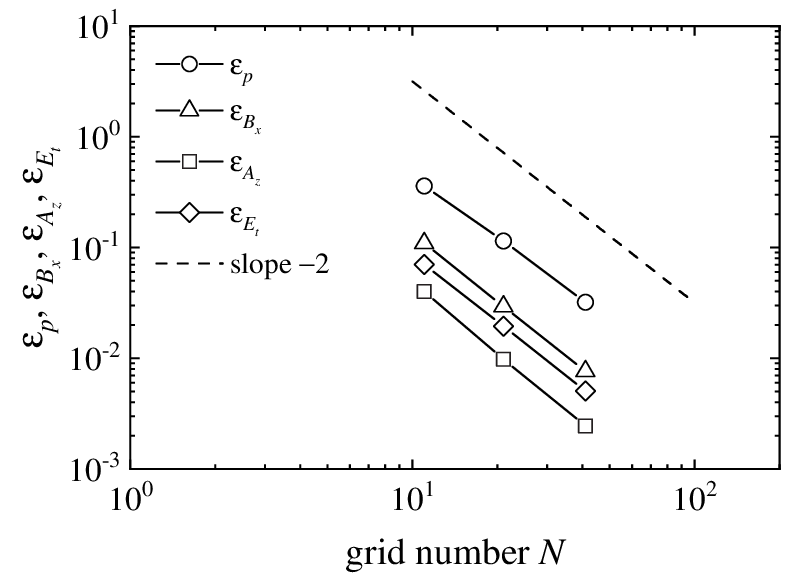} \\
{\small (b) $Re_m=10$}
\end{center}
\end{minipage}
\caption{Maximum errors of pressure, magnetic flux density, 
and magnetic vector potential, and relative error of total energy 
using nonuniform grid: 
$Re = 100$, $Al = 1$, $Re_m = 1, 10$.}
\label{vp2d1_error_n}
\end{figure}

For $Re_m = 1$ and $10$, the maximum errors, $\varepsilon_{p}$, $\varepsilon_{B_x}$, 
and $\varepsilon_{A_z}$, of the pressure, magnetic flux density, 
and magnetic vector potential, respectively 
as well as the relative error, $\varepsilon_{E_t} = |(\langle E_t \rangle - \langle E_t \rangle_e)/\langle E_t \rangle_e|$, 
of the total energy are shown in Figs. \ref{vp2d1_error} and \ref{vp2d1_error_n}, 
where $\langle E_t \rangle$ is the total amount of total energy 
and the subscript $e$ represents the exact solution. 
The total amount is obtained via volume integration. 
The dashed line is a straight line with a slope of $-2$. 
The error decreases as the number of grid points $N$ increases, 
indicating second-order convergence. 
As $Re_m$ increases, the error level increases, 
but the convergence with respect to the number of grid points does not change. 
For the nonuniform grid, the grid resolution increases in the region 
where the gradient of distribution is large; 
therefore, the error is lower than in the uniform grid.

%------------------------------------------------------------------------------
% Figure 1
%------------------------------------------------------------------------------
%------------------------------------------------------------------------------

%------------------------------------------------------------------------------
% Figure 2
%------------------------------------------------------------------------------
%------------------------------------------------------------------------------

%------------------------------------------------------------------------------
% Figure 3
%------------------------------------------------------------------------------
%------------------------------------------------------------------------------

%------------------------------------------------------------------------------
% Figure 4
%------------------------------------------------------------------------------
%------------------------------------------------------------------------------

The maximum divergence errors of velocity and magnetic flux density in this analysis 
are $3.86\times 10^{-11}$ and $2.42\times 10^{-6}$, respectively, for the uniform grid 
and $4.30\times 10^{-11}$ and $1.16\times 10^{-6}$, respectively, for the nonuniform grid. 
In the case of steady calculations, the number of Newton iterations is set to one; 
hence, the maximum divergence error of the magnetic flux density 
is greater than the error at the initial value.

%++++++++++++++++++++++++++++++++++++++++++++++++++++++++++++++++++++++++++++++
\subsection{Two-dimensional flow}
%++++++++++++++++++++++++++++++++++++++++++++++++++++++++++++++++++++++++++++++

In this analysis, a two-dimensional viscous MHD flow is considered. 
As a model of the steady flow and magnetic fields, 
the vector potential $\bm{\Psi}$ and magnetic vector potential $\bm{A}$ 
representing the two-dimensional flow and magnetic fields are given as
%------------------------------------------------------------------------------
\begin{equation}
  \Psi_x = - \frac{1}{2} x^2, \quad 
  \Psi_y = - \frac{1}{2} y^2, \quad 
  \Psi_z = - \frac{1}{2} x + (x + y) z,
  \label{stream_vector_potential_3}
\end{equation}
%------------------------------------------------------------------------------
and 
%------------------------------------------------------------------------------
\begin{equation}
  A_x = - \frac{1}{2} x^2, \quad 
  A_y = - \frac{1}{2} y^2, \quad 
  A_z = (x + y) z,
  \label{magnetic_vector_potential_3}
\end{equation}
%------------------------------------------------------------------------------
respectively. 
Similar magnetic vector potentials were used in \citep{Salah_et_al_2001}. 
Equations (\ref{stream_vector_potential_3}) and (\ref{magnetic_vector_potential_3}) 
also satisfy the divergence-free condition discretely. 
Using Eqs. (\ref{stream_vector_potential_3}) and (\ref{magnetic_vector_potential_3}), 
the velocity and magnetic flux density are obtained from 
the relations of $\bm{u} = \nabla \times \bm{\Psi}$ 
and $\bm{B} = \nabla \times \bm{A}$ as follows:
%------------------------------------------------------------------------------
\begin{equation}
  u = z, \quad v = \frac{1}{2} - z, \quad w = 0,
  \label{velocity_3}
\end{equation}
\begin{equation}
  B_x = z, \quad B_y = - z, \quad B_z = 0.
  \label{magnetic_flux_3}
\end{equation}
%------------------------------------------------------------------------------
Equations (\ref{velocity_3}) and (\ref{magnetic_flux_3}) 
automatically satisfy the divergence-free conditions, $\nabla \cdot \bm{u} = 0$ 
and $\nabla \cdot \bm{B} = 0$, respectively. 
The computational domain is a cube with side $L$. 
The velocity and magnetic flux density at $z = L$ are denoted as $U$ and $B$, respectively, 
and Eqs. (\ref{velocity_3}) and (\ref{magnetic_flux_3}) are nondimensionalized 
by these reference values. 
From Eq. (\ref{navier-stokes}) for the steady flow with an applied magnetic field, 
the following exact solution for pressure is obtained:
%------------------------------------------------------------------------------
\begin{equation}
  p = - \frac{1}{Al^2} z^2.
  \label{pressure_3}
\end{equation}
%------------------------------------------------------------------------------
Eq. (\ref{vector_potential}) yields 
$\partial \psi/\partial x = \partial \psi/\partial y = -1/Re_m$. 
Additionally, solving the equation for $A_z$ gives $\partial \psi/\partial z = - z/2$. 
Therefore, assuming a steady field, the electric potential $\psi$ can be obtained as
%------------------------------------------------------------------------------
\begin{equation}
  \psi = - \frac{1}{Re_m} (x + y) - \frac{z^2}{4}.
  \label{electric_potential_3}
\end{equation}
%------------------------------------------------------------------------------

The computational region is $[0, L]$ in each coordinate direction. 
As the initial values satisfying the divergence-free condition 
for velocity and magnetic flux density should be given, 
exact solutions for velocity, pressure, and magnetic flux density are given 
as initial values for viscous analysis. 
Therefore, in this problem, 
the author examined the method of solving the magnetic vector 
and electric potentials 
and confirm that the divergence-free condition of velocity and magnetic flux density is maintained. 
In inviscid analysis, exact solutions for all dependent variables are given as initial values. 
At the boundaries in the $x$- and $y$-directions, 
periodic boundary conditions are applied to all dependent variables. 
At the boundary in the $z$-direction, the Dirichlet conditions are imposed 
for the velocity, magnetic flux density, and magnetic vector potential, 
and the electric potential gradient is given using Ohm's and Ampere's laws. 
In this calculation, uniform and nonuniform grids of $N \times N \times 2$ are used, 
and $N = 11$, $21$, and $41$. 
A nonuniform grid is generated using the following function:
%------------------------------------------------------------------------------
\begin{equation}
  z_k = \frac{1}{2} \left[ 
  \frac{\tanh (\alpha_z r)}{\tanh (\alpha_z)} + 1 \right], 
  \quad r = 2 \frac{k-1}{N-1} - 1,
\end{equation}
%------------------------------------------------------------------------------
where $\alpha_z = 1$. 
The maximum ratio of grid width $\Delta z_k$ is $\Delta z_{k-1}/\Delta z_{k} = 1.30$. 
The grids in the $x$- and $y$-directions are evenly spaced. 
The reference values used for nondimensionalization are 
$l_\mathrm{ref} = L$, $u_\mathrm{ref} = U$, $t_\mathrm{ref} = L/U$, $B_\mathrm{ ref} = B$, 
$A_\mathrm{ref} = B L$, and $\psi_\mathrm{ref} = U L B$. 
The Courant number is defined as $\mbox{CFL} = \Delta t U/\Delta z_\mathrm{min}$ 
using the reference velocity $U$ and the minimum grid width $\Delta z_\mathrm{min}$. 
First, an inviscid analysis is performed using the uniform grid with $N = 41$ grid points 
to confirm the energy conservation property. 
Although this flow field is not periodic, it is steady; 
hence, the total energy and magnetic helicity are kept constant. 
This analysis sets the Courant number CFL = 0.5 and the time step $\Delta t/(L/U) = 0.0125$. 
In viscous analysis, the Reynolds number is set to be $Re = 100$, 
the Alfven number is $Al = 1$, and the magnetic Reynolds number is $Re_m = 1$. 
The Courant numbers CFL = 0.2 and $\mbox{CFL} = 0.14-0.16$ are set for the calculations 
using the uniform and nonuniform grids, respectively.

\begin{figure}[!t]
\begin{minipage}{0.5\linewidth}
\begin{center}
\includegraphics[trim=0mm 0mm 0mm 0mm, clip, width=70mm]{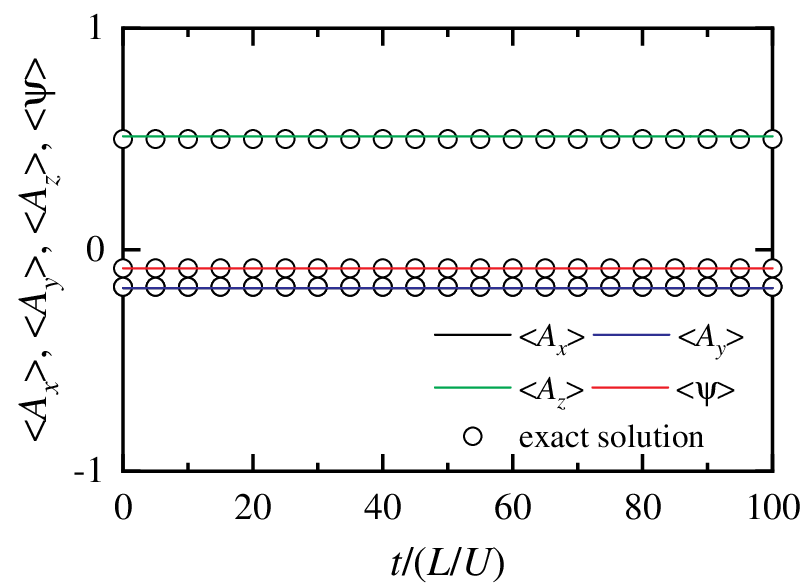} \\
{\small (a) $A_x$, $A_y$, $A_z$, $\psi$}
\end{center}
\end{minipage}
\begin{minipage}{0.5\linewidth}
\begin{center}
\includegraphics[trim=0mm 0mm 0mm 0mm, clip, width=70mm]{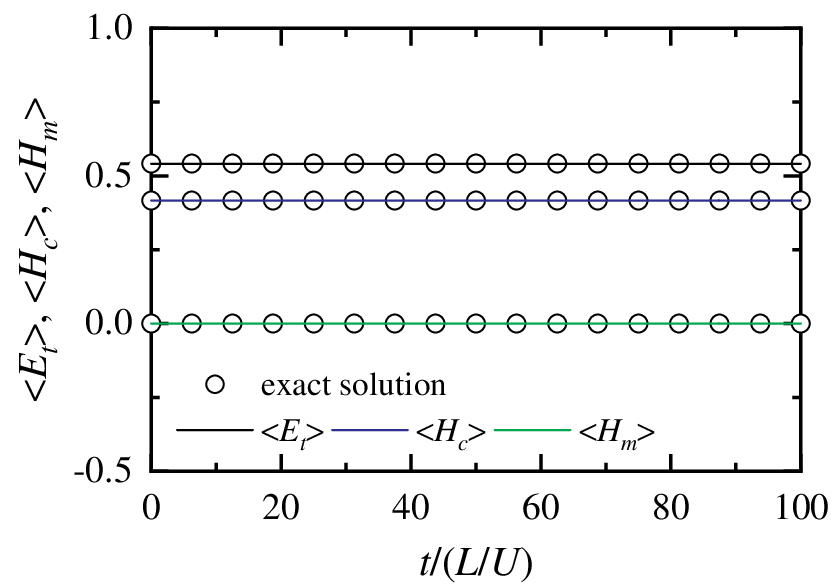} \\
{\small (b) $E_t$, $H_c$. $H_m$}
\end{center}
\end{minipage}
\caption{Total amounts of magnetic vector potential ($A_x$, $A_y$, $A_z$), 
electric potential ($\psi$), total energy ($E_t$), cross-helicity ($H_c$), 
and magnetic helicity ($H_m$): 
$Re = Re_m = \infty$, $Al = 1$, $N = 41$ (uniform grid).}
\label{vp3d1_invis_sum}
\end{figure}

Figure \ref{vp3d1_invis_sum} (a) shows the total amount, 
$\langle A_x \rangle$, $\langle A_y \rangle$, and $\langle A_z \rangle$, 
of the magnetic vector potentials at $Re = Re_m = \infty$. 
Figure \ref{vp3d1_invis_sum} (b) shows the total amounts, 
$\langle E_t \rangle$, $\langle H_c \rangle$, and $\langle H_m \rangle$, 
of total energy, cross-helicity, and magnetic helicity, respectively. 
The total amount is obtained via volume integration. 
All total amounts are kept constant. 
In addition, the calculated results agree well with the exact solutions.

%------------------------------------------------------------------------------
% Figure 5
%------------------------------------------------------------------------------
%------------------------------------------------------------------------------

%------------------------------------------------------------------------------
% Figure 6
%------------------------------------------------------------------------------
\begin{figure}[!t]
\begin{minipage}{0.5\linewidth}
\begin{center}
\includegraphics[trim=0mm 0mm 0mm 0mm, clip, width=70mm]{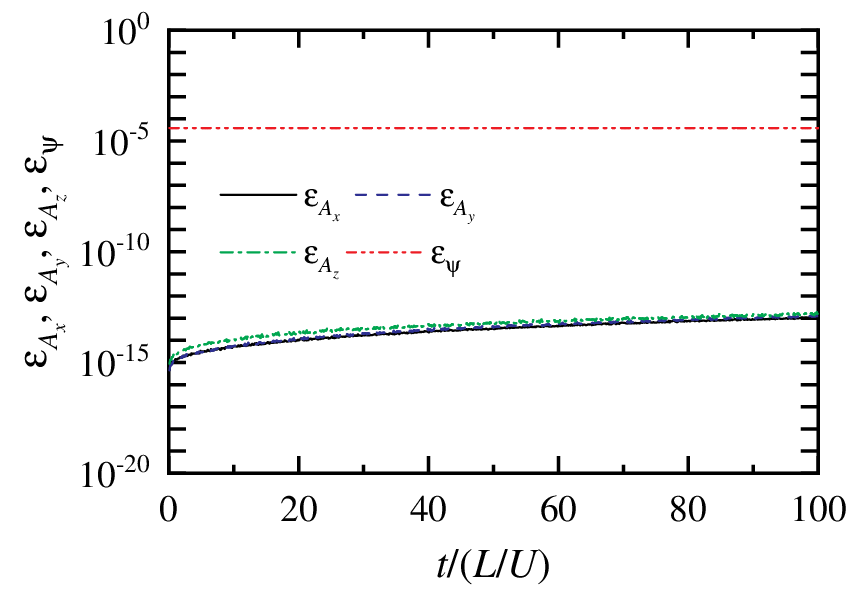} \\
{\small (a) $A_x$, $A_y$, $A_z$, $\psi$}
\end{center}
\end{minipage}
\begin{minipage}{0.5\linewidth}
\begin{center}
\includegraphics[trim=0mm 0mm 0mm 0mm, clip, width=70mm]{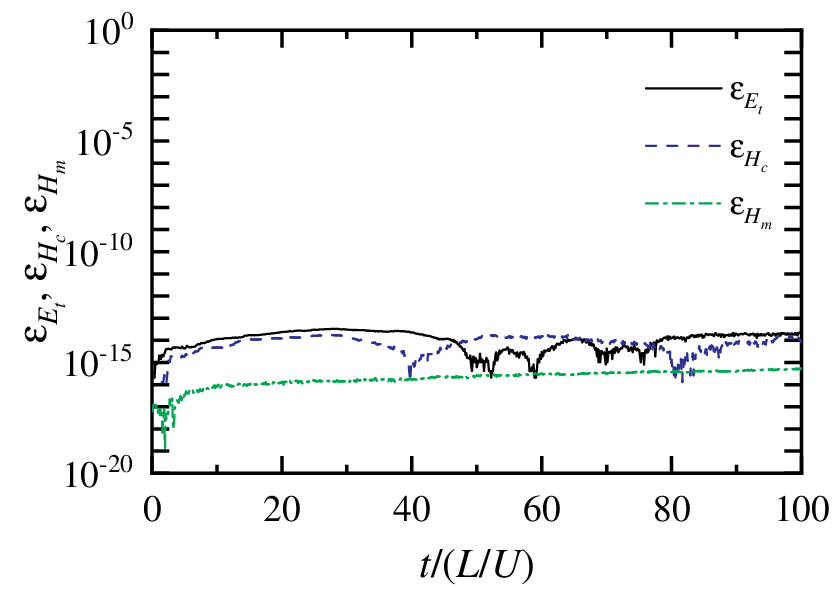} \\
{\small (b) $E_t$, $H_c$. $H_m$}
\end{center}
\end{minipage}
\caption{Maximum errors of magnetic vector potential ($A_x$, $A_y$, $A_z$) and 
electric potential ($\psi$), 
and relative errors of total energy ($E_t$), 
cross-helicity ($H_c$), and magnetic helicity ($H_m$): 
$Re = Re_m = \infty$, $Al = 1$, $N = 41$ (uniform grid).}
\label{vp3d1_invis_error}
\end{figure}
%------------------------------------------------------------------------------

%------------------------------------------------------------------------------
% Figure 7
%------------------------------------------------------------------------------
\begin{figure}[!t]
\begin{minipage}{0.5\linewidth}
\begin{center}
\includegraphics[trim=0mm 0mm 0mm 0mm, clip, width=70mm]{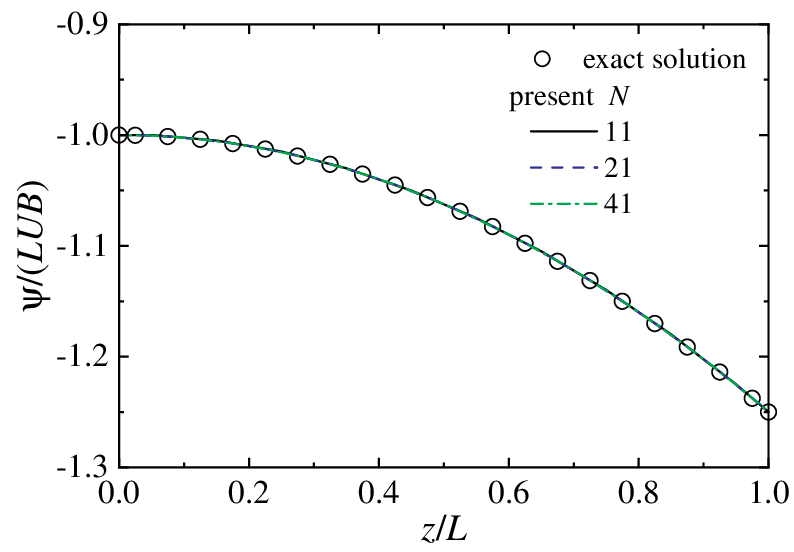} \\
{\small (a) $\psi$}
\end{center}
\end{minipage}
\begin{minipage}{0.5\linewidth}
\begin{center}
\includegraphics[trim=0mm 0mm 0mm 0mm, clip, width=70mm]{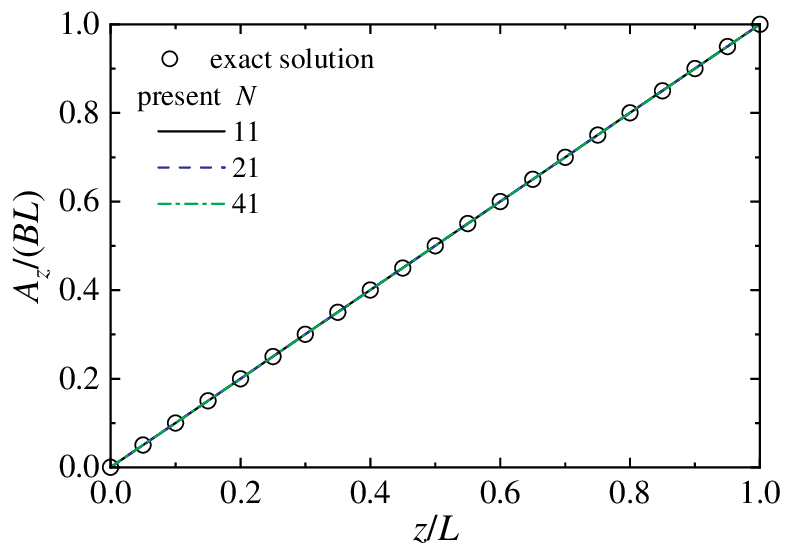} \\
{\small (b) $A_z$}
\end{center}
\end{minipage}
\caption{Distributions of electric potential ($\psi$) 
and magnetic vector potential ($A_z$) at $x/L = 0.5$ and $y/L = 0.5$ using uniform grid: 
$Re = 100$, $Al = 1$, $Re_m = 1$.}
\label{vp3d1_psi&Az}
\end{figure}
%------------------------------------------------------------------------------

%------------------------------------------------------------------------------
% Figure 8
%------------------------------------------------------------------------------
\begin{figure}[!t]
\begin{minipage}{0.5\linewidth}
\begin{center}
\includegraphics[trim=0mm 0mm 0mm 0mm, clip, width=70mm]{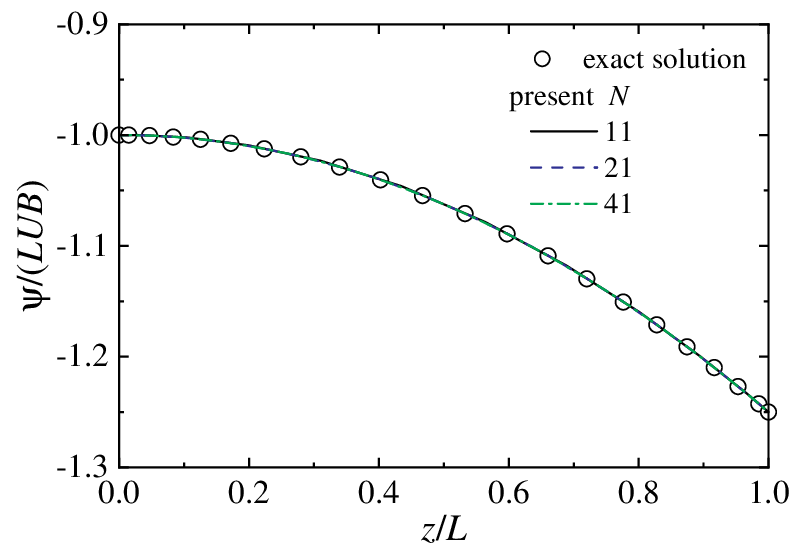} \\
{\small (a) $\psi$}
\end{center}
\end{minipage}
\begin{minipage}{0.5\linewidth}
\begin{center}
\includegraphics[trim=0mm 0mm 0mm 0mm, clip, width=70mm]{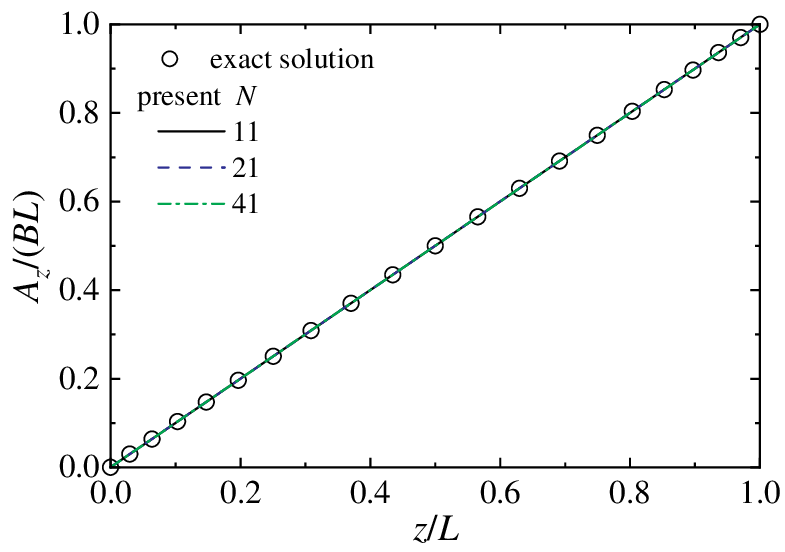} \\
{\small (b) $A_z$}
\end{center}
\end{minipage}
\caption{Distributions of electric potential ($\psi$) 
and magnetic vector potential ($A_z$) at $x/L = 0.5$ and $y/L = 0.5$ using nonuniform grid: 
$Re = 100$, $Al = 1$, $Re_m = 1$.}
\label{vp3d1_psi&Az_n}
\end{figure}
%------------------------------------------------------------------------------

Figure \ref{vp3d1_invis_error} (a) shows the maximum errors, 
$\varepsilon_{A_x}$, $\varepsilon_{A_y}$, $\varepsilon_{A_z}$, and $\varepsilon_{\psi}$, 
for the magnetic vector and electric potentials. 
The relative errors, 
$\varepsilon_{E_t} = |(\langle E_t \rangle - \langle E_t \rangle_0)/\langle E_t \rangle_0|$, 
$\varepsilon_{H_c} = |(\langle H_c \rangle - \langle H_c \rangle_0)/\langle H_c \rangle_0|$, 
and $\varepsilon_{H_m} = |(\langle H_m \rangle - \langle H_m \rangle_0)/\langle H_m \rangle_\mathrm{max}|$, 
for total energy, cross-helicity, and magnetic helicity are shown in Fig. \ref{vp3d1_invis_error} (b). 
Here, the subscript $0$ represents the initial value, 
and the subscript max represents the maximum value of the initial value. 
As the total amount of magnetic helicity is zero, 
the relative error is defined using the maximum value. 
The electric potential error changes of the order of $10^{-5}$. 
As time passes, the error in the magnetic vector potential increases slightly but retains a low value. 
The relative error of magnetic helicity changes of the order of $10^{-16}$, 
and the relative errors of total energy and cross-helicity change on the order of $10^{-15}-10^{-14}$. 
These results show that the total energy, cross-helicity, and magnetic helicity are kept constant 
even after long calculations. 
The maximum divergence errors of velocity, magnetic flux density, 
and magnetic vector potential in the inviscid analysis are $1.31\times 10^{-14}$, 
$1.20\times 10^{-11}$, and $1.39\times 10^{-14}$, respectively.

%------------------------------------------------------------------------------
% Figure 9
%------------------------------------------------------------------------------
\begin{figure}[!t]
\begin{minipage}{0.5\linewidth}
\begin{center}
\includegraphics[trim=0mm 0mm 0mm 0mm, clip, width=70mm]{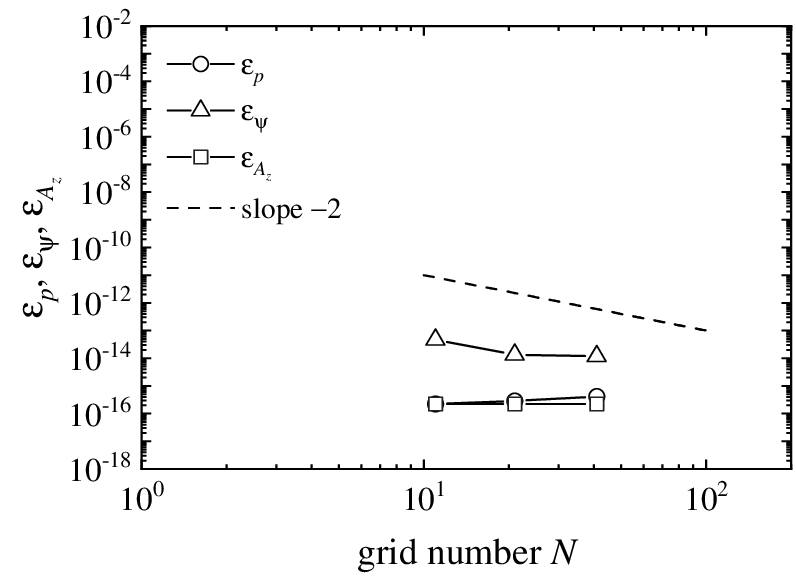} \\
{\small (a) $p$, $\psi$, $A_z$}
\end{center}
\end{minipage}
\begin{minipage}{0.5\linewidth}
\begin{center}
\includegraphics[trim=0mm 0mm 0mm 0mm, clip, width=70mm]{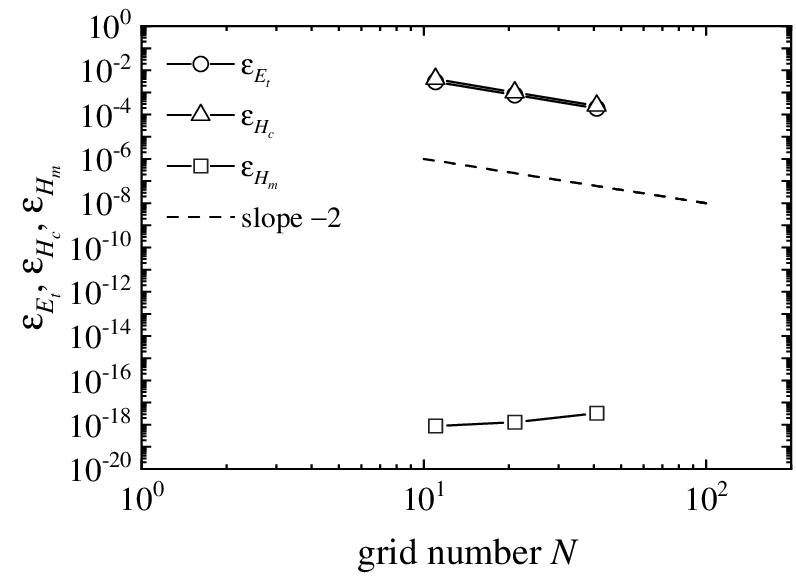} \\
{\small (b) $E_t$, $H_c$, $H_m$}
\end{center}
\end{minipage}
\caption{Maximum errors of pressure ($p$), electric potential ($\psi$), 
and magnetic vector potential ($A_z$) 
and relative errors of total energy ($E_t$), cross-helicity ($H_c$), 
and magnetic helicity ($H_m$) using uniform grid: 
$Re = 100$, $Al = 1$, $Re_m = 1$.}
\label{vp3d1_error}
\end{figure}
%------------------------------------------------------------------------------

%------------------------------------------------------------------------------
% Figure 10
%------------------------------------------------------------------------------
\begin{figure}[!t]
\begin{minipage}{0.5\linewidth}
\begin{center}
\includegraphics[trim=0mm 0mm 0mm 0mm, clip, width=70mm]{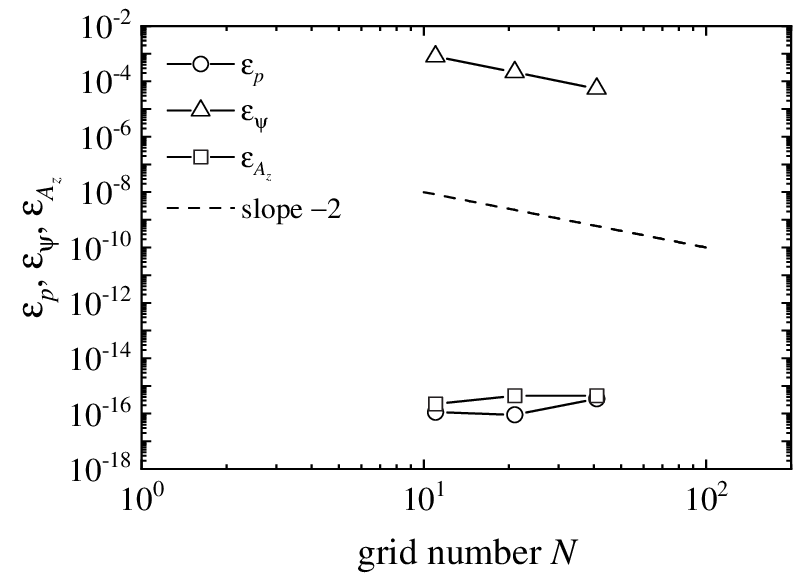} \\
{\small (a) $p$, $\psi$, $A_z$}
\end{center}
\end{minipage}
\begin{minipage}{0.5\linewidth}
\begin{center}
\includegraphics[trim=0mm 0mm 0mm 0mm, clip, width=70mm]{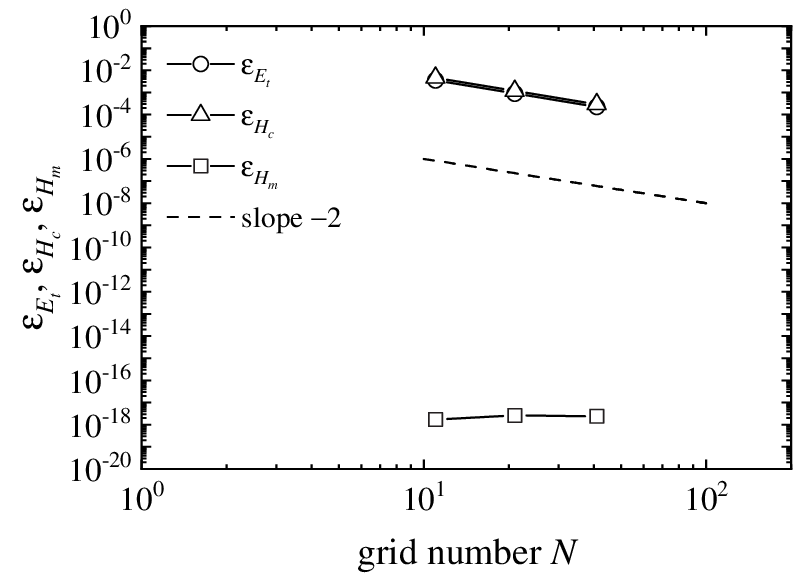} \\
{\small (b) $E_t$, $H_c$, $H_m$}
\end{center}
\end{minipage}
\caption{Maximum errors of pressure ($p$), electric potential ($\psi$), 
and magnetic vector potential ($A_z$) 
and relative errors of total energy ($E_t$), cross-helicity ($H_c$), 
and magnetic helicity ($H_m$) using nonuniform grid: 
$Re = 100$, $Al = 1$, $Re_m = 1$.}
\label{vp3d1_error_n}
\end{figure}
%------------------------------------------------------------------------------

Subsequently, the results of the viscous analysis are presented. 
Figures \ref{vp3d1_psi&Az} and \ref{vp3d1_psi&Az_n} show 
the distributions of the electric potential $\psi$ and magnetic vector potential $A_z$ 
at $x/L = 0.5$ and $y/L = 0.5$. 
There are no differences in the results based on the number of grid points $N$. 
This result supports the exact solution. 
Moreover, the results for the uniform and nonuniform grids do not differ. 
Evidently, the method for obtaining the magnetic vector 
and electric potentials is appropriate. 

The maximum errors for pressure $\varepsilon_{p}$, 
electric potential $\varepsilon_{\psi}$, 
and magnetic vector potentials $\varepsilon_{A_z}$, 
as well as relative errors for total energy 
$\varepsilon_{E_t} = |(\langle E_t \rangle - \langle E_t \rangle_e)/\langle E_t \rangle_e|$, 
cross-helicity 
$\varepsilon_{H_c} = |(\langle H_c \rangle - \langle H_c \rangle_e)/ \langle H_c \rangle_e|$, 
and magnetic helicity 
$\varepsilon_{H_m} = |(\langle H_m \rangle - \langle H_m \rangle_e)/\langle H_m \rangle_\mathrm{max}|$ 
are shown in Fig. \ref{vp3d1_error}. 
Here, the subscript $e$ represents the exact solution, 
and the subscript $\mathrm{max}$ represents the maximum value of the exact solution. 
The errors in the pressure and magnetic vector potential remain at the level of rounding errors 
irrespective of the number of grid points. 
The electric potential error is of the order of $10^{-14}$. 
As the number of grid points $N$ increases, 
the errors of total energy and cross-helicity decrease, 
indicating second-order convergence. 
The magnetic helicity error is at a very low level. 
Figure \ref{vp3d1_error_n} shows the errors obtained using the nonuniform grid. 
The electric potential error increases and the calculation accuracy decreases. 
The errors in total energy, cross-helicity, and magnetic helicity are 
at the same level as the results using the uniform grid, 
and the convergence of the solutions for the uniform and nonuniform grids 
does not differ significantly.

%------------------------------------------------------------------------------
% Figure 11
%------------------------------------------------------------------------------
\begin{figure}[!t]
\begin{minipage}{0.5\linewidth}
\begin{center}
\includegraphics[trim=0mm 0mm 0mm 0mm, clip, width=70mm]{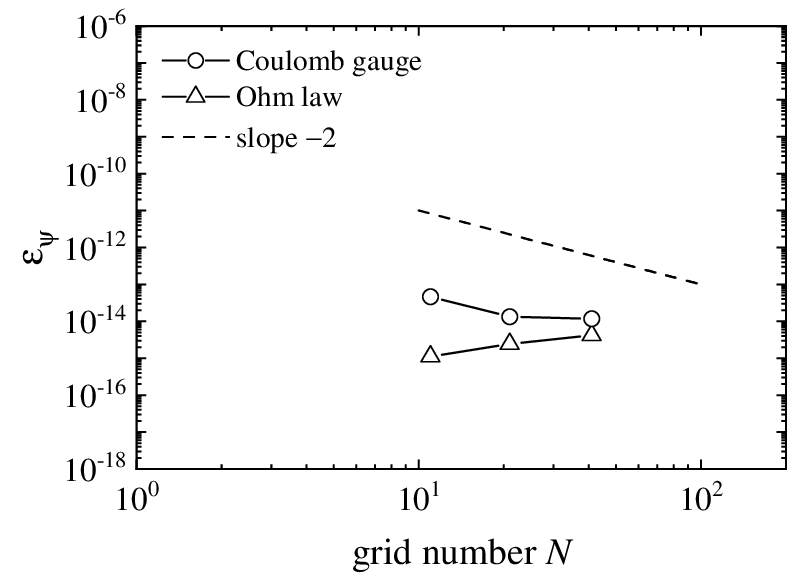} \\
{\small (a) Uniform grid}
\end{center}
\end{minipage}
\begin{minipage}{0.5\linewidth}
\begin{center}
\includegraphics[trim=0mm 0mm 0mm 0mm, clip, width=70mm]{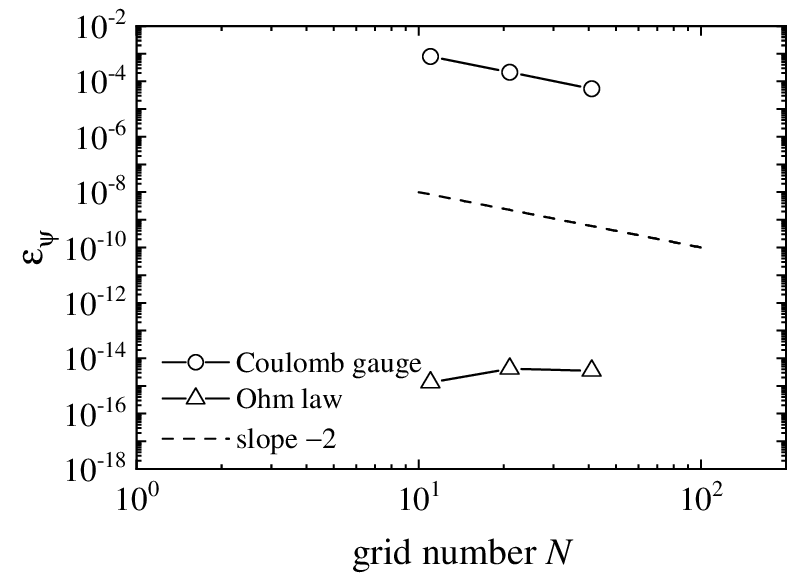} \\
{\small (b) Nonuniform grid}
\end{center}
\end{minipage}
\caption{Maximum errors of electric potential using Coulomb gauge and Ohm's law: 
$Re = 100$, $Al = 1$, $Re_m = 1$.}
\label{vp3d1_error_ohm}
\end{figure}
%------------------------------------------------------------------------------

The difference in the error depending on the method for calculating the electric potential is confirmed. 
Figure \ref{vp3d1_error_ohm} shows the maximum error of the electric potential 
obtained by solving the charge conservation law using Ohm's law. 
For uniform grids, the error is almost the same as 
the maximum error of the electric potential obtained using the Coulomb gauge. 
The method for calculating the electric potential does not introduce variation. 
For nonuniform grids, the error due to the method of obtaining the electric potential 
from Ohm's law is almost the same as for the uniform grid. 
Conversely, when the Coulomb gauge is used, the error increases. 
As shown in Fig. \ref{vp3d1_psi&Az_n}, 
the electric potential distribution is consistent with the exact solution, 
therefore, the level of this error is considered low; 
and second-order convergence is obtained.

The maximum divergence errors of the velocity, magnetic flux density, 
and magnetic vector potential in the viscous analysis remain at the level of rounding error, 
irrespective of the grid. 
Although the velocity and magnetic fields are two-dimensional in this model, 
the electric potential, magnetic vector potential, and magnetic helicity have 
three-dimensional distributions. 
Even in such MHD flow field, the accuracy and convergence of this numerical method 
are found to be appropriate.

%++++++++++++++++++++++++++++++++++++++++++++++++++++++++++++++++++++++++++++++
\subsection{Hartmann flow}
%++++++++++++++++++++++++++++++++++++++++++++++++++++++++++++++++++++++++++++++

The Hartmann flow is a model of MHD flow for which an analytical solution exists. 
The validity of the propsed numerical method is verified by analyzing the Hartmann flow 
and comparing the approximate solution with the exact solution. 
A uniform magnetic field perpendicular to the wall is applied to a laminar flow between parallel plates. 
The origin is placed at the center, between the plates. 
The $x$- and $y$-axes are set horizontally and vertically, respectively, to the flow, 
and the $z$-axis is perpendicular to the $x$--$y$ plane. 
The half length of the height between the plates is $h$. 
The upper and lower walls are impermeable, and the fluid flows between the plates. 
The average velocity of the developed flow is $U$.

The computational region is set to $10h$, $2h$, and $2h$ in the $x$-, 
$y$-, and $z$-directions, respectively. 
To confirm whether a fully developed flow can be obtained, 
the velocity field is developed by applying a pressure gradient 
from the stationary state with a uniform magnetic field applied. 
Velocity, magnetic vector potential, and electric potential are set to zero. 
Regarding the boundary conditions, a no-slip condition is given at the wall 
for the velocity. 
An insulating condition is imposed for the electric potential, 
and a uniform magnetic flux density $B$ is applied in the positive direction of the $y$-axis. 
The Dirichlet condition is imposed for the magnetic vector potential. 
At the boundary in the $x$-direction, periodic boundary conditions are applied 
to all dependent variables. 
A pressure gradient obtained from the exact solution is given to drive the flow. 
In the $z$-direction, periodic boundary conditions are applied to velocity, 
pressure, magnetic flux density, and magnetic vector potential. 
Additionally, the insulation condition is imposed, 
the electric field $E_z$ in the $z$-direction is obtained from Ohm's and Ampere's laws, 
and the gradient of electric potential is given.

The simulation is based on a $51 \times N \times 11$ nonuniform grid, 
with $N$ taking values of 51, 101, and 201. 
The results do not change 
even if the number of grid points in the $x$-direction changes. 
The minimum grid widths for each grid are $2.12 \times 10^{-3}$, 
$1.03 \times 10^{-3}$, and $5.18 \times 10^{-4}$, respectively. 
The reference values used in this calculation are as follows: 
the length is $l_\mathrm{ref} = h$, velocity is $u_\mathrm{ref} = U$, 
time is $t_\mathrm{ref} = h/U$, magnetic flux density is $B_\mathrm{ref} = B$, 
magnetic vector potential is $A_\mathrm{ref} = B h$, 
and electric potential is $\psi_\mathrm{ref} = U h B$. 
The Reynolds number is set as $Re = 10^3$, 
the Alfv\'{e}n number $Al = 1$, and the magnetic Reynolds number $Re_m = 10$. 
The Hartmann number at this time is $Ha = 10^2$. 
For all grids, the time step is set to $\Delta t/(h/U) = 0.1$. 
The Courant number is defined as $\mbox{CFL} = \Delta t V/\Delta y_\mathrm{min}$ 
using the representative velocity $U$ and the minimum grid width $\Delta y_\mathrm{min}$. 
The Courant numbers are CFL = 47.2, 97.3, and 193.0 at $N = 51$, $101$, and $201$, respectively. 
The Courant numbers using local velocities are CFL = 0.17, 0.34, and 0.68, respectively.

For the velocity $u$, magnetic flux density $B_x$, pressure $p$, 
electric field $E_z$, current density $j_z$, Lorentz forces $F_x$, $F_y$, 
and magnetic vector potential $A_z$, 
the present results are compared with the analytical solutions in \citep{Moreau_1990}. 
The nondimensionalized exact solutions are given as
%------------------------------------------------------------------------------
\begin{equation}
  u = \frac{Ha}{Ha - \tanh(Ha)} 
  \left[ 1-\frac{\cosh (Ha y)}{\cosh (Ha)} \right],
\end{equation}
\begin{equation}
  p = P_0(x) - \frac{1}{Al^2} \frac{1}{2} B_x^{2}, \quad 
  P_0(x) = \frac{Ha^2 \tanh(Ha)}{Re [Ha - \tanh (Ha)]} x,
\end{equation}
\begin{equation}
  B_x = Re_m \frac{\tanh(Ha)}{Ha - \tanh (Ha)} 
  \left[ \frac{\sinh (Ha y)}{\sinh (Ha)} - y \right],
\end{equation}
\begin{equation}
  E_z = - 1,
\end{equation}
\begin{equation}
  J_z = - Re_m \frac{\tanh (Ha)}{Ha - \tanh (Ha)} 
  \left[ \frac{Ha \cosh (Ha y)}{\sinh (Ha)} - 1 \right],
\end{equation}
\begin{equation}
  F_x = - \frac{1}{Al^2} J_z B_y,
\end{equation}
\begin{equation}
  F_y = \frac{1}{Al^2} J_z B_x,
\end{equation}
\begin{equation}
  A_z = - x +Re_m \frac{\tanh (Ha)}{Ha - \tanh (Ha)}
  \left[ \frac{\cos (Ha y)}{Ha \sinh(Ha)} - \frac{1}{2} y^{2} \right].
\end{equation}
%------------------------------------------------------------------------------

\begin{figure}[!t]
\begin{minipage}{0.325\linewidth}
\begin{center}
\includegraphics[trim=0mm 0mm 0mm 0mm, clip, width=50mm]{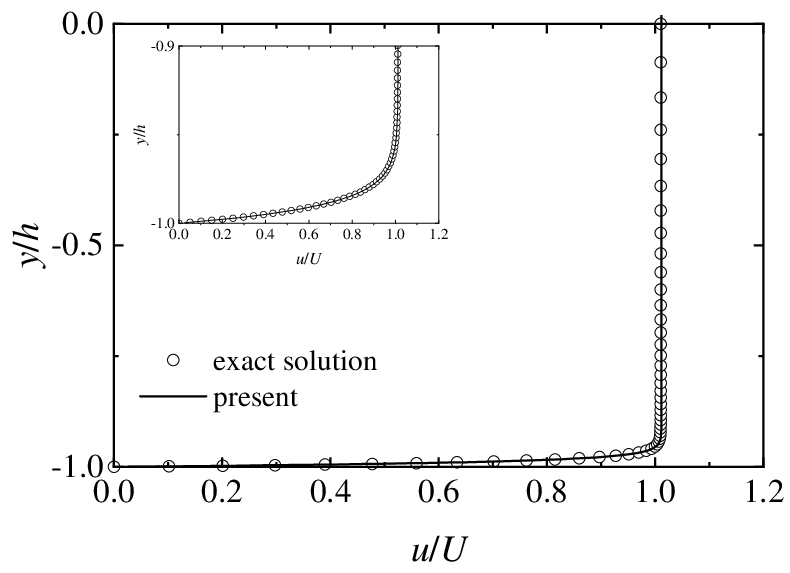} \\
{\small (a) $u$}
\end{center}
\end{minipage}
\begin{minipage}{0.325\linewidth}
\begin{center}
\includegraphics[trim=0mm 0mm 0mm 0mm, clip, width=50mm]{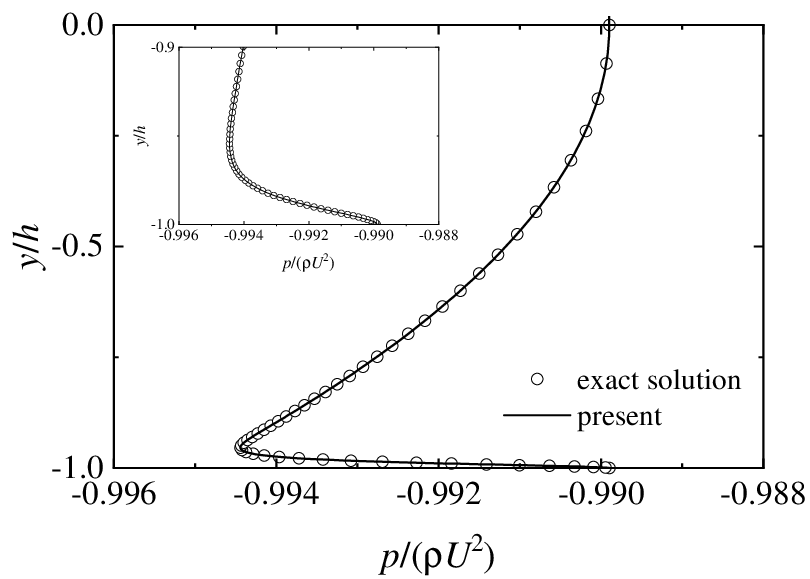} \\
{\small (b) $p$}
\end{center}
\end{minipage}
\begin{minipage}{0.325\linewidth}
\begin{center}
\includegraphics[trim=0mm 0mm 0mm 0mm, clip, width=50mm]{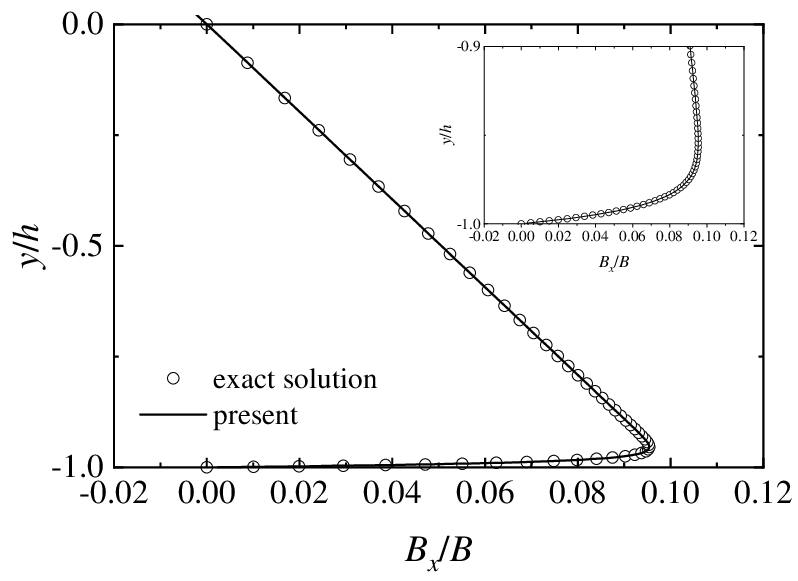} \\
{\small (c) $B_x$}
\end{center}
\end{minipage}
\\

\begin{minipage}{0.325\linewidth}
\begin{center}
\includegraphics[trim=0mm 0mm 0mm 0mm, clip, width=50mm]{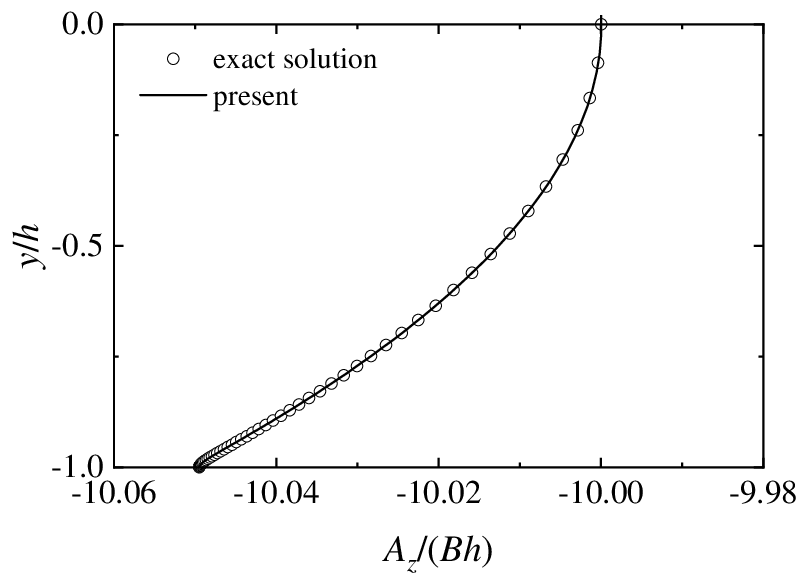} \\
{\small (d) $A_z$}
\end{center}
\end{minipage}
\begin{minipage}{0.325\linewidth}
\begin{center}
\includegraphics[trim=0mm 0mm 0mm 0mm, clip, width=50mm]{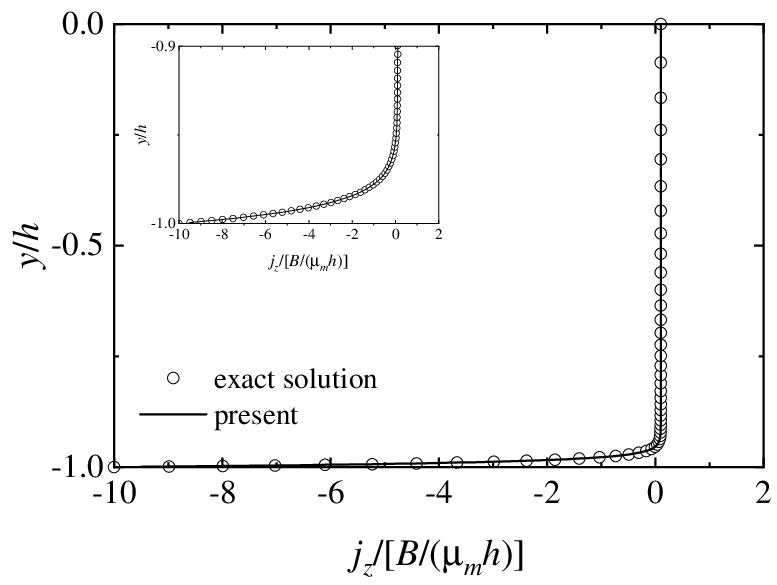} \\
{\small (e) $j_z$}
\end{center}
\end{minipage}
\begin{minipage}{0.325\linewidth}
\begin{center}
\includegraphics[trim=0mm 0mm 0mm 0mm, clip, width=50mm]{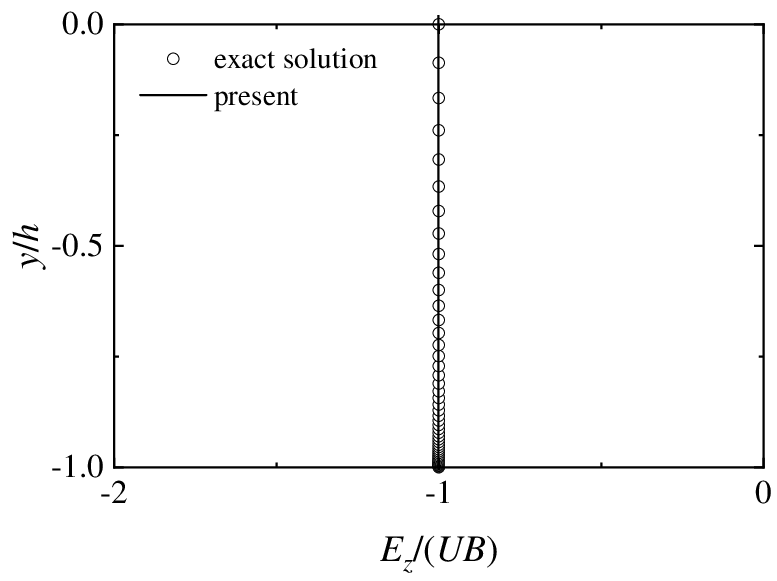} \\
{\small (f) $E_z$}
\end{center}
\end{minipage}
\\

\begin{minipage}{0.5\linewidth}
\begin{center}
\includegraphics[trim=0mm 0mm 0mm 0mm, clip, width=50mm]{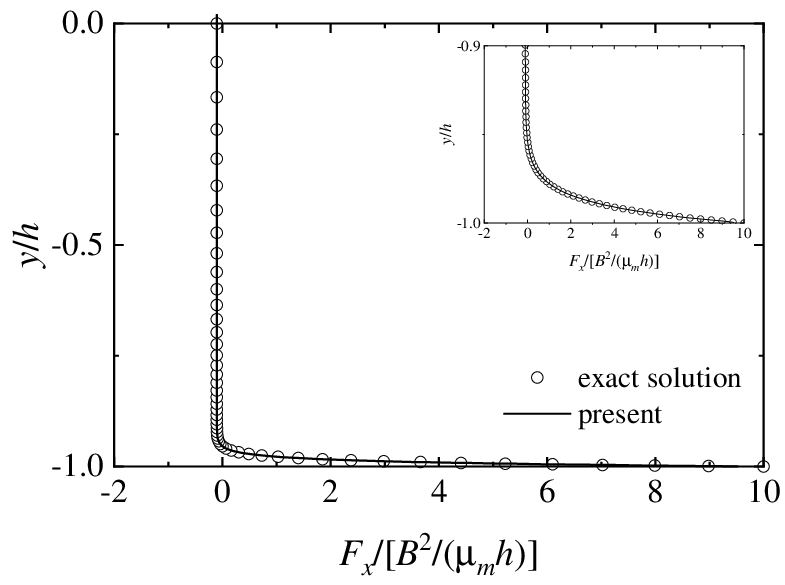} \\
{\small (g) $F_x$}
\end{center}
\end{minipage}
\begin{minipage}{0.5\linewidth}
\begin{center}
\includegraphics[trim=0mm 0mm 0mm 0mm, clip, width=50mm]{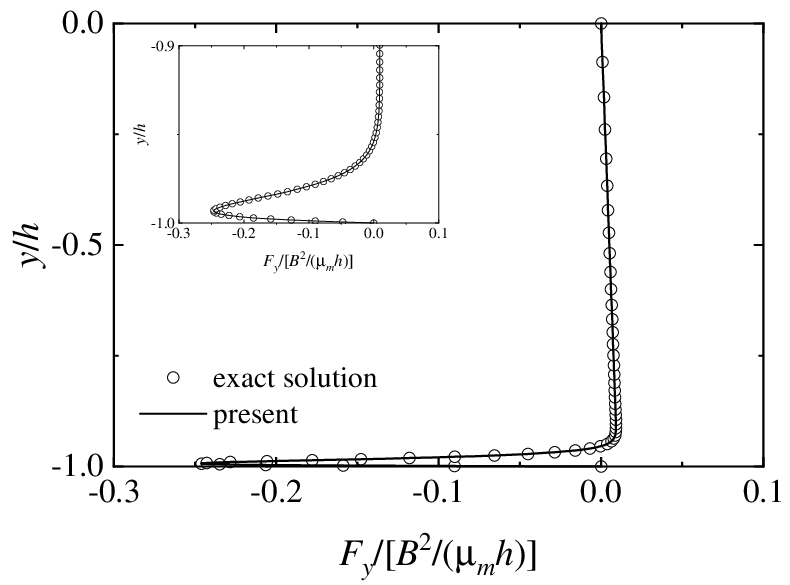} \\
{\small (h) $F_y$}
\end{center}
\end{minipage}
\caption{Distributions of velocity, pressure, magnetic flux density, 
magnetic vector potential, electric potential, current density, and 
Lorentz force at $x/h = 5$ and $z/h = 1$: 
$Re = 10^3$, $Al = 1$, $Re_m = 10$, $N = 201$.}
\label{hartmann_dist}
\end{figure}

Figure \ref{hartmann_dist} shows the distribution at $x/h = 5$ and $z/h = 1$ 
for the velocity $u$, pressure $p$, magnetic flux density $B_x$, 
magnetic vector potential $A_z$, current density $j_z$, electric field $E_z$, 
and Lorentz forces $F_x$ and $F_y$. 
The flow and magnetic fields are fully developed. 
For all distributions, the calculated values support the exact solutions. 
The thin Hartmann layer can be accurately captured by this computational method.

%------------------------------------------------------------------------------
% Figure 12
%------------------------------------------------------------------------------
%------------------------------------------------------------------------------

%------------------------------------------------------------------------------
% Figure 13
%------------------------------------------------------------------------------
\begin{figure}[!t]
\begin{minipage}{0.5\linewidth}
\begin{center}
\includegraphics[trim=0mm 0mm 0mm 0mm, clip, width=70mm]{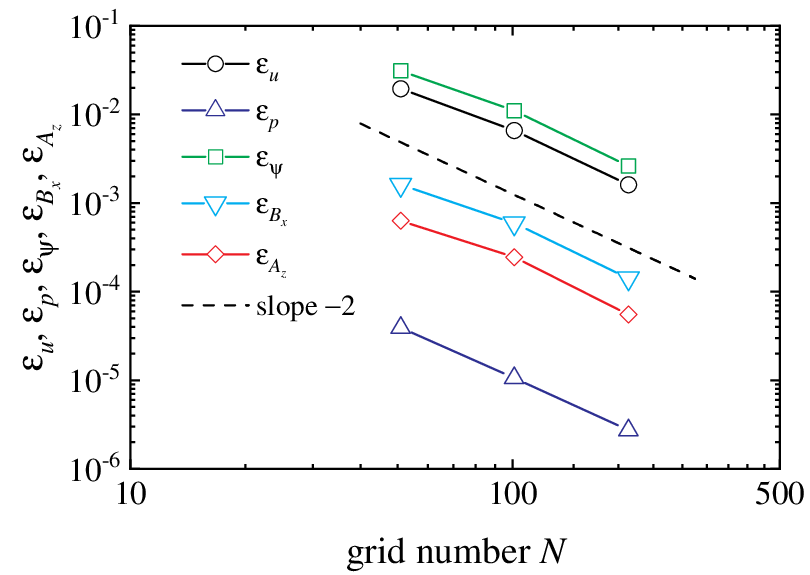} \\
{\small (a) $u$, $p$, $\psi$, $B_x$, $A_z$}
\end{center}
\end{minipage}
\begin{minipage}{0.5\linewidth}
\begin{center}
\includegraphics[trim=0mm 0mm 0mm 0mm, clip, width=70mm]{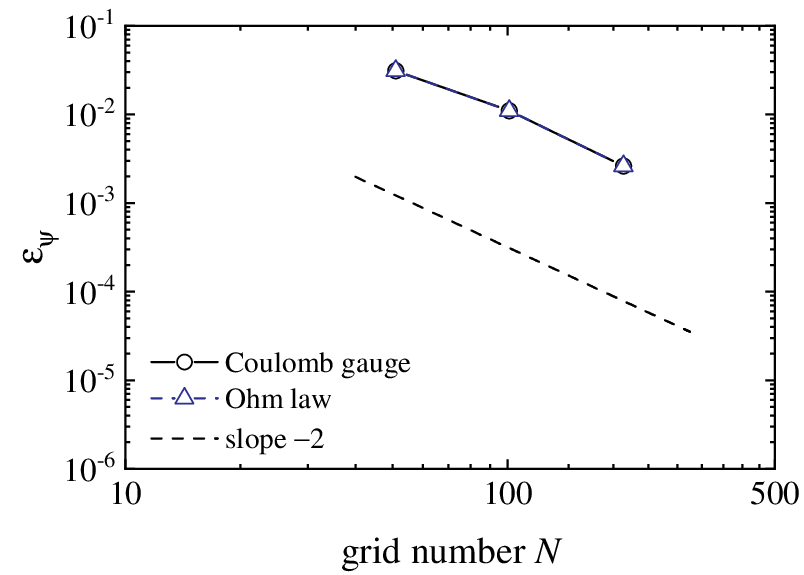} \\
{\small (b) $\psi$}
\end{center}
\end{minipage}
\caption{Maximum errors of velocity ($u$), pressure ($p$), electric potential ($\psi$), 
magnetic flux density ($B_x$), and magnetic vector potential ($A_z$): 
$Re = 10^3$, $Al = 1$, $Re_m = 10$.}
\label{hartmann_error}
\end{figure}
%------------------------------------------------------------------------------

Figure \ref{hartmann_error} (a) shows the maximum errors, 
$\varepsilon_{u}$, $\varepsilon_{p}$, and $\varepsilon_{\psi}$, 
$\varepsilon_{B_x}$, and $\varepsilon_{A_z}$, 
for the velocity, pressure, electric potential, magnetic flux density, 
and magnetic vector potential. 
As the number of grid points $N$ increases, 
the error decreases with a slope of $-2$, 
demonstrating the second-order convergence of this numerical method. 
Figure \ref{hartmann_error} (b) shows the maximum error of electric potential 
obtained by solving the charge conservation law using Ohm's law. 
The error is almost the same as that of the electric potential obtained 
using the Coulomb gauge, 
and the method for calculating the electric potential does not introduce variation.

The maximum divergence errors of velocity, magnetic flux density, 
and magnetic vector potential in this analysis are $6.00\times 10^{-14}$, 
$4.44\times 10^{-10}$ and $4.00\times 10^{-14}$, respectively. 
Even without giving the exact solution as an initial condition, 
the divergence-free condition is maintained until the solution converges.

%++++++++++++++++++++++++++++++++++++++++++++++++++++++++++++++++++++++++++++++
\subsection{Taylor decaying vortex}
%++++++++++++++++++++++++++++++++++++++++++++++++++++++++++++++++++++++++++++++

the accuracy and convergence of the present numerical method were also verified 
in high Reynolds number flows with the decaying of total energy. 
The Taylor decaying vortex was analyzed, 
and the approximate solutions were compared with the exact solution \citep{Taylor_1923}. 
The solution to the Taylor decaying vortex problem 
under the magnetic field is given as
%------------------------------------------------------------------------------
\begin{align}
  \Psi_z &= \frac{1}{k} \cos(k x) \cos(k y) e^{-\frac{2 k^2}{Re}t}, \\
  A_z &= \frac{1}{k} \cos(k x) \cos(k y) e^{-\frac{2 k^2}{Re_m}t}, \\
  p &= - \frac{1}{4} \left[ \cos(2 k x) + \cos(2 k y) \right] e^{-\frac{4 k^2}{Re}t} 
        \nonumber \\
        & \quad 
        + \frac{1}{4 Al^2} \left[4 \cos^2(k x) \cos^2(k y) - 1 \right] 
          e^{-\frac{4 k^2}{Re_m}t},
\end{align}
%------------------------------------------------------------------------------
where $k = 2\pi$. 
$\Psi_z$ and $A_z$ are the stream function and magnetic vector potential, respectively. 
The velocities are calculated as $u = \partial \Psi_z/\partial y$ 
and $v = -\partial \Psi_z/\partial x$. 
The magnetic flux densities are calculated 
as $B_x = \partial A_z/\partial y$ and $B_y = -\partial A_z/\partial x$. 
These equations are nondimensionalized by the maximum values, $U$ and $B$, 
of the velocity and magnetic flux density, respectively, 
and the wavelength, $L$, of the periodic vortex. 

The computational region is $L \times L$, and 
the length in the $z$-direction is the grid spacing. 
The exact solution is given as the initial condition, 
and the periodic boundary is set as the boundary condition. 
A uniform grid with $N \times N \times 2$ is used. 
$N$ is the number of grid points in the $x$- and $y$-directions. 
Similarly to the existing study \citep{Yanaoka_2023}, 
$N = 11$, $21$, $41$, and $81$ are used to investigate the convergence of the numerical solutions 
against the number of grid points. 
The reference values used in this calculation are 
$l_\mathrm{ref} = L$, $u_\mathrm{ref} = U$, $t_\mathrm{ref} = L/U$, 
$B_\mathrm{ref} = B$, $A_\mathrm{ref} = B L$, and $\psi_\mathrm{ref} = U L B$. 
The calculation conditions are the same as in \citep{Liu&Wang_2001}, 
and $Re = 10^4$, $Re_m = 50$, and $Al = 1$. 
The condition of $Re = 10^2$ and $Re_m = 1$ is also considered 
to determine the decaying tendency of the vortex. 
The time step is fixed at $\Delta t/(L/U) = 0.001$, 
and the approximate value at $t/(L/U) = 0.5$, 
when the strength of the vortex is halved, is compared with the exact solution. 
The same time step as that in \citep{Liu&Wang_2001} is used. 
The Courant number is defined as $\mbox{CFL} = \Delta t U/\Delta x$ 
using the maximum velocity $U$ and grid spacing $\Delta x$. 
As the time step is fixed, the Courant number varies with the grid width, 
and the Courant numbers are $\mbox{CFL} = 0.01-0.08$. 
In an inviscid analysis, the time step of $\Delta t/(L/U) = 0.025$ 
with a Courant number of $\mbox{CFL} = 1.0$ is used.

%------------------------------------------------------------------------------
% Figure 14
%------------------------------------------------------------------------------
\begin{figure}[!t]
\begin{minipage}{0.5\linewidth}
\begin{center}
\includegraphics[trim=0mm 0mm 0mm 0mm, clip, width=70mm]{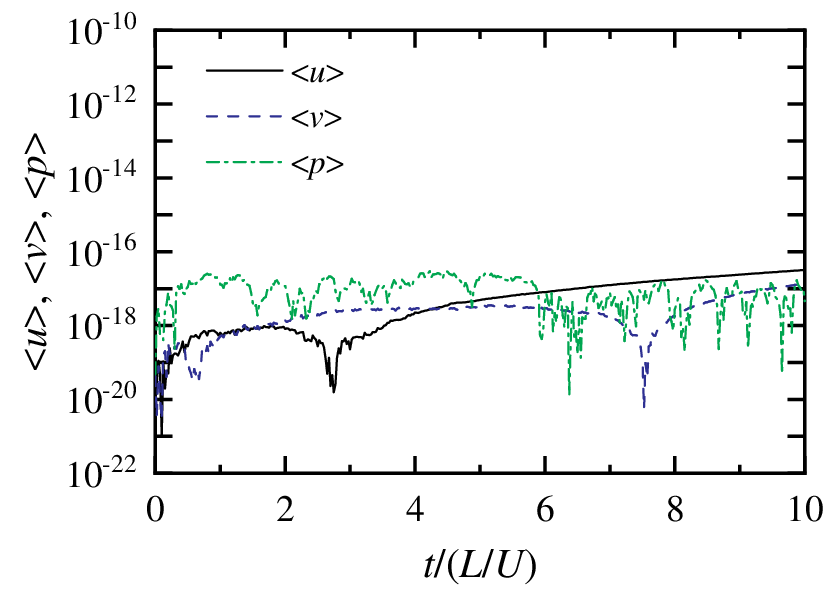} \\
{\small (a) $u$, $v$, $p$}
\end{center}
\end{minipage}
\begin{minipage}{0.5\linewidth}
\begin{center}
\includegraphics[trim=0mm 0mm 0mm 0mm, clip, width=70mm]{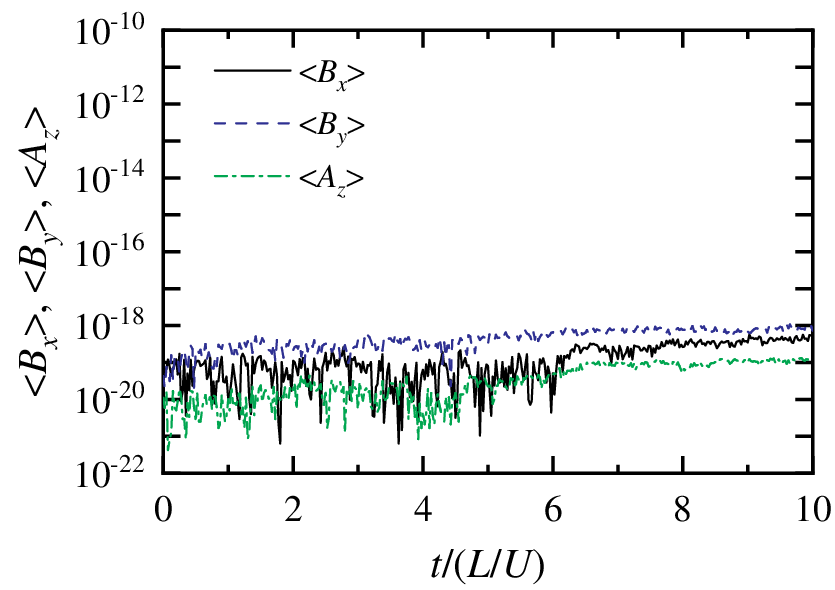} \\
{\small (b) $B_x$, $B_y$, $A_z$}
\end{center}
\end{minipage}
\caption{Total amounts of velocity ($u$, $v$), pressure ($p$), 
magnetic flux density ($B_x$, $B_y$), and magnetic vector potential ($A_z$): 
$Re = Re_m = \infty$, $Al = 1$, $N = 41$.}
\label{decay2d_sum_u&B}
\end{figure}
%------------------------------------------------------------------------------

As this computational model is a periodic flow, 
the total amounts of velocity and magnetic flux density are conserved for $Re = Re_m = \infty$. 
Figure \ref{decay2d_sum_u&B} shows the total amounts, 
$\langle u \rangle$, $\langle v \rangle$, 
$\langle B_x \rangle$, and $\langle B_y \rangle$, 
of the velocity and magnetic flux density. 
The total amounts, $\langle p \rangle$ and $\langle A_z \rangle$, 
of the pressure and magnetic vector potential are shown in Fig. \ref{decay2d_sum_u&B}. 
The results are obtained using $N = 41$. 
From the volume integral of the exact solution, each total amount is zero. 
All the total amounts remain at low levels, 
indicating good conservation of velocity and magnetic flux density.

\begin{figure}[!t]
\begin{minipage}{0.325\linewidth}
\begin{center}
\includegraphics[trim=0mm 0mm 0mm 0mm, clip, width=50mm]{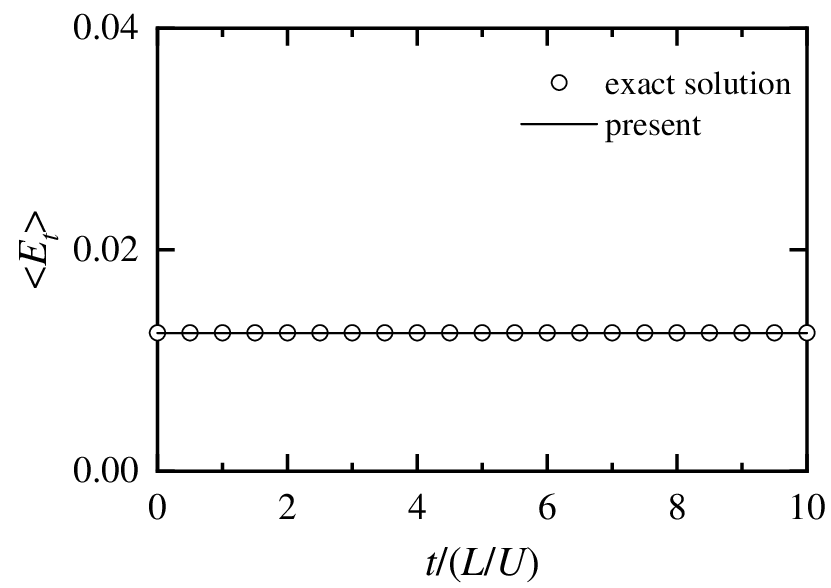} \\
{\small (a) $E_t$}
\end{center}
\end{minipage}
\begin{minipage}{0.325\linewidth}
\begin{center}
\includegraphics[trim=0mm 0mm 0mm 0mm, clip, width=50mm]{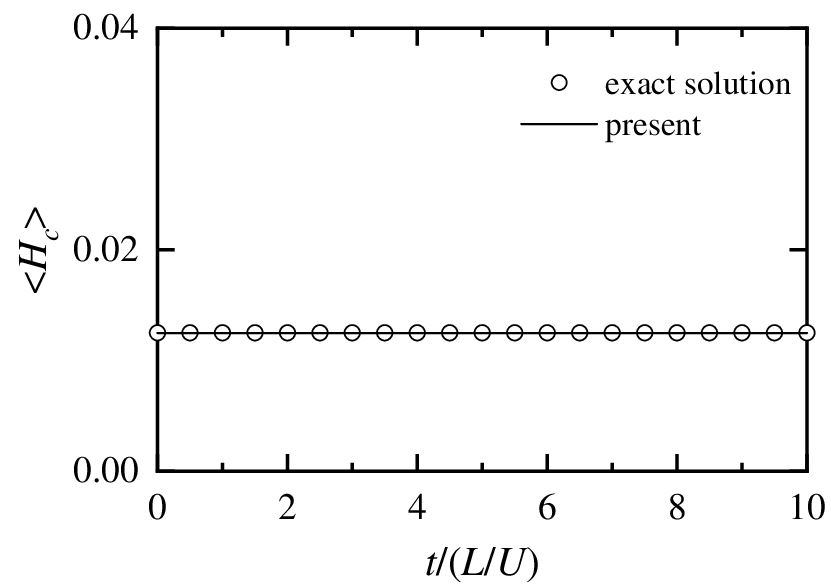} \\
{\small (b) $H_c$}
\end{center}
\end{minipage}
\begin{minipage}{0.325\linewidth}
\begin{center}
\includegraphics[trim=0mm 0mm 0mm 0mm, clip, width=50mm]{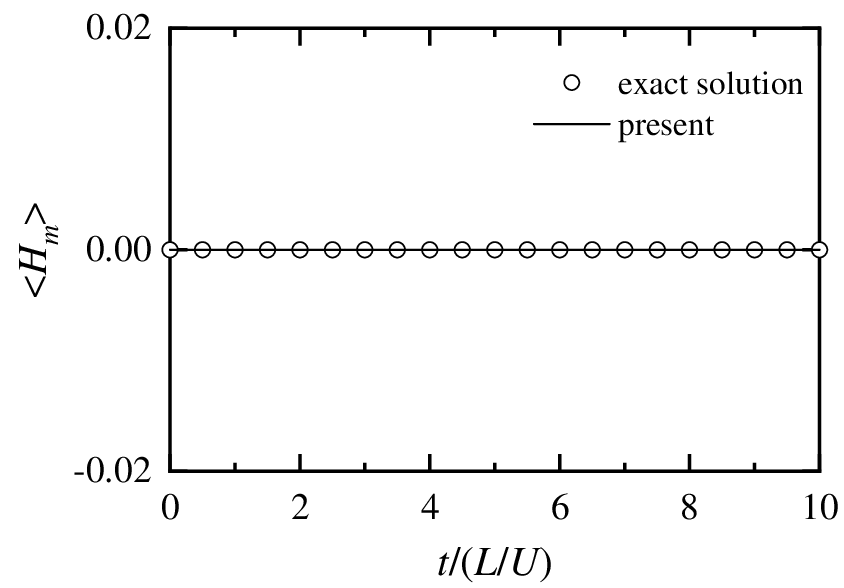} \\
{\small (c) $H_m$}
\end{center}
\end{minipage}
\caption{Total amounts of total energy ($E_t$), cross-helicity ($H_c$), 
and magnetic helicity ($H_m$): $Re = Re_m = \infty$, $Al = 1$, $N = 41$.}
\label{decay2d_sum_Et&Hc&Hm}
\end{figure}

The total amounts, $\langle E_t \rangle$, $\langle H_c \rangle$, 
and $\langle H_m \rangle$, of total energy, cross-helicity, 
and magnetic helicity are shown in Fig. \ref{decay2d_sum_Et&Hc&Hm}. 
This approximate solution supports the exact solution, 
and the energy is conserved. 
Magnetic helicity is congenitally conserved in two-dimensional flow and magnetic fields. 
The magnetic helicity in this calculation remains zero, 
and no unphysical behavior such as the generation of magnetic helicity appears.

%------------------------------------------------------------------------------
% Figure 15
%------------------------------------------------------------------------------
%------------------------------------------------------------------------------

%------------------------------------------------------------------------------
% Figure 16
%------------------------------------------------------------------------------
\begin{figure}[!t]
\begin{minipage}{0.5\linewidth}
\begin{center}
\includegraphics[trim=0mm 0mm 0mm 0mm, clip, width=70mm]{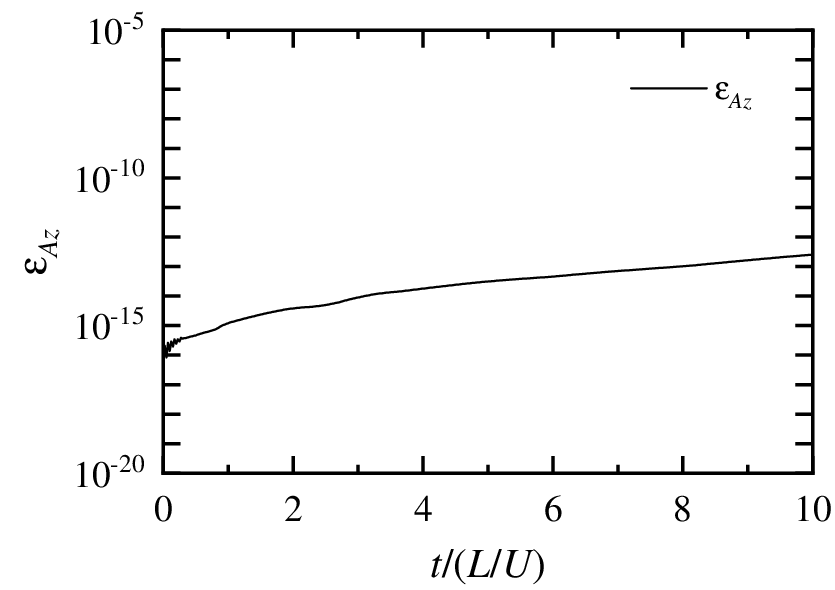} \\
{\small (a) $A_z$}
\end{center}
\end{minipage}
\begin{minipage}{0.5\linewidth}
\begin{center}
\includegraphics[trim=0mm 0mm 0mm 0mm, clip, width=70mm]{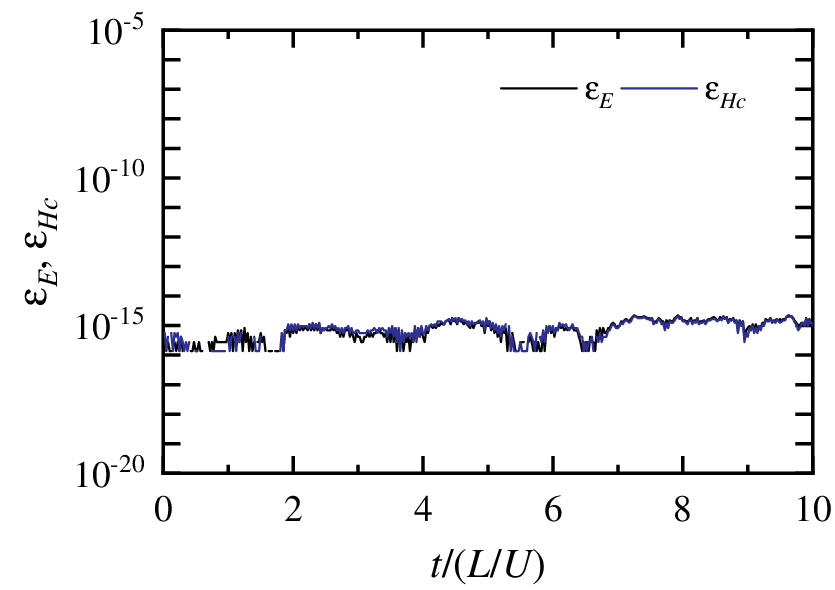} \\
{\small (b) $E_t$, $H_c$}
\end{center}
\end{minipage}
\caption{Maximum error of magnetic vector potential ($A_z$), 
and relative errors of total energy ($E_t$) and cross-helicity ($H_c$): 
$Re = Re_m = \infty$, $Al = 1$, $N = 41$.}
\label{decay2d_error}
\end{figure}
%------------------------------------------------------------------------------

Figure \ref{decay2d_error} (a) shows the maximum error, $\varepsilon_{A_z}$, 
of the magnetic vector potential. 
The error remains low. 
Figure \ref{decay2d_error} (b) shows the relative errors, 
$\varepsilon_{E_t} = |(\langle E_t \rangle - \langle E_t \rangle_e)/\langle E_t \rangle_e|$ 
and $\varepsilon_{H_c} = |(\langle H_c \rangle - \langle H_c \rangle_e)/\langle H_c \rangle_e|$, 
of the total amounts of total energy and cross-helicity, respectively. 
The subscript $e$ represents the exact solution. 
The error remains at the level of rounding errors. 
This numerical method achieves excellent energy conservation properties. 
The absolute error of magnetic helicity is zero. 
The maximum divergence errors of velocity and magnetic flux density 
in the inviscid analysis are $1.39\times 10^{-14}$ and $3.54\times 10^{-13}$, respectively.

%------------------------------------------------------------------------------
% Figure 17
%------------------------------------------------------------------------------
\begin{figure}[!t]
\begin{minipage}{0.5\linewidth}
\begin{center}
\includegraphics[trim=0mm 0mm 0mm 0mm, clip, width=70mm]{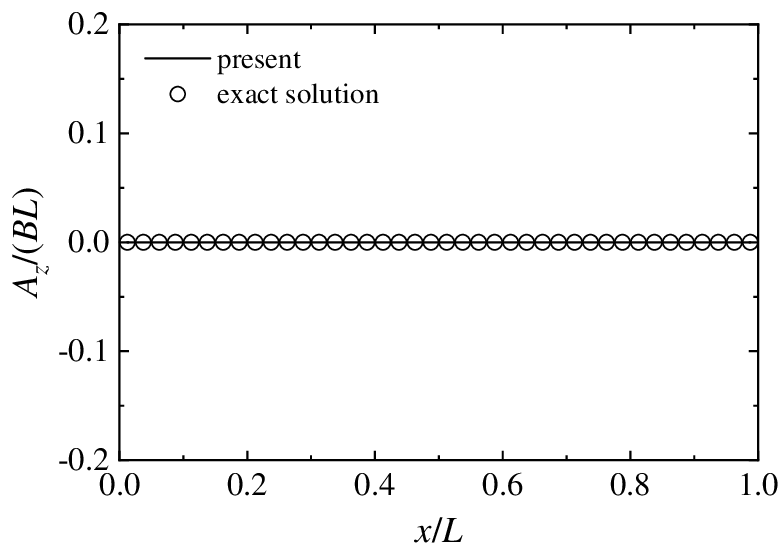} \\
{\small (a) $Re = 10^2$, $Re_m = 1$}
\end{center}
\end{minipage}
\begin{minipage}{0.5\linewidth}
\begin{center}
\includegraphics[trim=0mm 0mm 0mm 0mm, clip, width=70mm]{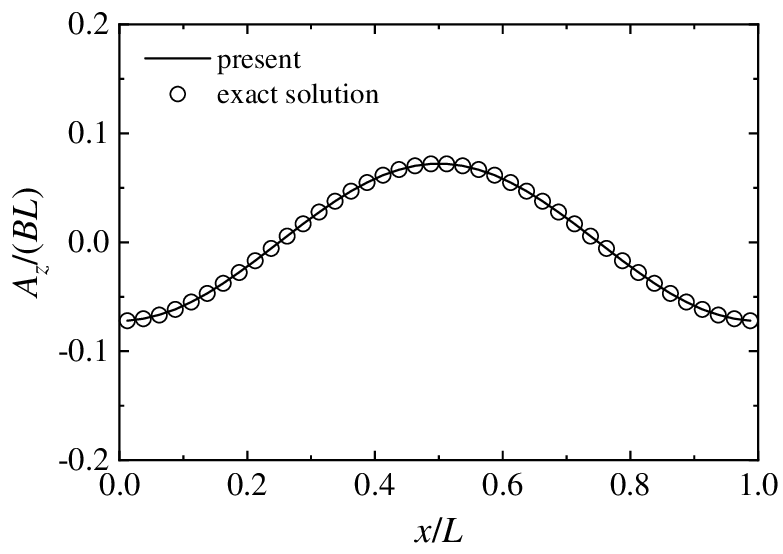} \\
{\small (b) $Re = 10^4$, $Re_m = 50$}
\end{center}
\end{minipage}
\caption{Distribution of magnetic vector potential at $y/L=0.5$: 
$t/(L/U) = 0.5$, $Re = 10^2, 10^4$, $Al = 1$, $Re_m = 1, 50$, $N = 41$.}
\label{decay2d_Az}
\end{figure}
%------------------------------------------------------------------------------

%------------------------------------------------------------------------------
% Figure 18
%------------------------------------------------------------------------------
\begin{figure}[!t]
\begin{minipage}{0.325\linewidth}
\begin{center}
\includegraphics[trim=0mm 0mm 0mm 0mm, clip, width=50mm]{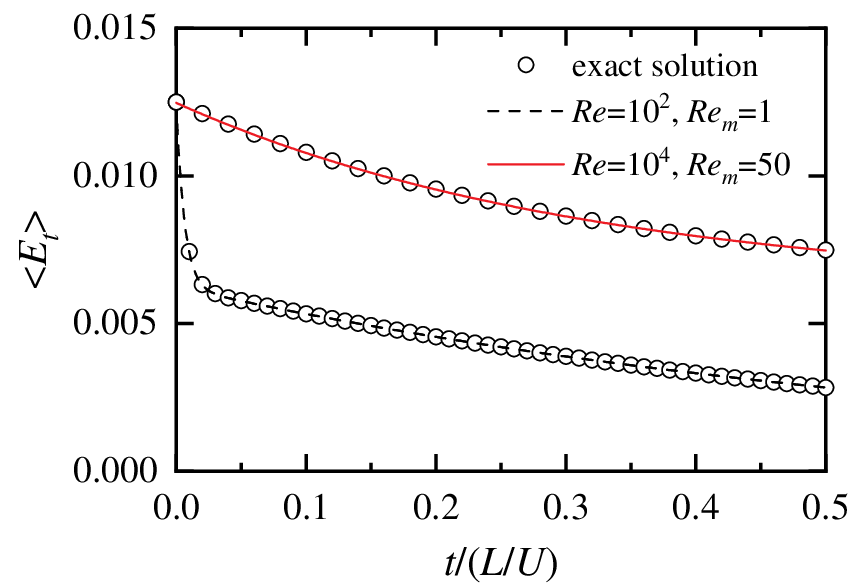} \\
{\small (a) $E_t$}
\end{center}
\end{minipage}
\begin{minipage}{0.325\linewidth}
\begin{center}
\includegraphics[trim=0mm 0mm 0mm 0mm, clip, width=50mm]{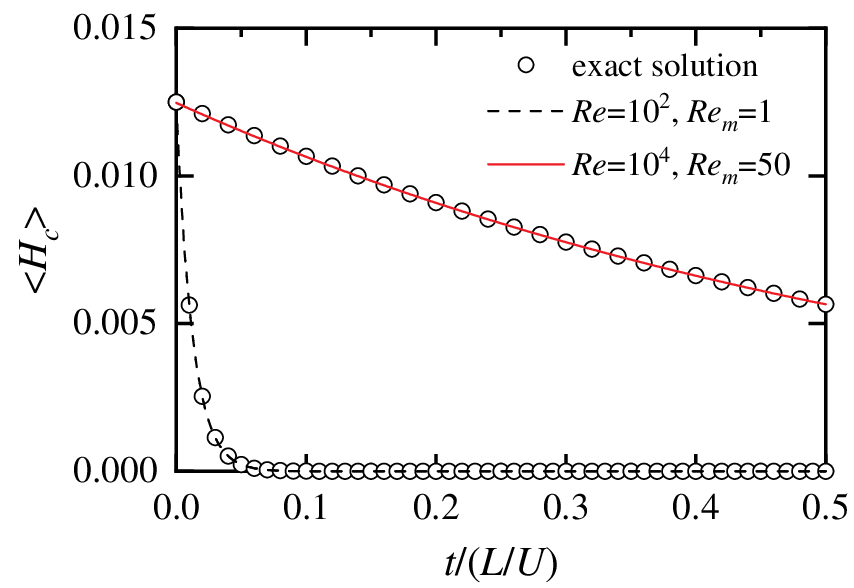} \\
{\small (b) $H_c$}
\end{center}
\end{minipage}
\begin{minipage}{0.325\linewidth}
\begin{center}
\includegraphics[trim=0mm 0mm 0mm 0mm, clip, width=50mm]{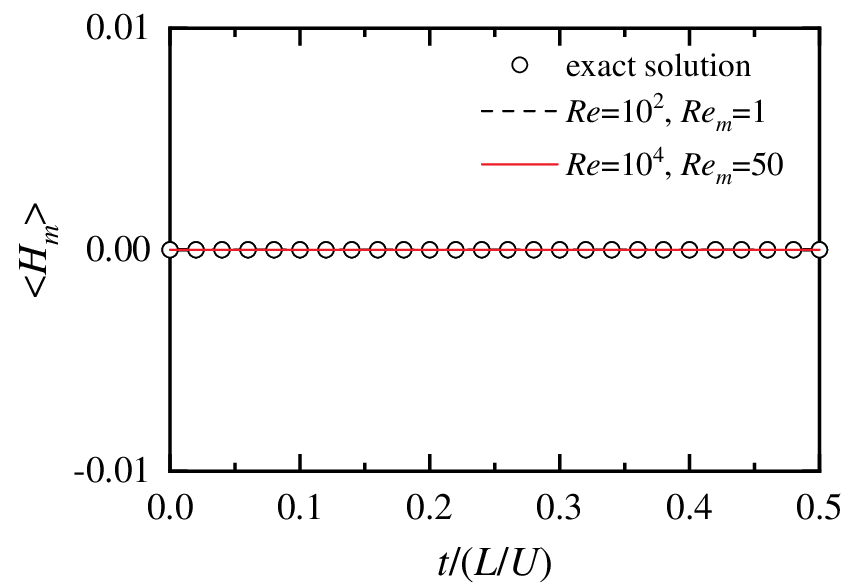} \\
{\small (c) $H_m$}
\end{center}
\end{minipage}
\caption{Total amounts of total energy ($E_t$), cross-helicity ($H_c$), 
and magnetic helicity ($H_m$): 
$Re = 10^2, 10^4$, $Al = 1$, $Re_m = 1, 50$, $N = 41$.}
\label{decay2d_Et&Hc&Hm}
\end{figure}
%------------------------------------------------------------------------------

For $Re = 10^2$, $10^4$ and $Re_m = 1$, $50$, 
the magnetic vector potential at $y/L = 0.5$ and $t/(L/U) = 0.5$ is compared with 
the exact solution in Fig. \ref{decay2d_Az}. 
This approximate solution supports the exact solution. 
Furthermore, the electric potential is kept constant. 
For $Re = 10^2$, $A_z$ is nearly zero because the vortex is damped by viscosity.

To investigate the trend of energy decaying, 
the total amounts, $\langle E_t \rangle$, $\langle H_c \rangle$, 
and $\langle H_m \rangle$, of total energy, cross-helicity, 
and magnetic helicity, respectively, for two conditions of $Re = 10^2$, $Re_m = 1$ and $Re = 10^4$, $Re_m = 50$ 
are shown in Fig. \ref{decay2d_Et&Hc&Hm}. 
At $Re = 10^2$, the total energy and cross-helicity decrease sharply. 
This approximate solution supports the exact solution, 
and the energy decay process is accurately captured. 
Regardless of the conditions, the total amount of magnetic helicity is kept at zero, 
which supports the exact solution. 
In a two-dimensional field, magnetic helicity is a priori conserved. 
As in the case of $Re = Re_m = \infty$, no nonphysical generation of magnetic helicity appears 
in this calculation, and conservation properties are not degraded. 

\begin{figure}[!t]
\begin{center}
\includegraphics[trim=0mm 0mm 0mm 0mm, clip, width=70mm]{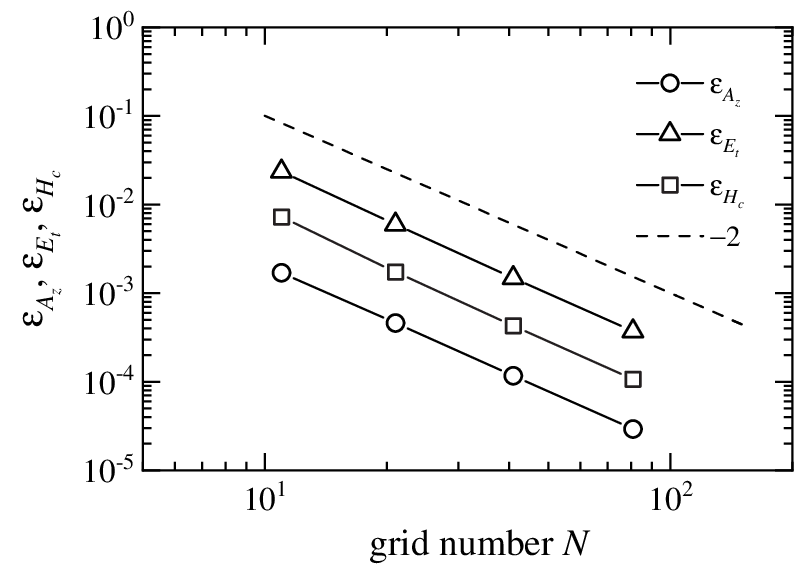} 
\end{center}
\vspace*{-1.0\baselineskip}
\caption{Errors of magnetic vector potential ($A_z$), total energy ($E_t$), 
and cross-helicity ($H_c$): 
$Re = 10^4$, $Re_m = 50$, $t/(L/U) = 0.5$.}
\label{decay2d_error_E_Hc}
\end{figure}

Figure \ref{decay2d_error_E_Hc} shows the maximum error, $\varepsilon_{A_z}$, 
of the magnetic vector potential 
and the relative errors, $\varepsilon_{E_t}$ and $\varepsilon_{H_c}$, 
of the total energy and cross-helicity, respectively. 
The error decreases with a slope of $-2$, 
and the present numerical method has second-order convergence.
The maximum divergence errors of velocity and magnetic flux density in the viscous analysis 
are $2.78\times 10^{-14}$ and $1.08\times 10^{-12}$, respectively.

%------------------------------------------------------------------------------
% Figure 19
%------------------------------------------------------------------------------
%------------------------------------------------------------------------------

%++++++++++++++++++++++++++++++++++++++++++++++++++++++++++++++++++++++++++++++
\subsection{Three-dimensional Taylor decaying vortex}
%++++++++++++++++++++++++++++++++++++++++++++++++++++++++++++++++++++++++++++++

Antuono \citep{Antuono_2020} found an analytical solution 
for a periodic three-dimensional decaying vortex 
using the method of Ethier and Steinman \citep{Ethier&Steinman_1994} as follows:
%------------------------------------------------------------------------------
\begin{subequations}
\begin{align}
  u_{1,2} &= \alpha \left[ 
    \sin(k x + \theta_{1,2}) \cos(k y + \phi_{1,2}) \sin(k z + \psi_{1,2}) 
  \right. \nonumber \\
  & \, \left. 
  - \cos(k z + \theta_{1,2}) \sin(k x + \phi_{1,2}) \sin(k y + \psi_{1,2}) 
  \right] e^{-3 \frac{k^2}{Re} t}, 
  \label{3D_Taylor_vortex_u} \\
  v_{1,2} &= \alpha \left[ 
    \sin(k y + \theta_{1,2}) \cos(k z + \phi_{1,2}) \sin(k x + \psi_{1,2}) 
  \right. \nonumber \\
  & \, \left. 
  - \cos(k x + \theta_{1,2}) \sin(k y + \phi_{1,2}) \sin(k z + \psi_{1,2}) 
  \right] e^{-3 \frac{k^2}{Re} t}, 
  \label{3D_Taylor_vortex_v} \\
  w_{1,2} &= \alpha \left[ 
    \sin(k z + \theta_{1,2}) \cos(k x + \phi_{1,2}) \sin(k y + \psi_{1,2}) 
  \right. \nonumber \\
  & \, \left. 
  - \cos(k y + \theta_{1,2}) \sin(k z + \phi_{1,2}) \sin(k x + \psi_{1,2}) 
  \right] e^{-3 \frac{k^2}{Re} t}, 
  \label{3D_Taylor_vortex_w} \\
  p_{1,2} &= p_0 - \frac{|\bm{u}_{1,2}|^2}{2}, \quad 
  p_0 = \frac{1}{2} e^{-6 \frac{k^2}{Re} t},
  \label{3D_Taylor_vortex_p}
\end{align}
\end{subequations}
%------------------------------------------------------------------------------
where $\alpha = 4 \sqrt{2}/(3 \sqrt{3})$, and 
$k = 2\pi$ is the nondimensionalized wavenumber. 
The above equation is nondimensionalized using the vortex wavelength $L$ 
and the maximum velocity $U$, 
with the Reynolds number defined as $Re = UL/\nu$. 
Subscripts 1 and 2 represent two solutions, which have similar distributions. 
The phases $\theta_{1,2}$, $\phi_{1,2}$, and $\psi_{1,2}$ are given as
%------------------------------------------------------------------------------
\begin{equation}
  \theta_1 = \psi_1 - \frac{5 \pi}{6}, \quad 
  \phi_1 = \psi_1 - \frac{\pi}{6}, \quad 
  \psi_1 = \cos^{-1} \left( \frac{R}{\sqrt{1 + R^2}} \right),
\end{equation}
\begin{equation}
  \theta_2 = \phi_1, \quad 
  \phi_2 = \theta_1, \quad 
  \psi_2 = \psi_1,
\end{equation}
%------------------------------------------------------------------------------
where $R$ is a parameter, 
and the value excluding the singularity value $R = \pm 1/\sqrt{3}$ is set. 
In this study, $R = 0$ is set the same as in \citep{Antuono_2020}. 

This flow of the three-dimensional Taylor decaying vortex 
is categorized as the Beltrami flow. 
Therefore, the velocity vector $\bm{u}$ and the vorticity vector $\bm{\omega}$ are parallel, 
and $\bm{u} \times \bm{\omega} = \bm{u} \times (\nabla \times \bm{u}) = 0$. 
Assuming that the velocity vector and magnetic flux density vector represent 
the Beltrami flow, the exact solution of the velocity and pressure 
of the three-dimensional Taylor decaying vortex is expressed 
by Eq. (\ref{3D_Taylor_vortex_u}) to Eq. (\ref{3D_Taylor_vortex_p}) 
even under an applied magnetic field. 
Using the three-dimensional Taylor decaying vortex model 
proposed by Antuono \citep{Antuono_2020}, 
the magnetic flux density in an MHD flow 
under an applied magnetic field is given as
%------------------------------------------------------------------------------
\begin{subequations}
\begin{align}
  B_{x 1,2} &= \alpha \left[ 
    \sin(k x + \theta_{1,2}) \cos(k y + \phi_{1,2}) \sin(k z + \psi_{1,2}) 
  \right. \nonumber \\
  & \, \left. 
  - \cos(k z + \theta_{1,2}) \sin(k x + \phi_{1,2}) \sin(k y + \psi_{1,2}) 
  \right] e^{-3 \frac{k^2}{Re_m} t}, 
  \label{3D_Taylor_vortex_mag_Bx} \\
  B_{y 1,2} &= \alpha \left[ 
    \sin(k y + \theta_{1,2}) \cos(k z + \phi_{1,2}) \sin(k x + \psi_{1,2}) 
  \right. \nonumber \\
  & \, \left. 
  - \cos(k x + \theta_{1,2}) \sin(k y + \phi_{1,2}) \sin(k z + \psi_{1,2}) 
  \right] e^{-3 \frac{k^2}{Re_m} t}, 
  \label{3D_Taylor_vortex_mag_By} \\
  B_{z 1,2} &= \alpha \left[ 
    \sin(k z + \theta_{1,2}) \cos(k x + \phi_{1,2}) \sin(k y + \psi_{1,2}) 
  \right. \nonumber \\
  & \, \left. 
  - \cos(k y + \theta_{1,2}) \sin(k z + \phi_{1,2}) \sin(k x + \psi_{1,2}) 
  \right] e^{-3 \frac{k^2}{Re_m} t},
  \label{3D_Taylor_vortex_mag_Bz}
\end{align}
\end{subequations}
%------------------------------------------------------------------------------
where the magnetic Reynolds number is defined as $Re_m = UL/\nu_m$. 
Because this three-dimensional magnetic flux density flow is the same 
as the Beltrami flow, 
the magnetic flux density vector $\bm{B}$ and current density vector $\bm{j}$ are parallel, 
and $\bm{B} \times \bm{j} = \bm{B} \times (\nabla \times \bm{B}) = 0$. 
Therefore, the Lorentz force does not act, and work is not done. 
Assuming that the kinematic viscosity and magnetic diffusivity are zero, 
the kinetic and magnetic energies are conserved in a periodic flow; 
the total energy is also conserved. 
As expected, when investigating the energy conservation property, 
we must analyze a three-dimensional problem 
in which energy conversion by the Lorentz force occurs; 
however, creating such a test problem is difficult. 
Although no Lorentz force is generated, 
this decaying vortex model is used as a benchmark test problem 
to verify the validity of the energy conservation property in the computational method.

As an initial condition for the velocity, 
the velocity must be given such that it discretely satisfies 
Eq. (\ref{continuity}). 
A vector potential defined at the midpoint of a cell edge satisfies 
the continuity equation at the cell center. 
Thus, the velocity also satisfies the continuity equation. 
The vector potentials $\Psi_x$, $\Psi_y$, and $\Psi_z$ are defined 
at $(i, j+1/2, k+1/2)$, $(i+1/2, j, k+1/2)$, and $(i+1/2, j+1/2, k)$, respectively. 
The vector potential is given as
%------------------------------------------------------------------------------
\begin{subequations}
\begin{align}
  \Psi_x &= \alpha 
  \sin(k y + \theta_{1,2}) \sin(k z + \phi_{1,2}) \sin(k x + \psi_{1,2}) 
  e^{-3 \frac{k^2}{Re} t}, 
  \label{flow_vector_potential_x} \\
  \Psi_y &= \alpha 
  \sin(k z + \theta_{1,2}) \sin(k x + \phi_{1,2}) \sin(k y + \psi_{1,2}) 
  e^{-3 \frac{k^2}{Re} t}, 
  \label{flow_vector_potential_y} \\
  \Psi_x &= \alpha 
  \sin(k x + \theta_{1,2}) \sin(k y + \phi_{1,2}) \sin(k z + \psi_{1,2}) 
  e^{-3 \frac{k^2}{Re} t}.
  \label{flow_vector_potential_z}
\end{align}
\end{subequations}
%------------------------------------------------------------------------------
The velocities given by Eqs. (\ref{3D_Taylor_vortex_u}) -- (\ref{3D_Taylor_vortex_w}) 
are calculated from the definition of $\bm{u} = \nabla \times \bm{\Psi}$.

Similarly, the magnetic vector potential is given as
%------------------------------------------------------------------------------
\begin{subequations}
\begin{align}
  A_x &= \frac{\alpha}{k} 
  \sin(k y + \theta_{1,2}) \sin(k z + \phi_{1,2}) \sin(k x + \psi_{1,2}) 
  e^{-3 \frac{k^2}{Re_m} t}, 
  \label{magnetic_vector_potential_x} \\
  A_y &= \frac{\alpha}{k} 
  \sin(k z + \theta_{1,2}) \sin(k x + \phi_{1,2}) \sin(k y + \psi_{1,2}) 
  e^{-3 \frac{k^2}{Re_m} t}, 
  \label{magnetic_vector_potential_y} \\
  A_z &= \frac{\alpha}{k} 
  \sin(k x + \theta_{1,2}) \sin(k y + \phi_{1,2}) \sin(k z + \psi_{1,2}) 
  e^{-3 \frac{k^2}{Re_m} t}.
  \label{magnetic_vector_potential_z}
\end{align}
\end{subequations}
%------------------------------------------------------------------------------
From the definition of $\bm{B} = \nabla \times \bm{A}$, 
the magnetic flux densities expressed by Eqs. (\ref{3D_Taylor_vortex_mag_Bx}) 
-- (\ref{3D_Taylor_vortex_mag_Bz}) are obtained. 
As this magnetic vector potential does not satisfy $\nabla \cdot \bm{A} = 0$, 
the electric potential cannot be obtained using the Coulomb gauge. 
Thus, applying the present numerical method, 
which simultaneously relaxes the magnetic vector and electric potentials, is unfeasible. 
From the magnetic vector potential equation (\ref{vector_potential}), 
the electric potential is obtained as follows:
%------------------------------------------------------------------------------
\begin{subequations}
\begin{align}
  \psi &= - \frac{\alpha}{Re_m} \left[ 
    \sin(k x + \theta_{1,2}) \sin(k y + \phi_{1,2}) \cos(k z + \psi_{1,2}) 
  \right. \nonumber \\
  & \quad \left. 
  + \sin(k y + \theta_{1,2}) \sin(k z + \phi_{1,2}) \cos(k x + \psi_{1,2}) 
  \right. \nonumber \\
  & \quad \left. 
  + \sin(k z + \theta_{1,2}) \sin(k x + \phi_{1,2}) \cos(k y + \psi_{1,2}) \right] 
  e^{-3 \frac{k^2}{Re_m} t}.
\end{align}
\end{subequations}
%------------------------------------------------------------------------------

To satisfy $\partial \nabla \cdot \bm{A}/\partial t = 0$, 
the magnetic vector potential is given as follows:
%------------------------------------------------------------------------------
\begin{subequations}
\begin{align}
  \tilde{A_x} &= \frac{\alpha}{3 k} \left[ 
      \cos(k x + \theta_{1,2}) \sin(k y + \phi_{1,2}) \cos(k z + \psi_{1,2}) 
  \right. \nonumber \\
  & \quad \left. 
  +   \sin(k z + \theta_{1,2}) \cos(k x + \phi_{1,2}) \cos(k y + \psi_{1,2}) 
  \right. \nonumber \\
  & \quad \left. 
  + 2 \sin(k y + \theta_{1,2}) \sin(k z + \phi_{1,2}) \sin(k x + \psi_{1,2}) 
  \right] e^{-3 \frac{k^2}{Re_m} t}, 
  \label{magnetic_vector_potential2_x} \\
  \tilde{A_y} &= \frac{\alpha}{3 k} \left[ 
      \cos(k y + \theta_{1,2}) \sin(k z + \phi_{1,2}) \cos(k x + \psi_{1,2}) 
  \right. \nonumber \\
  & \quad \left. 
  +   \sin(k x + \theta_{1,2}) \cos(k y + \phi_{1,2}) \cos(k z + \psi_{1,2}) 
  \right. \nonumber \\
  & \quad \left. 
  + 2 \sin(k z + \theta_{1,2}) \sin(k x + \phi_{1,2}) \sin(k y + \psi_{1,2}) 
  \right] e^{-3 \frac{k^2}{Re_m} t}, 
  \label{magnetic_vector_potential2_y} \\
  \tilde{A_z} &= \frac{\alpha}{3 k} \left[ 
      \cos(k z + \theta_{1,2}) \sin(k x + \phi_{1,2}) \cos(k y + \psi_{1,2}) 
  \right. \nonumber \\
  & \quad \left. 
  +   \sin(k y + \theta_{1,2}) \cos(k z + \phi_{1,2}) \cos(k x + \psi_{1,2}) 
  \right. \nonumber \\
  & \quad \left. 
  + 2 \sin(k x + \theta_{1,2}) \sin(k y + \phi_{1,2}) \sin(k z + \psi_{1,2}) 
  \right] e^{-3 \frac{k^2}{Re_m} t}.
  \label{magnetic_vector_potential2_z}
\end{align}
\end{subequations}
%------------------------------------------------------------------------------
From the definition of $\bm{B} = \nabla \times \tilde{\bm{A}}$, 
the magnetic flux densities expressed by Eqs. (\ref{3D_Taylor_vortex_mag_Bx}) 
-- (\ref{3D_Taylor_vortex_mag_Bz}) are obtained. 
The electric potential gradient in the magnetic vector potential equation (\ref{vector_potential}) 
is $\nabla \psi = 0$; hence, the electric potential yields an arbitrary constant. 
This magnetic vector potential satisfies $\nabla \cdot \tilde{\bm{A}} = 0$ analytically, 
but not discretely. 
In the following, the magnetic vector potential values expressed 
in Eqs. (\ref{magnetic_vector_potential2_x}) -- (\ref{magnetic_vector_potential2_z}) are used. 
The superscript $\tilde{\, \cdot \,}$ is omitted below. 
The following vector potential $\bm{C}$ is defined to obtain the magnetic vector potential, 
ensuring that $\nabla \cdot \bm{A} = 0$ is discretely satisfied:
%------------------------------------------------------------------------------
\begin{subequations}
\begin{align}
  C_x &= \frac{\alpha}{3 k^2} \left[ 
      \sin(k x + \theta_{1,2}) \cos(k y + \phi_{1,2}) \sin(k z + \psi_{1,2}) 
  \right. \nonumber \\
  & \quad \left. 
  -   \cos(k z + \theta_{1,2}) \sin(k x + \phi_{1,2}) \sin(k y + \psi_{1,2}) 
  \right] e^{-3 \frac{k^2}{Re_m} t}, 
  \label{magnetic_vector_potential_Cx} \\
  C_y &= \frac{\alpha}{3 k^2} \left[ 
      \sin(k y + \theta_{1,2}) \cos(k z + \phi_{1,2}) \sin(k x + \psi_{1,2}) 
  \right. \nonumber \\
  & \quad \left. 
  -   \cos(k x + \theta_{1,2}) \sin(k y + \phi_{1,2}) \sin(k z + \psi_{1,2}) 
  \right] e^{-3 \frac{k^2}{Re_m} t}, 
  \label{magnetic_vector_potential_Cy} \\
  C_z &= \frac{\alpha}{3 k^2} \left[ 
      \sin(k z + \theta_{1,2}) \cos(k x + \phi_{1,2}) \sin(k y + \psi_{1,2}) 
  \right. \nonumber \\
  & \quad \left. 
  -   \cos(k y + \theta_{1,2}) \sin(k z + \phi_{1,2}) \sin(k x + \psi_{1,2}) 
  \right] e^{-3 \frac{k^2}{Re_m} t}.
  \label{magnetic_vector_potential_Cz}
\end{align}
\end{subequations}
%------------------------------------------------------------------------------
The vector potentials $C_x$, $C_y$, and $C_z$ are defined 
at $(i, j+1/2, k+1/2)$, $(i+1/2, j, k+1/2)$, and $(i+1/2, j+1/2, k)$, respectively. 
Using the definition of $\bm{A} = \nabla \times \bm{C}$, 
if the magnetic vector potential $\bm{A}$ is discretely calculated 
from the vector potential $\bm{C}$, 
the constraint $\nabla \cdot \bm{A} = 0$ is discretely satisfied at the cell center.

The computational region is a cube with one side $L$. 
As initial conditions, 
the vector potential in Eqs. (\ref{flow_vector_potential_x}) 
-- (\ref{flow_vector_potential_z}), 
pressure in Eq. (\ref{3D_Taylor_vortex_p}), 
and magnetic vector potential in Eqs. (\ref{magnetic_vector_potential_Cx}) 
-- (\ref{magnetic_vector_potential_Cz}) are given. 
Periodic boundary conditions are set as boundary conditions. 
A uniform grid of $N \times N \times N$ with $N = 41$ is used for the calculation. 
For error evaluation and grid dependency verification, 
grids with dimensions of $N = 11$, $21$, and $81$ are used. 
The reference values used in this calculation are 
$l_\mathrm{ref} = L$, $u_\mathrm{ref} = U$, $t_\mathrm{ref} = L/U$, 
$B_\mathrm{ref} = B$, $A_\mathrm{ref} = B L$, and $\psi_\mathrm{ref} = U L B$. 
First, to confirm the energy conservation properties of the present numerical method, 
an ideal inviscid MHD flow for $Re = Re_m = \infty$ is analyzed. 
The Courant number is defined as $\mbox{CFL} = \Delta t U/\Delta x$ 
using the reference velocity $U$ and grid width $\Delta x$. 
The calculation is performed up to the time $t/(L/U) = 10$ under the condition 
that the Courant number is CFL = 0.4. 
Subsequently, regarding the calculation conditions for viscous analysis, 
the Reynolds numbers are set to $Re = 10^2$ and $10^4$, 
referring to the existing research on two-dimensional Taylor decaying vortex 
\citep{Liu&Wang_2001, Yanaoka_2023}. 
The magnetic Reynolds numbers are $Re_m = 1$ and $50$, 
and the Alfv\'{e}n number is $Al = 1$. 
At $Re = 10^2$, the vortex decays quickly; therefore, the time step must be reduced. 
Therefore, the Courant number is set to CFL = 0.02. 
For $Re = 10^4$, the Courant number is CFL = 0.1.

Further, the method for extracting low- or high-pressure regions is explained herein. 
If the pressure distribution is concentric around a vortex tube, 
the vortex tube can be identified by displaying the isosurface of the pressure. 
However, when the vortex tube and shear layer coexist, 
the pressure changes owing to the two structures; 
thus, extracting only the vortex tube is not possible. 
As the radius of a thin vortex tube is small, 
the thin vortex tube can be identified by visualizing a vortex tube 
with a large curvature. 
Therefore, by calculating the curvature of an equipressure surface 
and displaying the isosurface with a large curvature, 
a vortex tube with a small radius of curvature can be identified. 
Now, the case where pressure is high at the center of a concentric circle 
and low at the periphery is considered. 
The curvature of the pressure isosurface can be defined as follows:
%------------------------------------------------------------------------------
\begin{equation}
  \kappa_p = - \nabla \cdot \hat{\bm{n}}, \quad 
  \hat{\bm{n}} = \frac{\bm{n}}{|\bm{n}|} = \frac{\nabla p}{|\nabla p|},
\end{equation}
%------------------------------------------------------------------------------
where $\hat{\bm{n}}$ is a unit normal vector on the isosurface. 
Therefore, a high-pressure region can be visualized 
by displaying the high value of $\kappa_p > 0$. 
Conversely, a low-pressure region can be extracted 
by showing the low value of $\kappa_p < 0$. 
$1/|\kappa_p|$ is the radius of curvature of the vortex tube.

%------------------------------------------------------------------------------
% Figure 20
%------------------------------------------------------------------------------
\begin{figure}[!t]
\begin{minipage}{0.325\linewidth}
\begin{center}
\includegraphics[trim=0mm 0mm 0mm 0mm, clip, width=50mm]{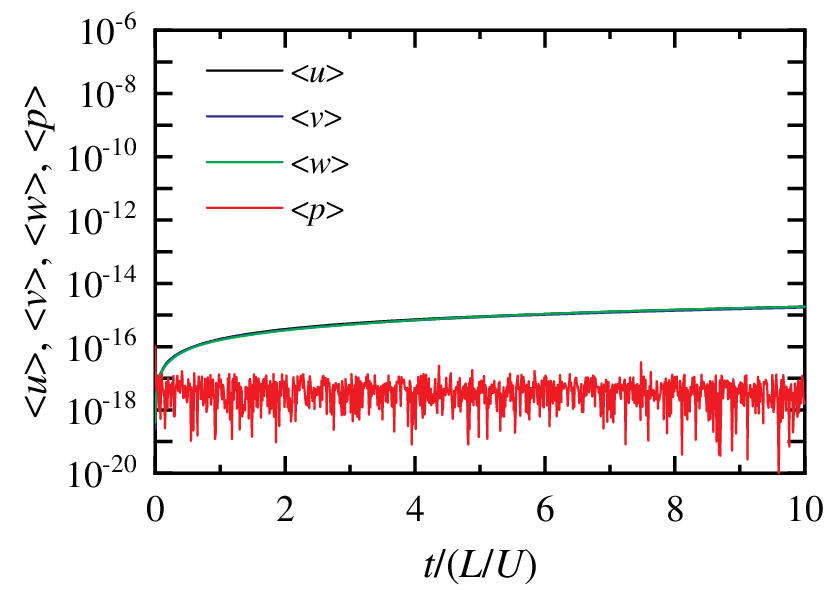} \\
{\small (a) $\bm{u}$, $p$}
\end{center}
\end{minipage}
\begin{minipage}{0.325\linewidth}
\begin{center}
\includegraphics[trim=0mm 0mm 0mm 0mm, clip, width=50mm]{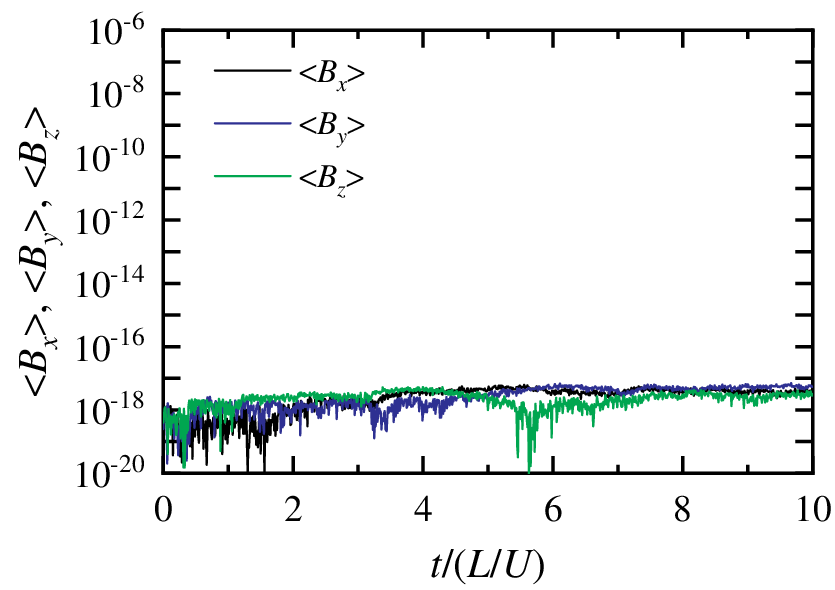} \\
{\small (b) $\bm{B}$}
\end{center}
\end{minipage}
\begin{minipage}{0.325\linewidth}
\begin{center}
\includegraphics[trim=0mm 0mm 0mm 0mm, clip, width=50mm]{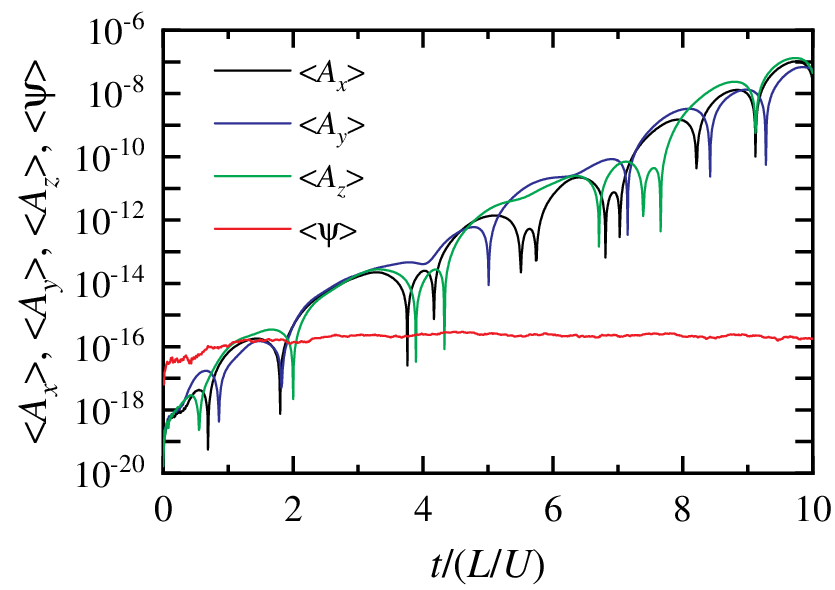} \\
{\small (c) $\bm{A}$}
\end{center}
\end{minipage}
\caption{Total amounts of velocity ($\bm{u}$), pressure ($p$), 
magnetic flux density ($\bm{B}$), 
and magnetic vector potential ($\bm{A}$): $Re = Re_m = \infty$, $N = 41$.}
\label{decay3d_invis_u_B_A}
\end{figure}
%------------------------------------------------------------------------------

%------------------------------------------------------------------------------
% Figure 21
%------------------------------------------------------------------------------
\begin{figure}[!t]
\begin{minipage}{0.325\linewidth}
\begin{center}
\includegraphics[trim=0mm 0mm 0mm 0mm, clip, width=50mm]{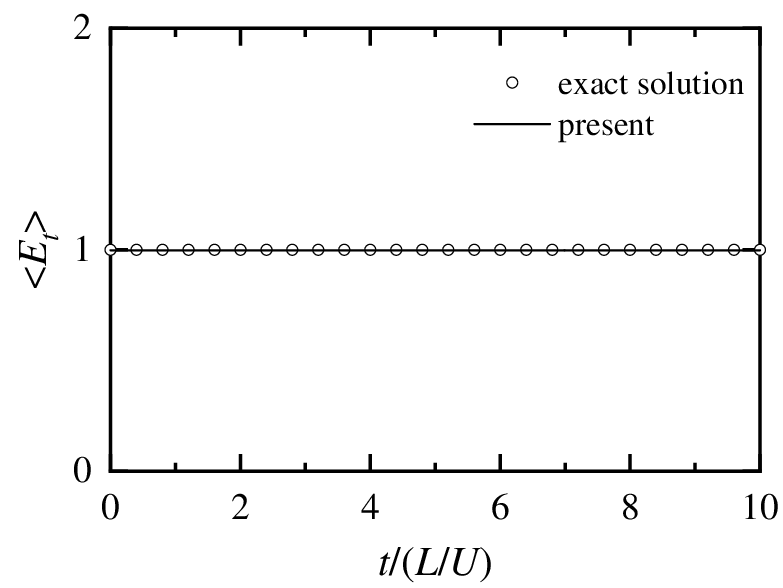} \\
{\small (a) $E_t$}
\end{center}
\end{minipage}
\begin{minipage}{0.325\linewidth}
\begin{center}
\includegraphics[trim=0mm 0mm 0mm 0mm, clip, width=50mm]{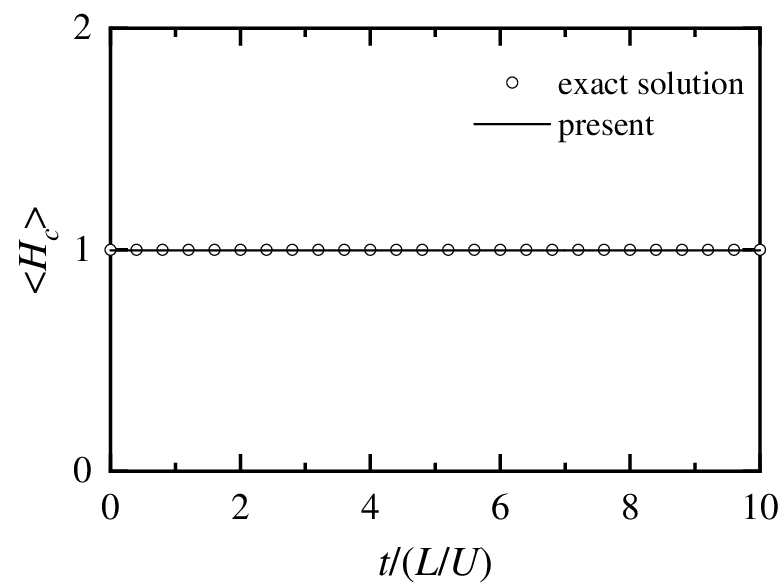} \\
{\small (b) $H_c$}
\end{center}
\end{minipage}
\begin{minipage}{0.325\linewidth}
\begin{center}
\includegraphics[trim=0mm 0mm 0mm 0mm, clip, width=50mm]{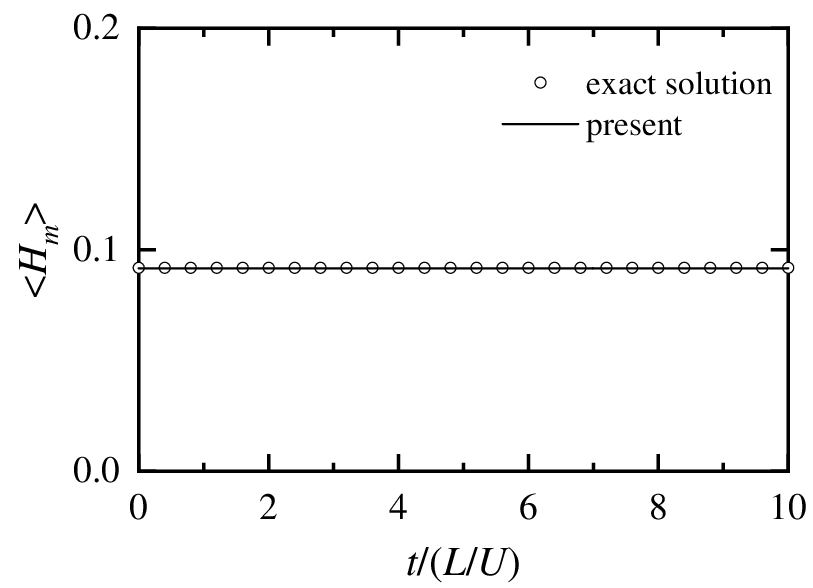} \\
{\small (c) $H_m$}
\end{center}
\end{minipage}
\caption{Total amounts of total energy ($E_t$), cross-helicity ($H_c$), 
and magnetic helicity ($H_m$):  $Re = Re_m = \infty$, $N = 41$.}
\label{decay3d_invis_Et_Hc_Hm}
\end{figure}
%------------------------------------------------------------------------------

%------------------------------------------------------------------------------
% Figure 22
%------------------------------------------------------------------------------
\begin{figure}[!t]
\begin{minipage}{0.5\linewidth}
\begin{center}
\includegraphics[trim=0mm 0mm 0mm 0mm, clip, width=70mm]{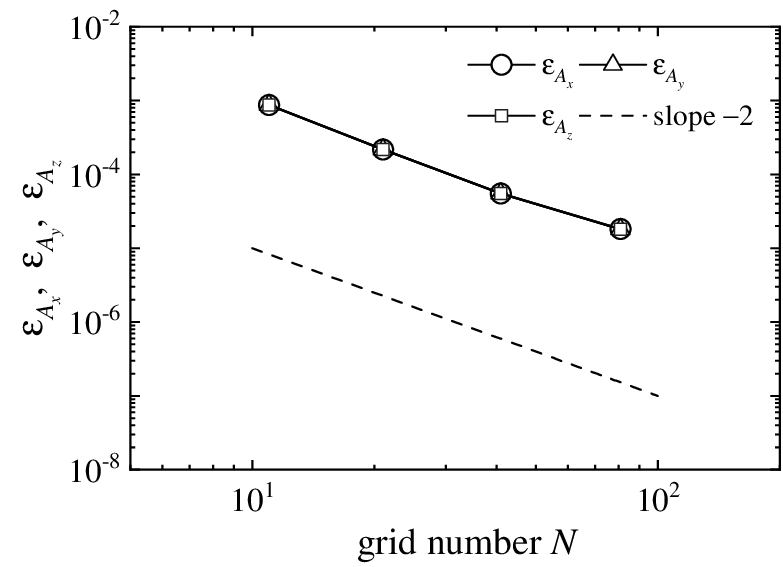} \\
{\small (a) $A_x$, $A_y$, $A_z$}
\end{center}
\end{minipage}
\begin{minipage}{0.5\linewidth}
\begin{center}
\includegraphics[trim=0mm 0mm 0mm 0mm, clip, width=70mm]{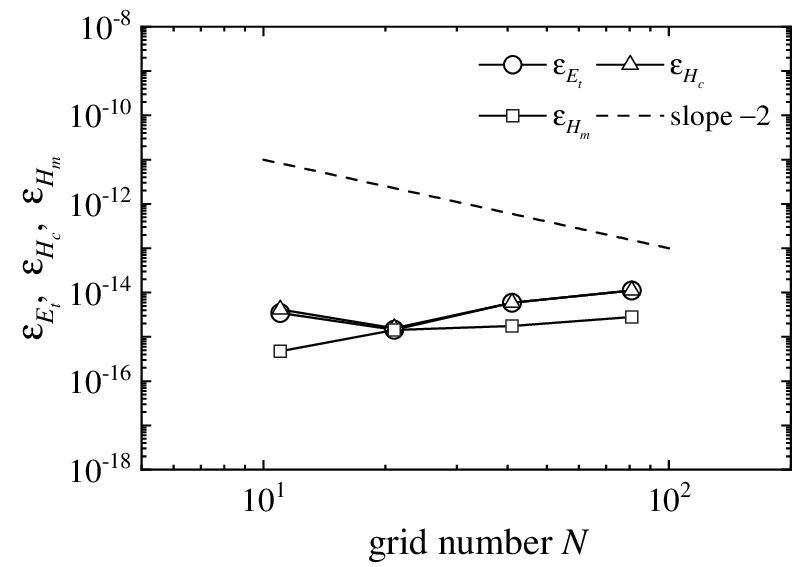} \\
{\small (b) $E_t$, $H_c$, $H_m$}
\end{center}
\end{minipage}
\caption{Errors of magnetic vector potential ($A_x$, $A_y$, $A_z$), 
total energy ($E_t$), cross-helicity ($H_c$), 
and  magnetic helicity ($H_m$):  $Re = Re_m = \infty$, $t/(L/U) = 10$.}
\label{decay3d_invis_error}
\end{figure}
%------------------------------------------------------------------------------

Figure \ref{decay3d_invis_u_B_A} shows the total amounts of 
velocity $\langle \bm{u} \rangle$, 
pressure $\langle p \rangle$, 
magnetic flux density $\langle \bm{B} \rangle$, 
magnetic vector potential $\langle \bm{A} \rangle$, 
and electric potential $\langle \psi \rangle$ for inviscid analysis. 
Owing to the periodic flow and magnetic fields, each total amount is preserved. 
Each total amount for velocity, pressure, and magnetic flux density varies 
only at the level of rounding error, 
demonstrating the excellent conservation properties in the proposed numerical method. 
Conversely, the total amount of magnetic vector potential increases with time. 
The volume integral of the exact solution of the magnetic vector 
and electric potentials is zero. 
The magnetic vector potential equation is not conservative; 
hence, the time variation of the total amount is affected 
by the discretization of the convective term. 
The convective term is zero analytically but not zero discretely. 
As the limitation of this computational method has been clarified, 
the discretization of the magnetic vector potential equation should be modified 
in the future.

%------------------------------------------------------------------------------
% Figure 23
%------------------------------------------------------------------------------
\begin{figure}[!t]
\begin{minipage}{0.325\linewidth}
\begin{center}
\includegraphics[trim=0mm 0mm 0mm 0mm, clip, width=50mm]{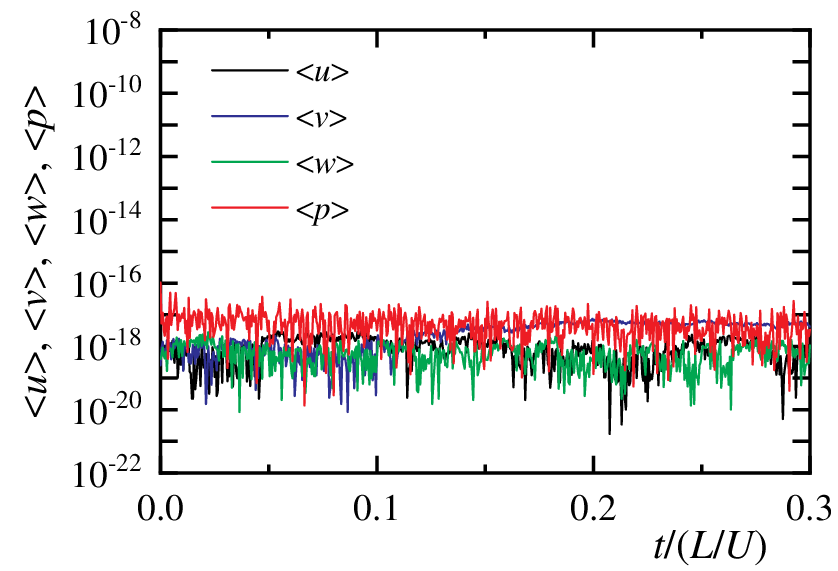} \\
{\small (a) $\bm{u}$, $p$}
\end{center}
\end{minipage}
\begin{minipage}{0.325\linewidth}
\begin{center}
\includegraphics[trim=0mm 0mm 0mm 0mm, clip, width=50mm]{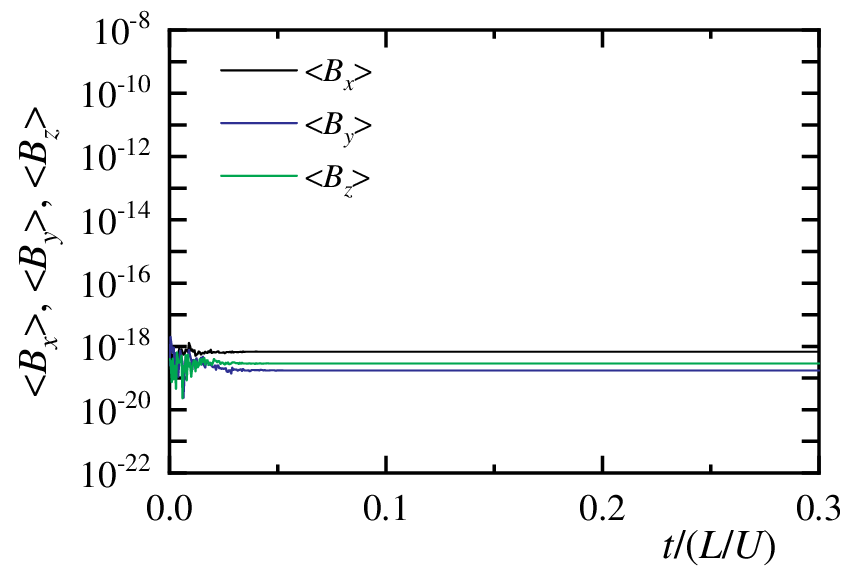} \\
{\small (b) $\bm{B}$}
\end{center}
\end{minipage}
\begin{minipage}{0.325\linewidth}
\begin{center}
\includegraphics[trim=0mm 0mm 0mm 0mm, clip, width=50mm]{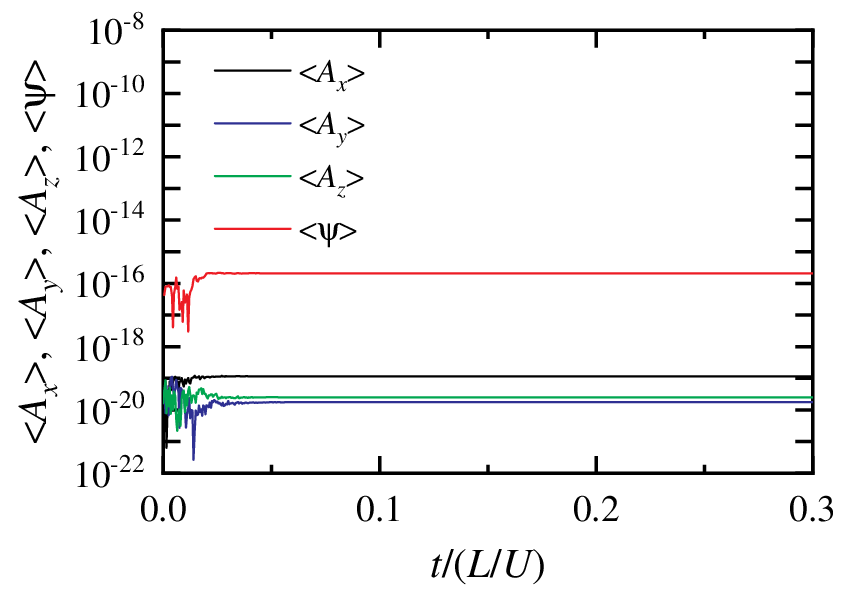} \\
{\small (c) $\bm{A}$}
\end{center}
\end{minipage}
\caption{Total amounts of velocity ($\bm{u}$), pressure ($p$), 
magnetic flux density ($\bm{B}$), and magnetic vector potential ($\bm{A}$): 
$Re = 10^2$, $Re_m = 1$, $N = 41$.}
\label{decay3d_Re100_u_B_A}
\end{figure}
%------------------------------------------------------------------------------

The total amounts, 
of total energy $\langle E_t \rangle$, 
cross-helicity $\langle H_c \rangle$, 
and magnetic helicity $\langle H_m \rangle$ 
are shown in Fig. \ref{decay3d_invis_Et_Hc_Hm}. 
This approximate solution supports the exact solution, 
showing excellent conservation properties. 

Figure \ref{decay3d_invis_error} (a) shows the maximum errors, 
$\varepsilon_{A_x}$, $\varepsilon_{A_y}$, and $\varepsilon_{A_z}$, 
of the magnetic vector potential at time $t/(L/U) = 10$. 
The error decreases with a slope of $-2$, 
indicating the second-order convergence. 
Figure \ref{decay3d_invis_error} (b) shows the relative errors, 
$\varepsilon_{E_t} = |(\langle E_t \rangle-\langle E_t \rangle_0)/\langle E_t \rangle_0|$, 
$\varepsilon_{H_c} = |(\langle H_c \rangle-\langle H_c \rangle_0)/\langle H_c \rangle_0|$, 
and $\varepsilon_{H_m} = |(\langle H_m \rangle-\langle H_m \rangle_0)/\langle H_m \rangle_0|$, 
of the total amounts of total energy, cross-helicity, and magnetic helicity 
The subscript $0$ represents the initial value. 
Regardless of the number of grid points, 
the error is at the level of rounding errors. 
Evidently, the energy conservation properties of this computational method are excellent. 
The maximum divergence errors of velocity, magnetic flux density, 
and magnetic vector potential in inviscid analysis are $2.78\times 10^{-14}$, 
$8.93\times 10^{-13}$, and $5.08\times 10^{-16}$ at time $t/(L/U) = 10$, respectively.

%------------------------------------------------------------------------------
% Figure 24
%------------------------------------------------------------------------------
\begin{figure}[!t]
\begin{center}
\includegraphics[trim=0mm 0mm 0mm 0mm, clip, width=70mm]{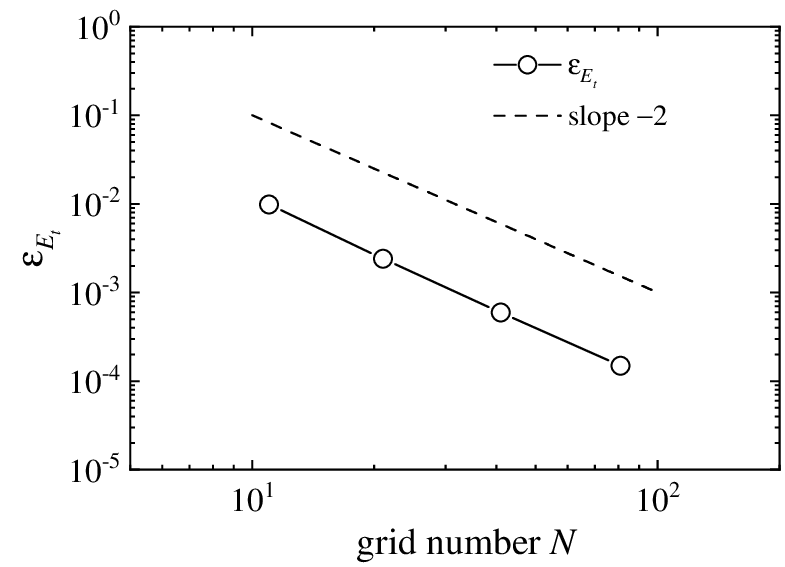} \\
\caption{Error of total energy: $Re = 10^2$, $Re_m = 1$, $t/(L/U) = 0.3$.}
\label{decay3d_error_Et_Re100_Rem1}
\end{center}
\end{figure}
%------------------------------------------------------------------------------

Figure \ref{decay3d_Re100_u_B_A} shows time variations of the total amounts of 
velocity, magnetic flux density, and magnetic vector potential 
for $Re = 10^2$ and $Re_m = 1$. 
The flow and magnetic fields are periodic; 
hence, each total amount is conserved and analytically zero. 
The total amount computed is at the level of rounding errors. 
The author confirmed that the total amount is within the level of rounding errors 
even at $Re = 10^4$ and $Re_m = 50$.

For $Re = 10^2$, Fig. \ref{decay3d_error_Et_Re100_Rem1} shows the relative error, 
$\varepsilon_{E_t}$, from the exact solution of the total energy 
at time $t/(L/U) = 0.3$ after the average kinetic energy is halved. 
The error decreases with a slope of $-2$, 
indicating that the numerical method has second-order accuracy.

%------------------------------------------------------------------------------
% Figure 25
%------------------------------------------------------------------------------
\begin{figure}[!t]
\begin{minipage}{0.48\linewidth}
\begin{center}
\includegraphics[trim=0mm 0mm 0mm 0mm, clip, width=70mm]{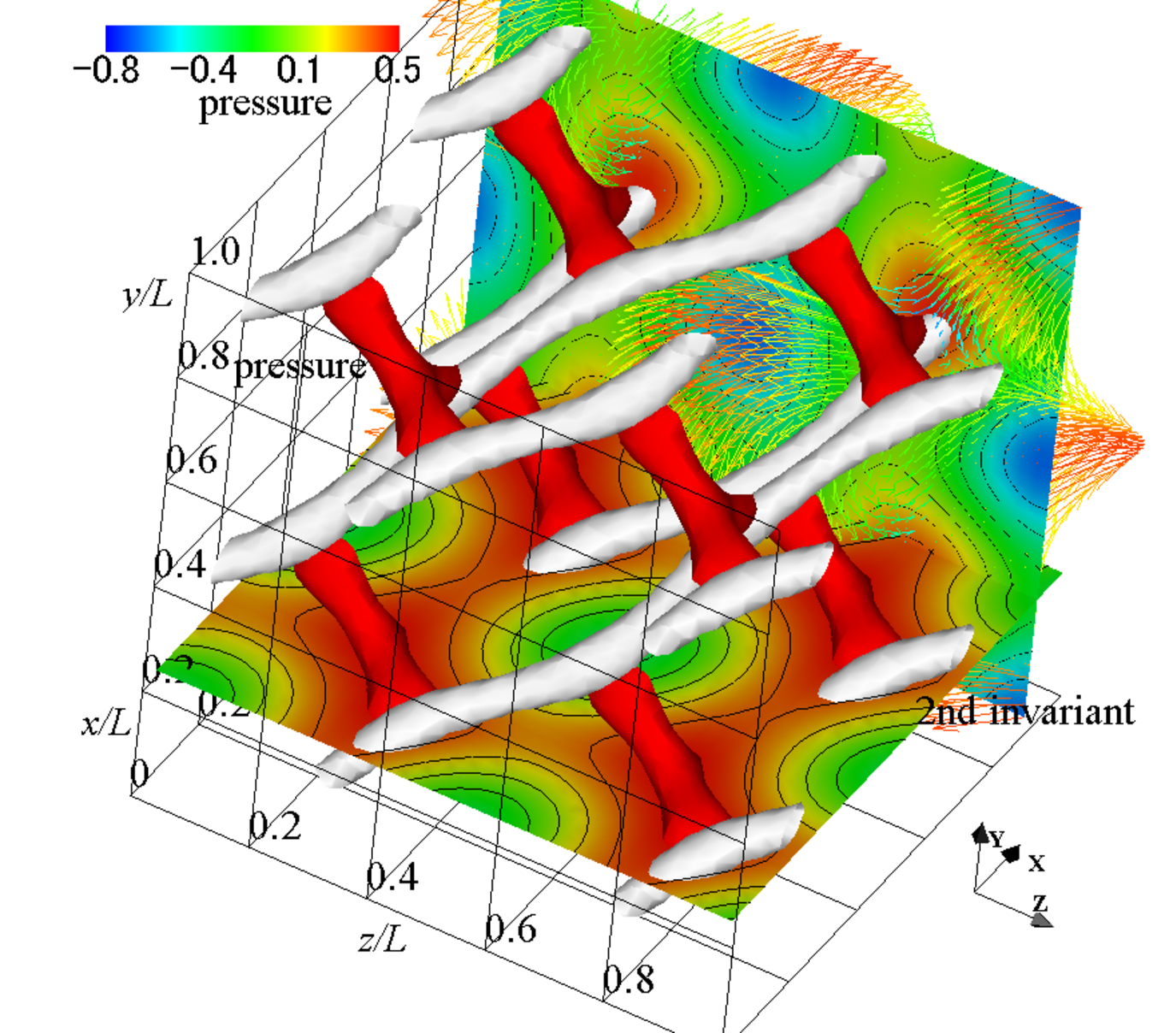} \\
{\small (a) Velocity vectors, pressure contour, isosurface of pressure, 
and isosurface of 2nd invariant of velocity gradient tensor}
\end{center}
\end{minipage}
\hspace{0.02\linewidth}
\begin{minipage}{0.48\linewidth}
\begin{center}
\includegraphics[trim=0mm 0mm 0mm 0mm, clip, width=70mm]{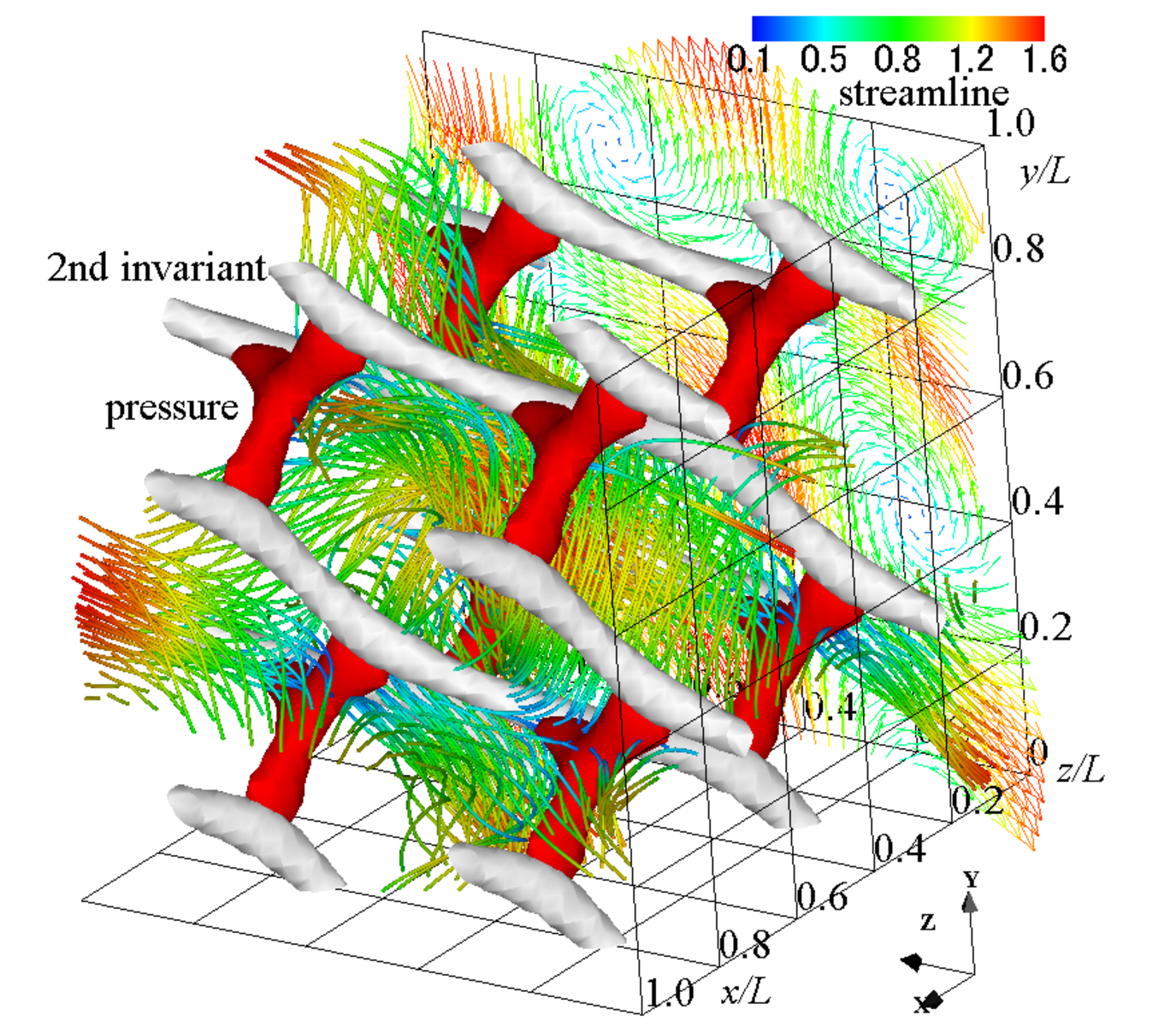} \\
{\small (b) Velocity vectors, streamlines, isosurface of pressure, 
and isosurface of 2nd invariant of velocity gradient tensor}
\vspace*{1.5\baselineskip}
\end{center}
\end{minipage}
\vspace*{0.5\baselineskip}
\begin{minipage}{0.48\linewidth}
\begin{center}
\includegraphics[trim=0mm 0mm 0mm 0mm, clip, width=70mm]{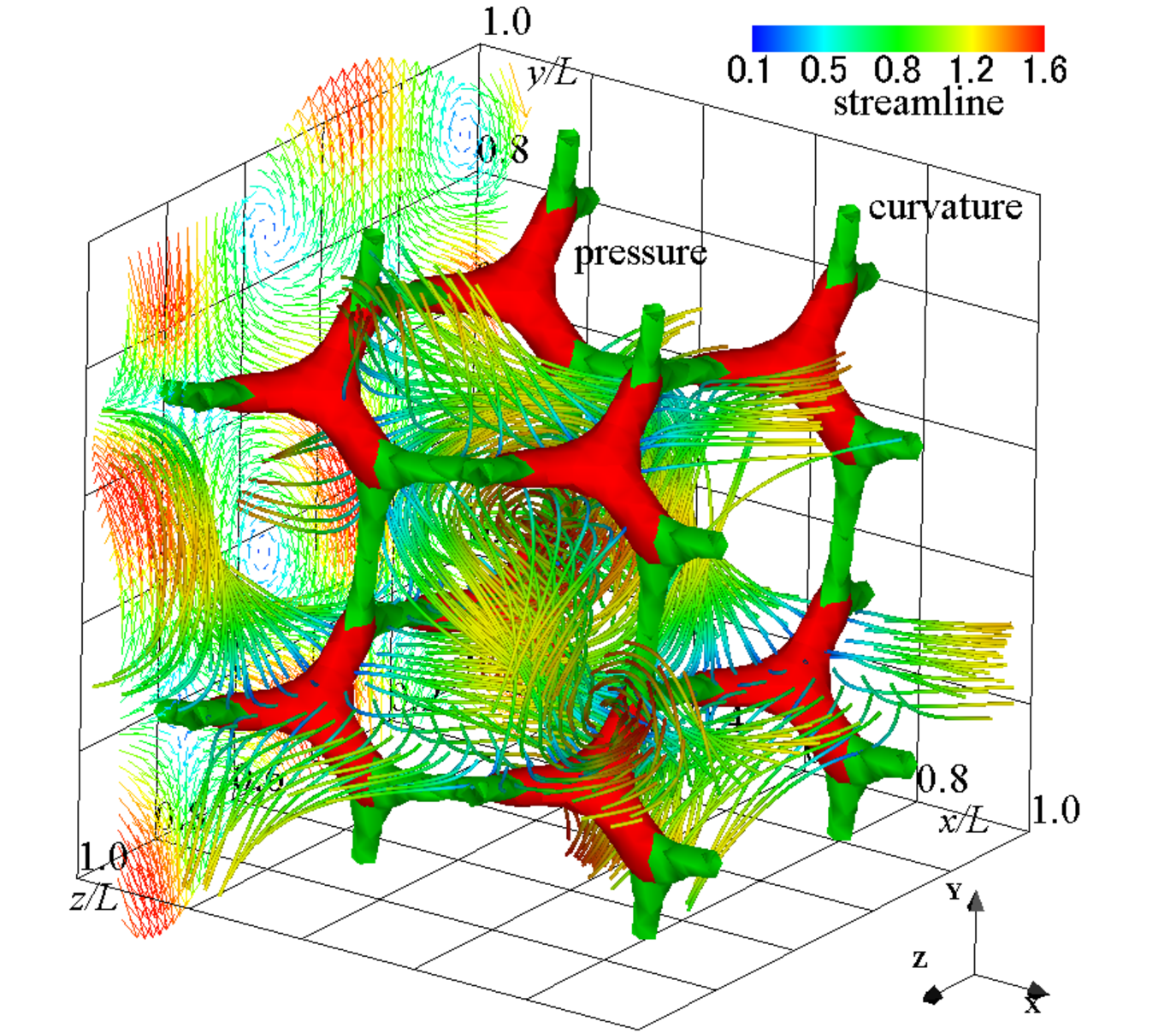} \\
{\small (c) Velocity vectors, streamlines, isosurface of pressure, 
and isosurface of the curvature of equipressure surface}
\end{center}
\end{minipage}
\begin{minipage}{0.48\linewidth}
\begin{center}
\includegraphics[trim=0mm 0mm 0mm 0mm, clip, width=70mm]{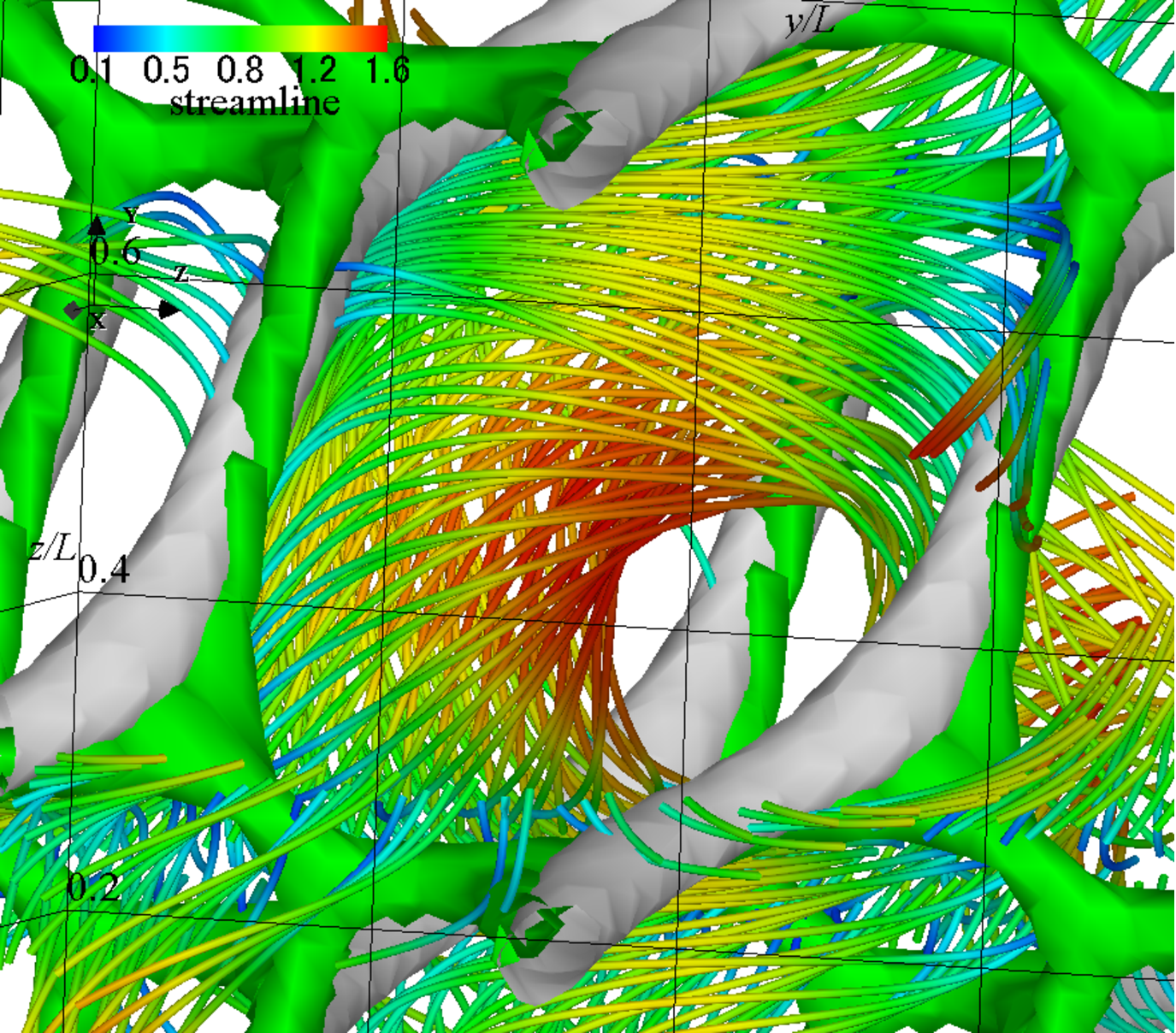} \\
{\small (d) Streamlines, isosurface of 2nd invariant of velocity gradient tensor, 
and isosurface of the curvature of equipressure surface}
\end{center}
\end{minipage}
%
%\vspace*{-0.5\baselineskip}
\caption{Velocity vectors, streamlines, pressure contour, 
and various isosurfaces for analytic solution: 
$Re = 10^4$, $Re_m = 50$, $t/(L/U) = 0.3$; 
The red, silver, and green isosurfaces show the pressure, 
2nd invariant of the velocity gradient tensor, 
and curvature of equipressure surface, respectively.}
\label{decay3d_flow_Re10000}
\end{figure}
%------------------------------------------------------------------------------

Subsequently, the tendency of decaying vortex for different calculation conditions 
was investigated. 
There are few research examples of this three-dimensional decaying vortex \citep{Antuono_2020}; 
hence, the structure of the flow and magnetic fields was examined in detail. 
Figure \ref{decay3d_flow_Re10000} shows the flow field for $Re = 10^4$ at time $t/(L/U) = 0.3$. 
The velocity vectors, streamlines, and the contour 
and isosurface of pressure are shown. 
The second invariant of the velocity gradient tensor and the curvature 
of an equipressure surface are also displayed in an isosurface form. 
Owing to the high Reynolds number, the vortex does not decay over time, 
and the distribution shown in Fig.\ref{decay3d_flow_Re10000} remains similar to the initial value distribution. 
The red isosurface of pressure shows the distribution of dimensionless pressure $p = 0.47$; 
the pressure field near a stagnation point is observed. 
The second invariant $Q$ of the velocity gradient tensor is a quantity 
that expresses the magnitude relationship between the strain rate 
and vorticity tensors. 
Structures with $Q < 0$ represent regions of high shear rates, 
where the viscous dissipation rate of kinetic energy is high. 
The second invariant of the velocity gradient tensor is $Q = -52$ 
and represents a tubular high-shear region where the strain rate tensor increases. 
Along the direction of the vector $(1, 1, 1)$, the tubular structure exists 
some distance away from the stagnation point. 
As observed in \citep{Antuono_2020}, 
the high-pressure region, which includes the stagnation point, has a Y-shaped structure. 
To extract the structure of a high-pressure area near a stagnation point, 
the author calculated the curvature of the isosurface of the pressure. 
The green isosurface represents its curvature, 
and the magnitude of the curvature is $\kappa_p = 48$. 
The high-pressure regions, which indicate the low-velocity areas 
with stagnation points, are connected in a mesh pattern, 
and a distorted cube structure appears. 
The isosurface of curvature envelops that of pressure, 
and the cube structure represents the structure of the pressure field. 
The pressure is low at the center of this cube structure; 
the streamline indicates a swirling flow occurs around the low-pressure region. 
The tubular high-shear structure passes through the high-pressure region. 
The pressure at the center of the tubular high-shear structure is higher than 
the central pressure of the decaying vortex. 
No clear rotational flow occurs around the tubular high-shear structure.

%------------------------------------------------------------------------------
% Figure 26
%------------------------------------------------------------------------------
\begin{figure}[!t]
\begin{minipage}{0.48\linewidth}
\begin{center}
\includegraphics[trim=0mm 0mm 0mm 0mm, clip, width=65mm]{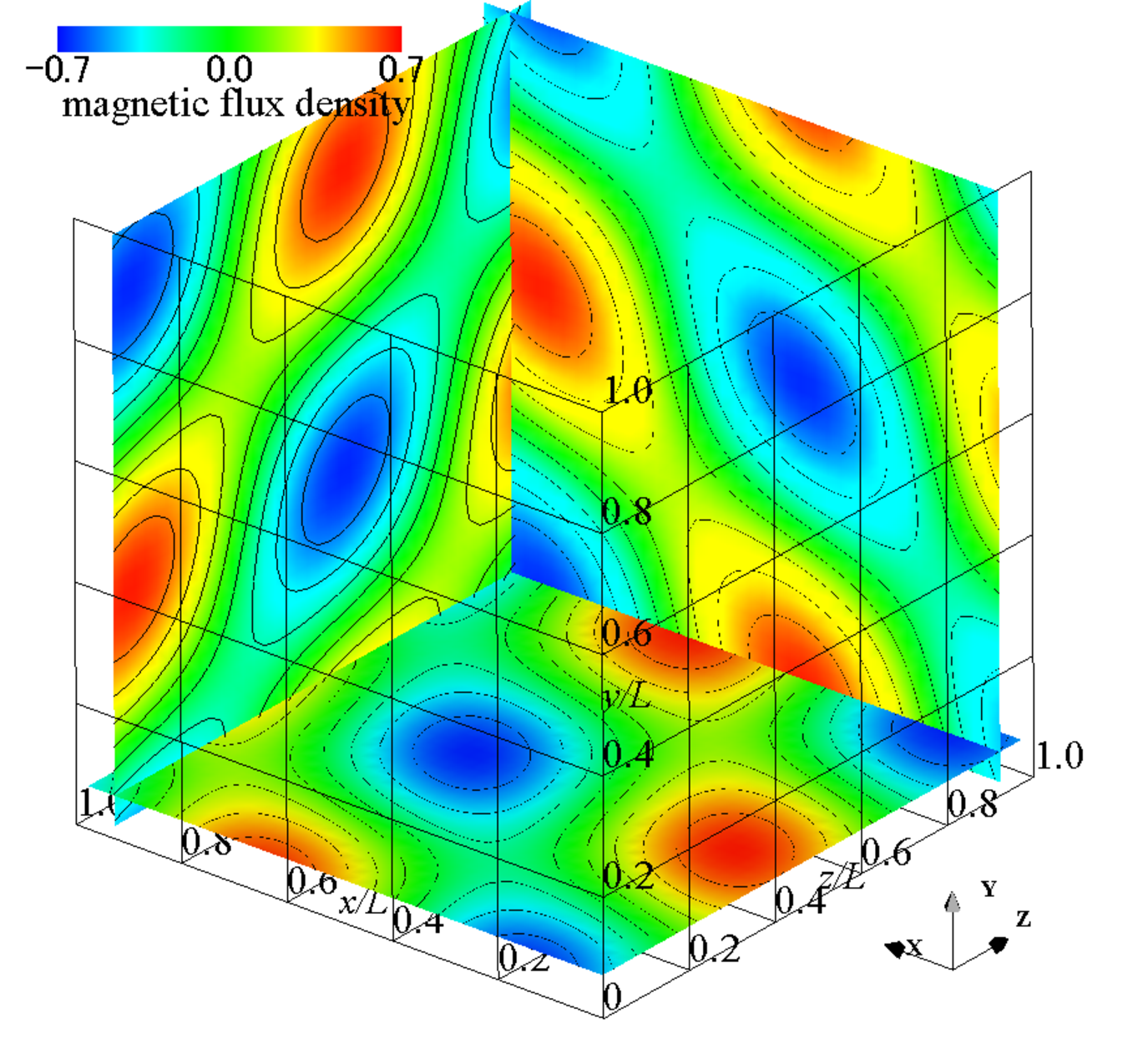} \\
{\small (a) Magnetic flux densities in $x$-, $y$-, and $z$-directions}
\end{center}
\end{minipage}
\hspace{0.02\linewidth}
\begin{minipage}{0.48\linewidth}
\begin{center}
\includegraphics[trim=0mm 0mm 0mm 0mm, clip, width=65mm]{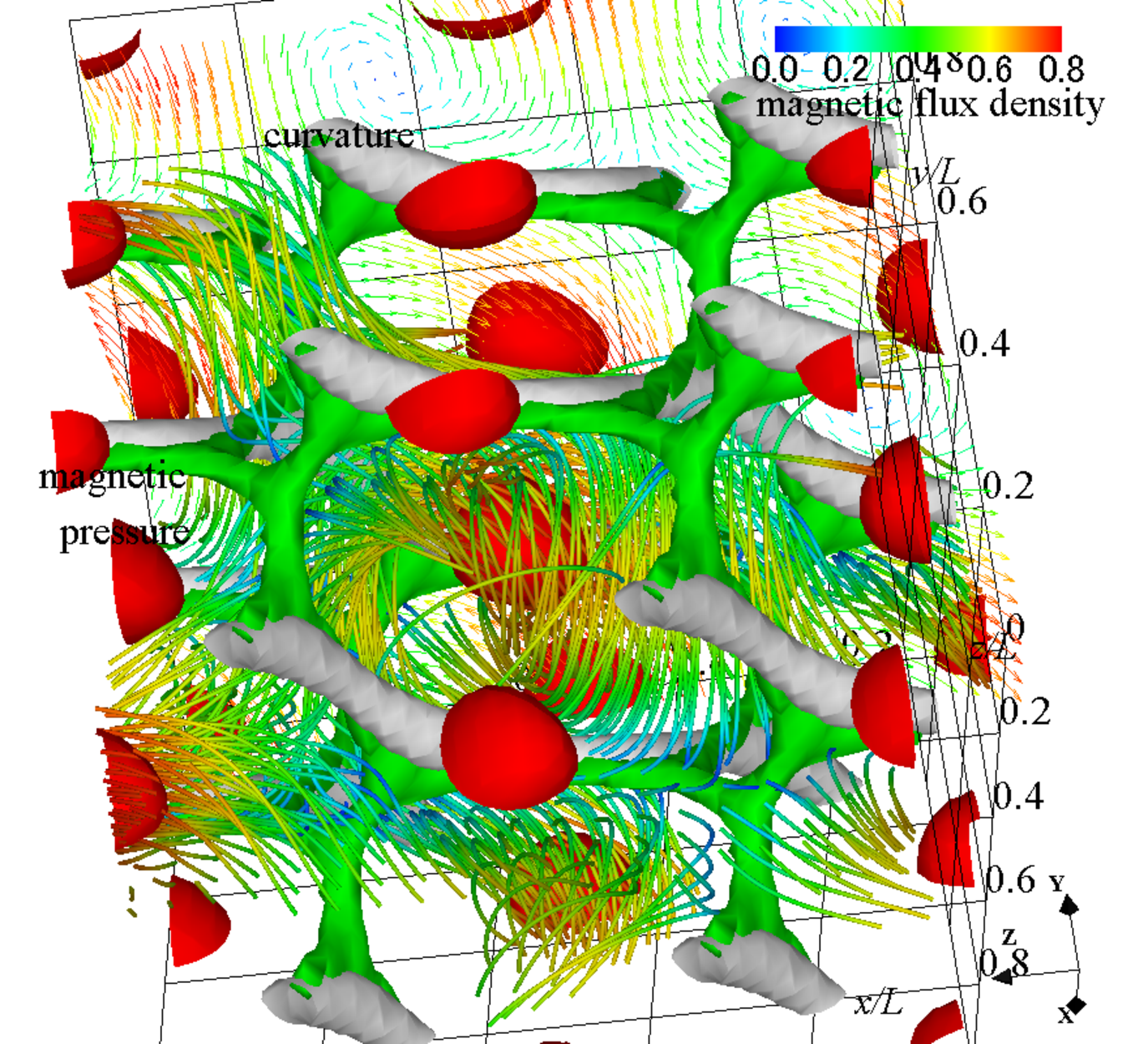} \\
{\small (b) Magnetic flux density lines, 
and isosurfaces of magnetic pressure, 2nd invariant of velocity gradient tensor, 
and curvature of magnetic pressure isosurface}
\end{center}
\end{minipage}
\begin{center}
\includegraphics[trim=0mm 0mm 0mm 0mm, clip, width=65mm]{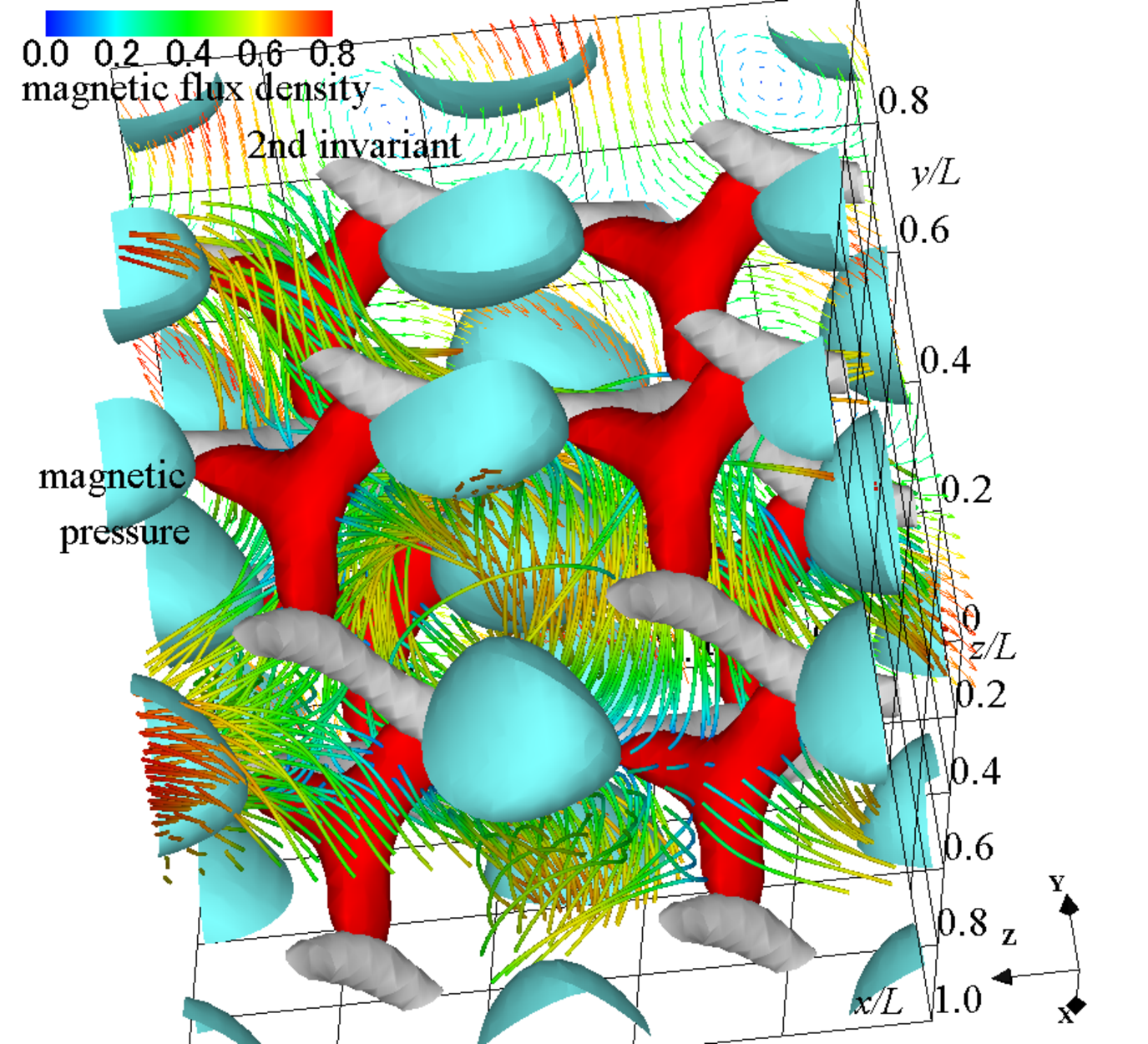} \\
{\small (c) Magnetic flux density lines, 
and isosurfaces of magnetic pressure, current density magnitude, 
and 2nd invariant of velocity gradient tensor}
\end{center}
%
%\vspace*{-1.2\baselineskip}
\caption{Contours of magnetic flux densities, magnetic flux density lines, 
and isosurfaces of magnetic pressure, current density magnitude, 
2nd invariant of velocity gradient tensor, and curvature of magnetic pressure isosurface: 
$Re = 10^4$, $Re_m = 50$, $t/(L/U) = 0.3$; 
The $y$--$z$, $z$--$x$, and $x$--$y$ planes show the contours of 
magnetic flux densities in the $x$-, $y$-, and $z$-directions, respectively. 
The red, blue, silver, and green isosurfaces show magnetic pressure, 
current density magnitude, 2nd invariant of velocity gradient tensor, 
and curvature of magnetic pressure isosurface, respectively.}
\label{decay3d_mag_flx_pres_Re10000_t03}
\end{figure}
%------------------------------------------------------------------------------

Figure \ref{decay3d_mag_flx_pres_Re10000_t03} (a) shows the distribution 
of the magnetic flux density at time $t/(L/U) = 0.3$. 
The $y$--$z$, $z$--$x$, and $x$--$y$ cross-sections show the distribution 
of magnetic flux densities in the $x$-, $y$-, and $z$-directions, respectively. 
At this time, the distinct periodicity of the magnetic flux density remains. 
Additionally, Figs. \ref{decay3d_mag_flx_pres_Re10000_t03} (b) and (c) show 
the magnetic flux lines, magnetic pressure, 
magnitude of the current density vector, second invariant of the velocity gradient tensor, 
and curvature of the isosurface of magnetic pressure. 
Here, the curvature of the isosurface of magnetic pressure is calculated 
to visualize the region of low magnetic pressure. 
The red isosurfaces show the distributions of magnetic pressure $P = 0.01$ and $P = 0.28$, 
and the light blue isosurface expresses the current density magnitude $j = 7.5$, 
indicating the occurrence of high current density. 
The silver isosurface shows the second invariant $Q = -52$ of the velocity gradient tensor 
and represents a tubular high-shear region. 
The green isosurface shows the curvature $\kappa = -50$ of the magnetic pressure isosurface, 
confirming the region of low magnetic pressure. 
The high magnetic pressure is shown in Fig. \ref{decay3d_mag_flx_pres_Re10000_t03} (b), 
and the low magnetic pressure and the magnitude of high current density 
are shown in Fig. \ref{decay3d_mag_flx_pres_Re10000_t03} (c). 
In Fig. \ref{decay3d_mag_flx_pres_Re10000_t03} (b), 
the magnetic pressure distribution of a distorted cubic structure 
appears so that the regions of low magnetic pressure are connected 
in a mesh pattern and the cubic structure surrounds the areas of high magnetic pressure. 
At this time, the attenuation of the velocity field is small; 
thus, a clear vortex structure is present. 
The magnetic flux density decays, but the magnetic flux lines are similar to 
the streamlines of the velocity field. 
The distribution of magnetic pressure shown in Fig. \ref{decay3d_mag_flx_pres_Re10000_t03} (c) is Y-shaped 
and the same as the shape of the pressure distribution in Fig. \ref{decay3d_flow_Re10000}. 
The dimensionless magnetic pressure corresponds to the dimensionless magnetic energy; 
therefore, the magnetic energy becomes high in the region 
where the magnetic pressure is high. 
The high current density occurs in a grid pattern, 
and the magnetic flux lines swirl to surround the high current density region. 
In the area of high current density, the magnetic pressure, 
namely, the magnetic energy, becomes high.

%------------------------------------------------------------------------------
% Figure 27
%------------------------------------------------------------------------------
\begin{figure}[!t]
\begin{minipage}{0.325\linewidth}
\begin{center}
\includegraphics[trim=0mm 0mm 0mm 0mm, clip, width=50mm]{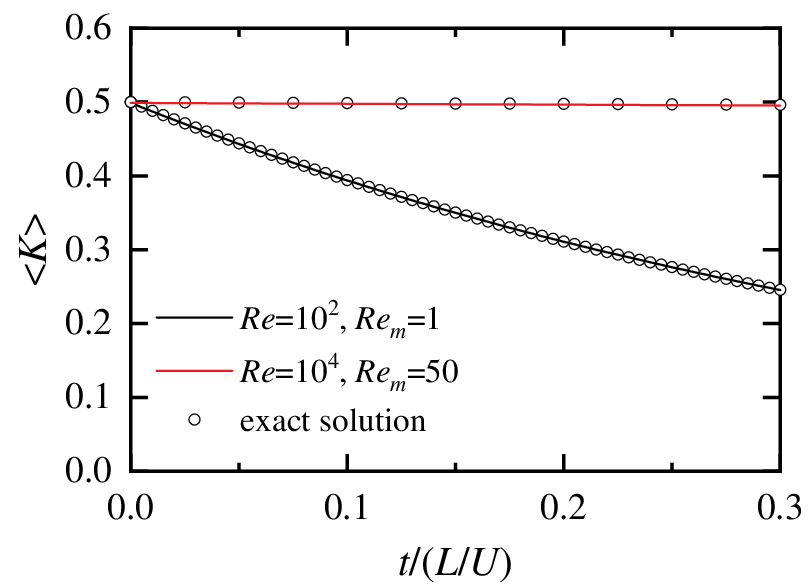} \\
{\small (a) $K$}
\end{center}
\end{minipage}
\begin{minipage}{0.325\linewidth}
\begin{center}
\includegraphics[trim=0mm 0mm 0mm 0mm, clip, width=50mm]{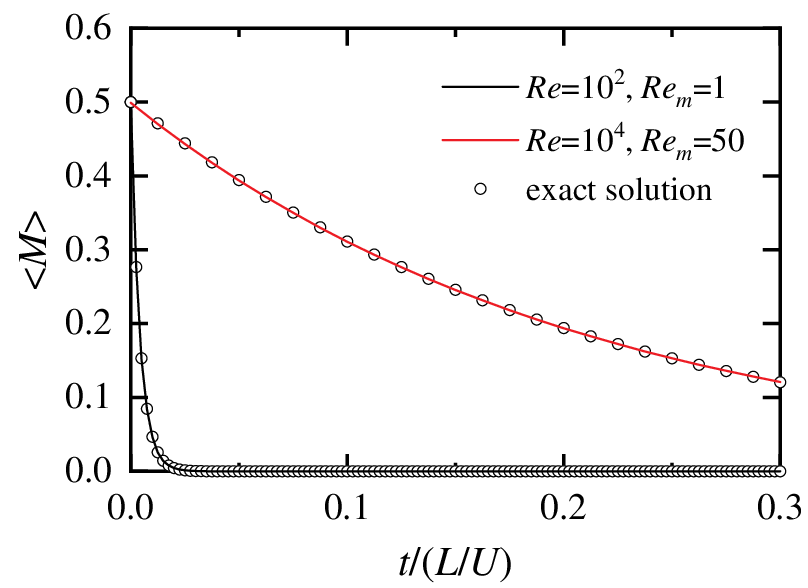} \\
{\small (b) $M$}
\end{center}
\end{minipage}
\begin{minipage}{0.325\linewidth}
\begin{center}
\includegraphics[trim=0mm 0mm 0mm 0mm, clip, width=50mm]{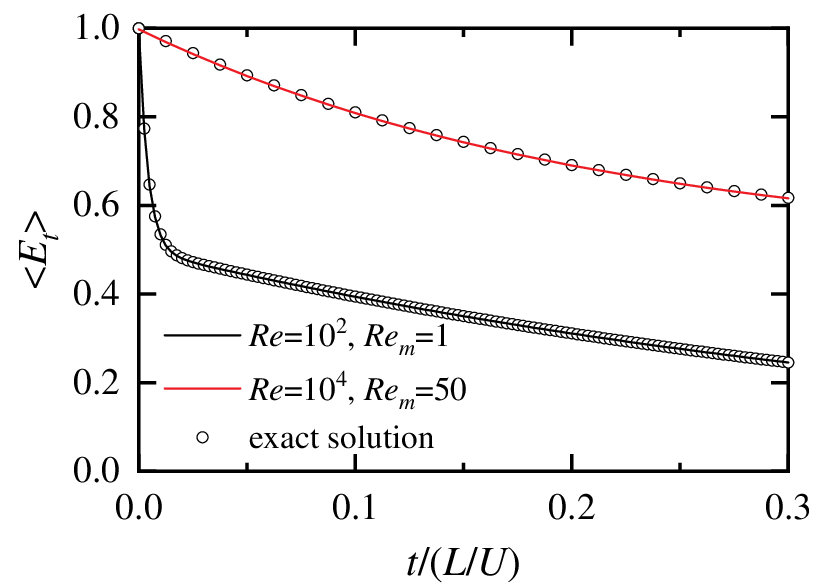} \\
{\small (c) $E_t$}
\end{center}
\end{minipage}

\vspace*{0.5\baselineskip}
\begin{minipage}{0.48\linewidth}
\begin{center}
\includegraphics[trim=0mm 0mm 0mm 0mm, clip, width=50mm]{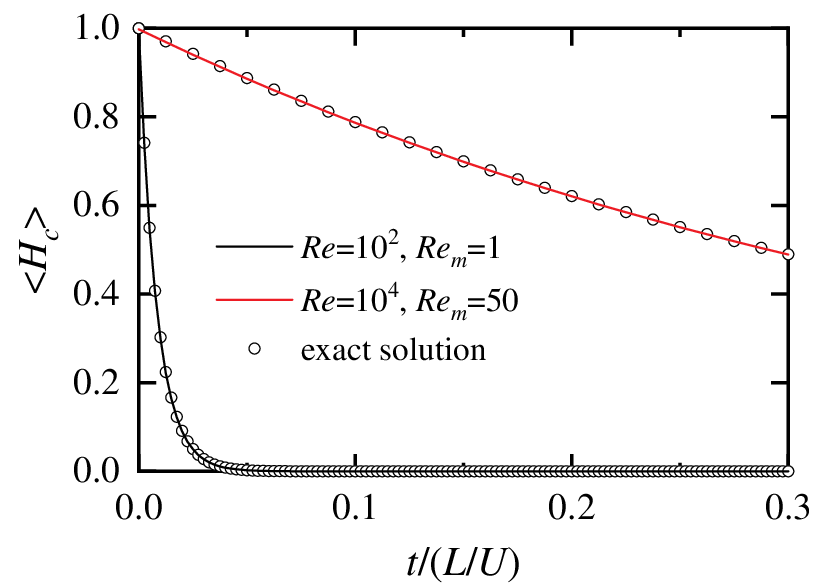} \\
{\small (d) $H_c$}
\end{center}
\end{minipage}
\begin{minipage}{0.48\linewidth}
\begin{center}
\includegraphics[trim=0mm 0mm 0mm 0mm, clip, width=50mm]{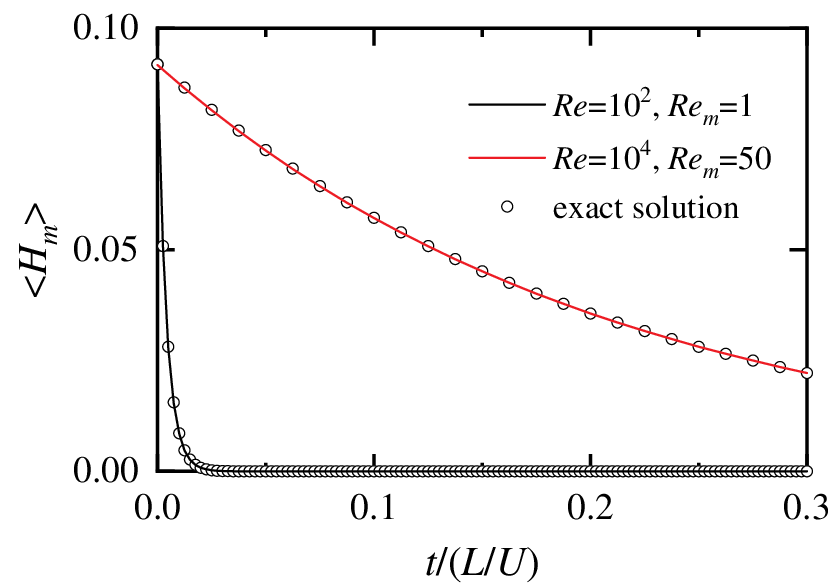} \\
{\small (e) $H_m$}
\end{center}
\end{minipage}
\caption{Time variations of kinetic energy ($K$), magnetic energy ($M$), 
total energy ($E_t$), cross-helicity ($H_c$), and magnetic helicity ($H_m$): 
$Re = 10^2$, $Re_m = 1$ and $Re = 10^4$, $Re_m = 50$.}
\label{decay3d_error_mag_energy}
\end{figure}
%------------------------------------------------------------------------------

Figure \ref{decay3d_error_mag_energy} shows the time variations in the kinetic, magnetic, 
and total energies, cross-helicity, and magnetic helicity for $Re = 10^4$ and $Re_m = 50$. 
For comparison, the results for $Re = 10^2$ and $Re_m = 1$ are also included. 
Under $Re = 10^4$ and $Re_m = 50$, there is almost no attenuation of the kinetic energy. 
Conversely, magnetic energy, total energy, cross-helicity, 
and magnetic helicity decay over time. 
For $Re = 10^2$, the magnetic energy decays early; 
moreover, the cross-helicity and magnetic helicity also decay rapidly. 
This approximate solution supports the exact solution, 
and the decaying process for energy and magnetic helicity is accurately captured.

\begin{figure}[!t]
\begin{minipage}{0.48\linewidth}
\begin{center}
\includegraphics[trim=0mm 0mm 0mm 0mm, clip, width=70mm]{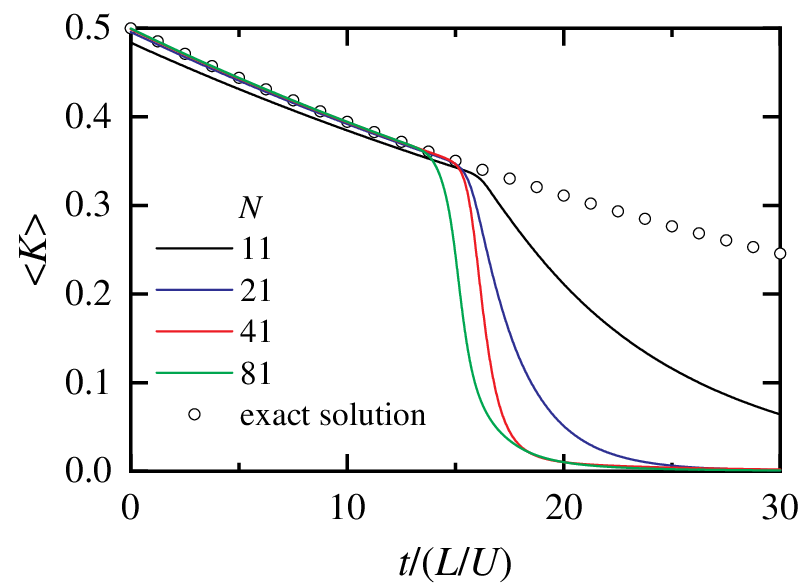} \\
{\small (a) $K$}
\end{center}
\end{minipage}
\hspace{0.02\linewidth}
\begin{minipage}{0.48\linewidth}
\begin{center}
\includegraphics[trim=0mm 0mm 0mm 0mm, clip, width=70mm]{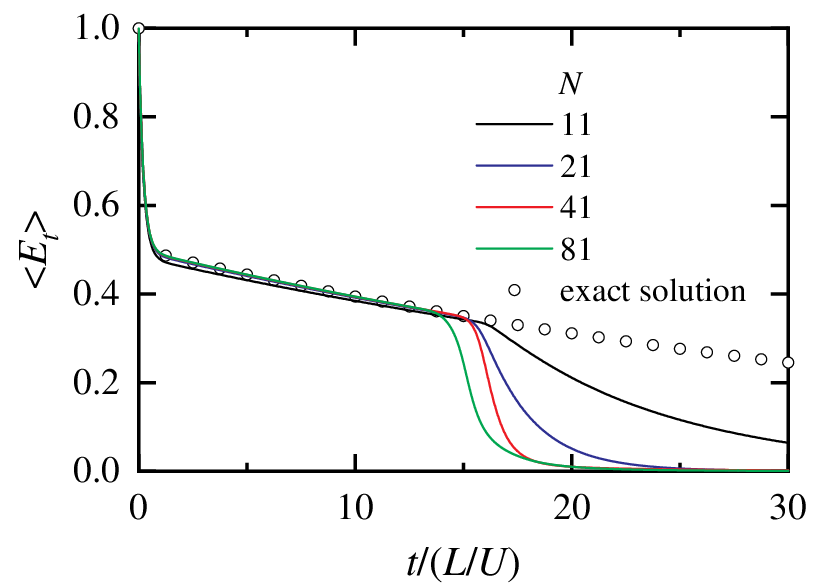} \\
{\small (b) $E_t$}
\end{center}
\end{minipage}
\caption{Time variations of kinetic energy ($K$) and totale energy ($E_t$) 
under applied magnetic field: $Re = 10^4$, $Re_m = 50$.}
\label{decay3d_error_mag_Re10000}
\end{figure}

Figure \ref{decay3d_error_mag_Re10000} shows the time variations in the kinetic 
and total energies for $Re = 10^4$ and $Re_m = 50$. 
Over time, the difference between the result obtained using each grid and 
the exact solution becomes apparent, 
and the kinetic and total energies decay sharply. 
Antuono \citep{Antuono_2020} reported that the difference between 
the analytic solution and calculation value suggested a transition to turbulent flow. 
When the flow transition occurred, the magnetic energy had sufficiently decreased. 
Therefore, under this condition, 
the influence of the magnetic field on the flow transition is considered minimal. 
As the number of grid points increases, 
the transition points approach the dimensionless time $t/(L/U) = 16$. 
However, these results do not show monotonic convergence with increasing grid points.

%------------------------------------------------------------------------------
% Figure 28
%------------------------------------------------------------------------------
%------------------------------------------------------------------------------

\begin{figure}[!t]
\begin{minipage}{0.48\linewidth}
\begin{center}
\includegraphics[trim=0mm 0mm 0mm 0mm, clip, width=65mm]{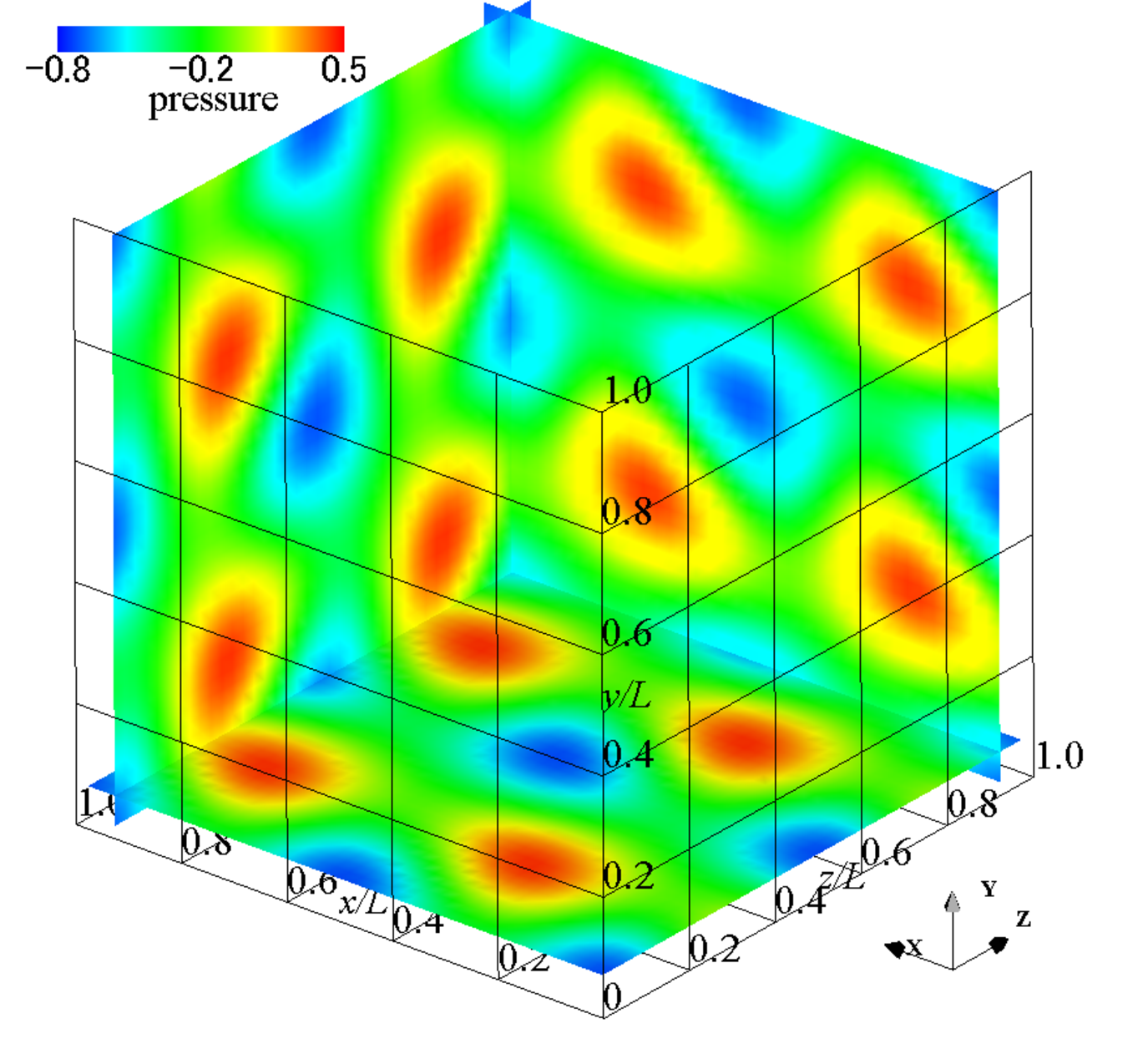} \\
{\small (a) Initial pressure}
\end{center}
\end{minipage}
\hspace{0.02\linewidth}
\begin{minipage}{0.48\linewidth}
\begin{center}
\includegraphics[trim=0mm 0mm 0mm 0mm, clip, width=65mm]{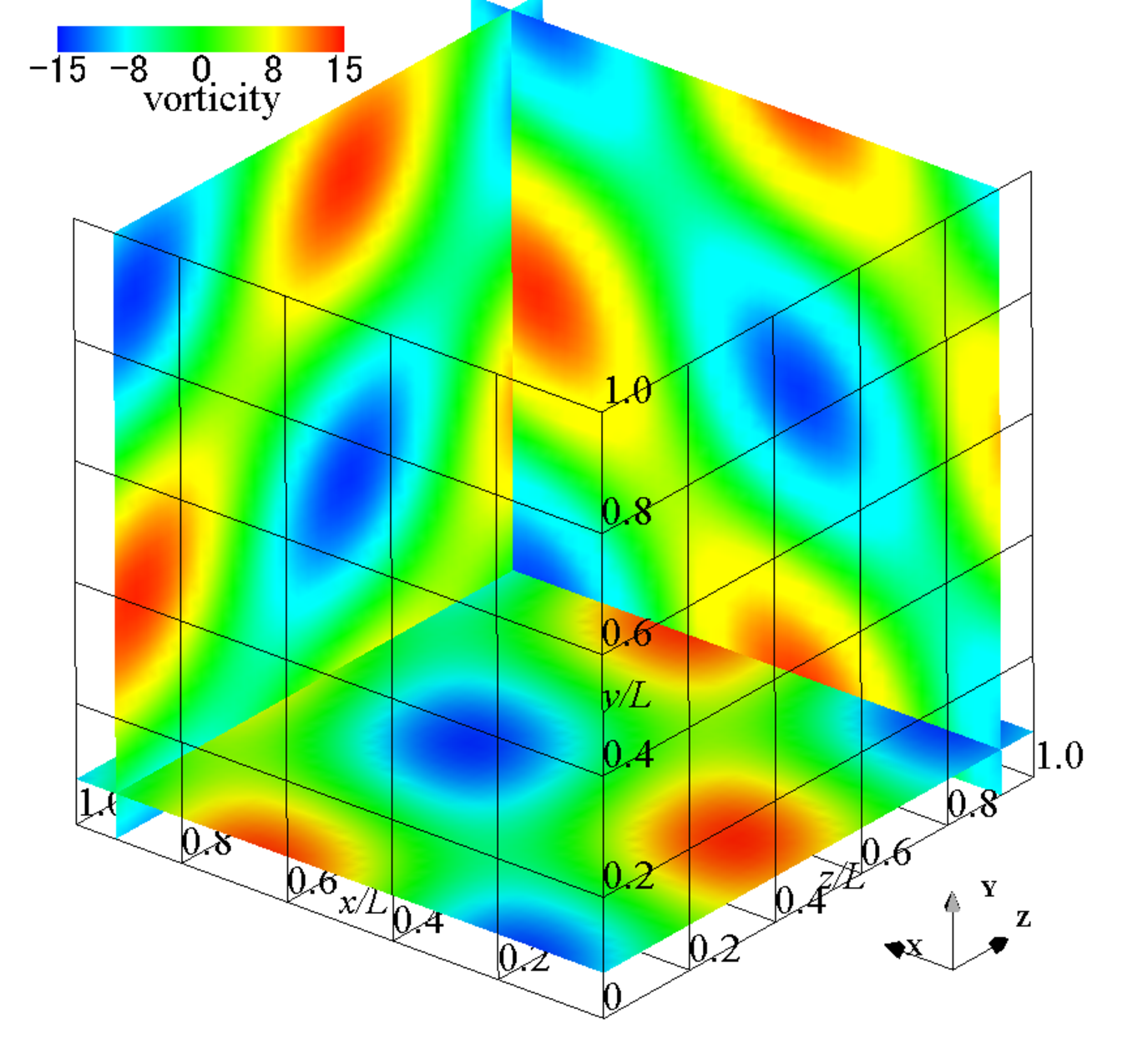} \\
{\small (b) Initial vorticity}
\end{center}
\end{minipage}
\begin{minipage}{0.48\linewidth}
\begin{center}
\includegraphics[trim=0mm 0mm 0mm 0mm, clip, width=65mm]{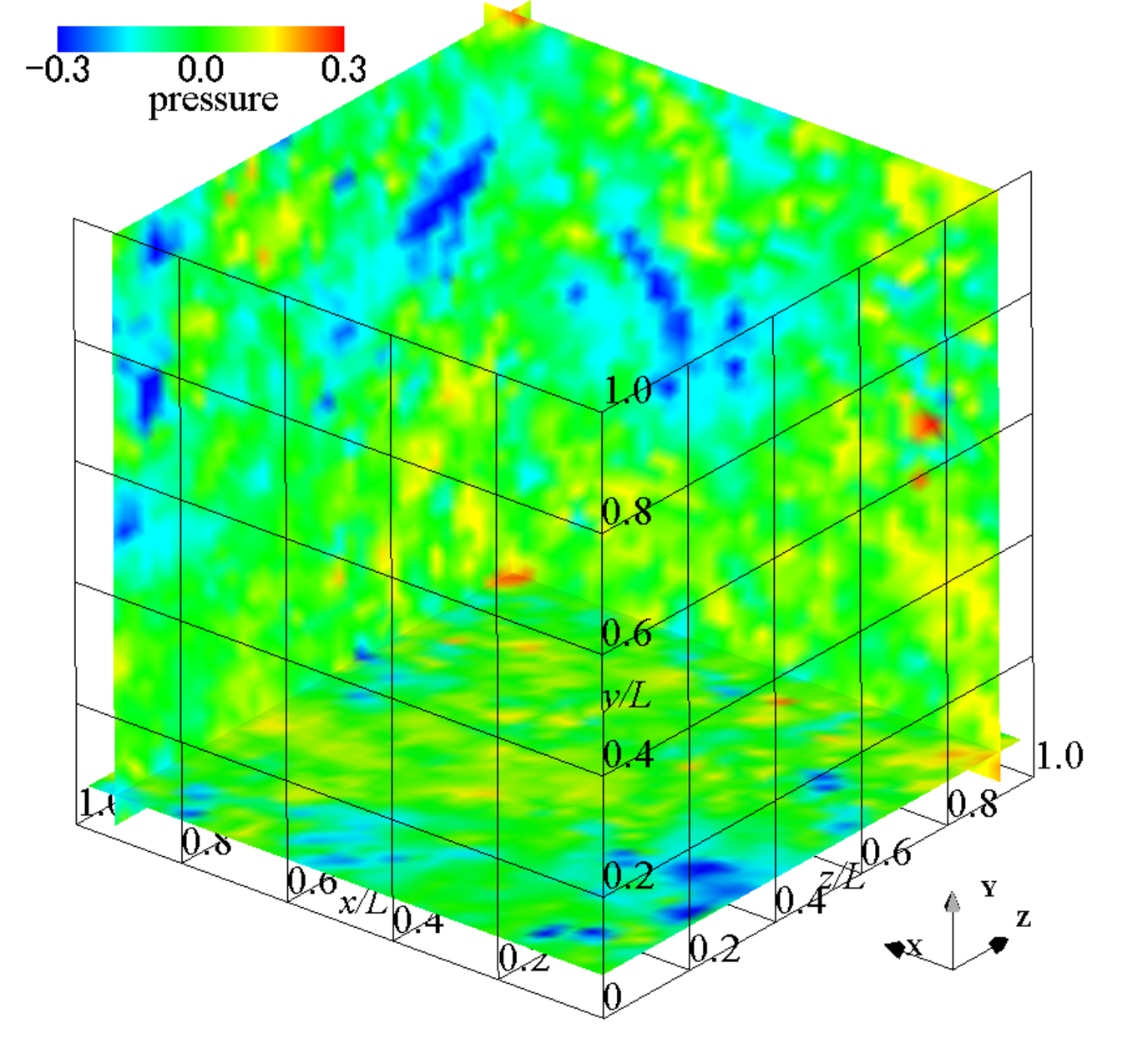} \\
{\small (c) Pressure}
\end{center}
\end{minipage}
\hspace{0.02\linewidth}
\begin{minipage}{0.48\linewidth}
\begin{center}
\includegraphics[trim=0mm 0mm 0mm 0mm, clip, width=65mm]{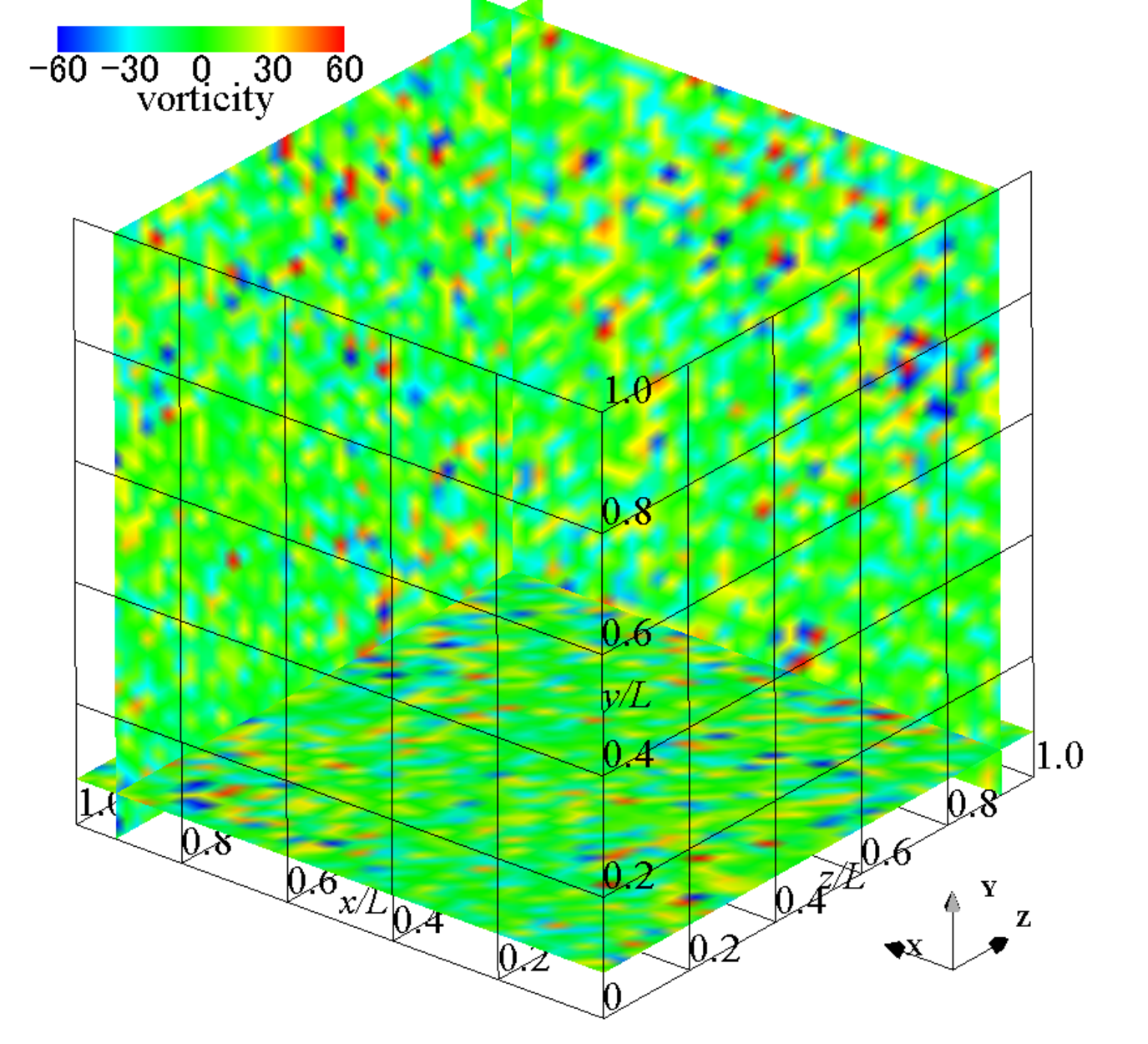} \\
{\small (d) Vorticity}
\end{center}
\end{minipage}
%
%\vspace*{-1.2\baselineskip}
\caption{Contours of pressure and vorticities under applied magnetic field: 
$Re = 10^4$, $Re_m = 50$, $t/(L/U) = 0$, 16.05: 
The $y$--$z$, $z$--$x$, and $x$--$y$ planes show the contours of 
vorticities in the $x$-, $y$-, and $z$-directions, respectively.}
\label{decay3d_pre_vor_mag_Re10000}
\end{figure}

The vorticity distribution at time $t/(L/U) = 0$ and $16.05$ is shown 
in Fig. \ref{decay3d_pre_vor_mag_Re10000}. 
The $y$--$z$, $z$--$x$, and $x$--$y$ cross-sections show the vorticity distributions 
in the $x$-, $y$-, and $z$-directions, respectively. 
In the initial state, a large-scale vortex exists; 
however, at $t/(L/U) = 16.05$, 
it is converted to small-scale vortex structures by a nonlinear effect 
and attenuated.

%------------------------------------------------------------------------------
% Figure 29
%------------------------------------------------------------------------------
%------------------------------------------------------------------------------

%------------------------------------------------------------------------------
% Figure 30
%------------------------------------------------------------------------------
\begin{figure}[!t]
\begin{center}
\includegraphics[trim=0mm 0mm 0mm 0mm, clip, width=70mm]{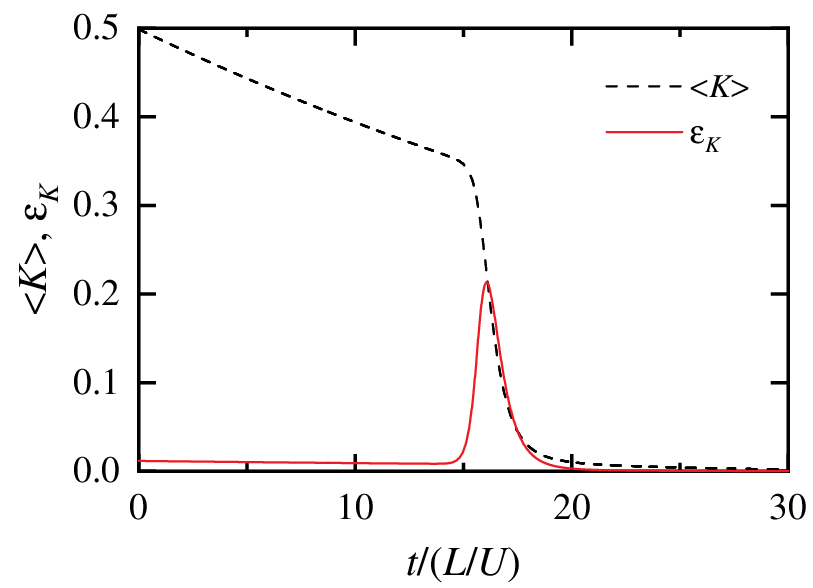}
\end{center}
\caption{Time variations of kinetic energy and energy dissipation
under applied magnetic field: $Re = 10^4$, $Re_m = 50$.}
\label{decay3d_energy_dissipation}
\end{figure}
%------------------------------------------------------------------------------

Figure \ref{decay3d_energy_dissipation} shows the time variation of 
the kinetic energy $\langle K \rangle$ 
and its dissipation rate $\varepsilon_K$. 
The dissipation rate is defined as $\varepsilon_K = -d \langle K \rangle/dt$. 
As the kinetic energy decays sharply, 
the dissipation rate increases and a maximum value appears. 
The pressure and vorticity distributions 
when the dissipation rate is maximum are shown 
in Figs. \ref{decay3d_pre_vor_mag_Re10000} (c) and (d). 
The vortex structure disappears with time owing to viscous dissipation, 
the induced magnetic field disappears, 
and the flow field asymptotically approaches the stationary state.

The maximum divergence errors of velocity, magnetic flux density, 
and magnetic vector potential in viscous analysis are $4.16\times 10^{-14}$, 
$3.50\times 10^{-12}$, and $1.30\times 10^{-15}$, at time $t/(L/U) = 0.3$, respectively.

%##############################################################################
\section{Conclusion}
\label{summary}
%##############################################################################

In this study, a method to simultaneously relax 
the magnetic vector and electric potentials for incompressible MHD flows was proposed 
using a conservative finite difference method 
that discretely conserves total energy. 
The conservation properties of magnetic helicity were also investigated 
in this numerical method. 
The results indicated that the equations of total energy, cross-helicity, 
and magnetic helicity could be discretely derived 
from the equations of momentum, magnetic flux density, and magnetic vector potential. 
In this numerical method, the Lorentz force is discretized 
to maintain the transformation  between conservative and nonconservative forms. 

Five types of flow models were analyzed 
and the accuracy and convergence of the proposed method were verified. 
The computational approach of the magnetic vector 
and electric potentials was further verified. 
In analyzing one- and two-dimensional flow models and Hartmann flow, 
the calculation accuracy and convergence were demonstrated 
by comparing the calculated results and exact solutions. 
Consequently, the validity of the numerical method was proved. 
Unsteady analyses of two- and three-dimensional decaying vortices were performed. 
The results showed that excellent conservation properties of total energy and cross-helicity 
were obtained in the ideal periodic inviscid MHD flow. 
Magnetic helicity was discretely preserved even in three-dimensional flow. 
For the ideal inviscid MHD flow in the three-dimensional decaying vortex model, 
the total amount of magnetic vector potential was kept at zero analytically 
but increased with time. 
The attenuation trends of the total energy, cross-helicity, 
and magnetic helicity in viscous flow supported the exact solution. 
Thus, the numerical method accurately captured the trend of decaying energy.

In conclusion, a numerical method for simultaneously relaxing 
velocity, pressure, density, and internal energy was developed. 
The present numerical method can be applied to compressible MHD flows. 
In the future, the effectiveness of this numerical method 
for compressible MHD flows at low Mach numbers will be studied.

%##############################################################################
\section*{Acknowledgment}
%##############################################################################

This research did not receive any specific grants from funding agencies 
in the public, commercial, or not-for-profit sectors. 
The author wishes to acknowledge the time and effort of everyone involved in this study.
I would like to thank Editage (www.editage.com) for English language editing.

%##############################################################################
%\section*{References}
%##############################################################################

%\nocite{*}
\bibliography{reference_mhd_flow_bibfile}
\end{document}